\documentclass[twocolumn]{aastex63}
\usepackage{amsmath}
\usepackage{graphicx}
\usepackage{appendix}
\usepackage{comment}
\usepackage{tabularx}
\usepackage{booktabs}
\usepackage{comment}
\usepackage{xcolor}
\usepackage{hyperref}
\usepackage{lineno}

\newcommand{\h}{\mathbf{h}}
\newcommand{\Msun}{\ensuremath{M_\odot}}
\newcommand{\Mzams}{\ensuremath{M_{\rm ZAMS}}}
\newcommand{\code}[1]{\texttt{#1}}
\newcommand{\mesa}{\code{MESA}}
\newcommand{\MESA}{\mesa}

\newlength{\apjcolwidth}
\newlength{\figwidth}
\newlength{\doublewide}
\begin{document}

\title{Observing intermediate-mass black holes and the upper--stellar-mass gap with LIGO and Virgo}

\shorttitle{Intermediate-mass black holes and the upper--stellar-mass gap}
\shortauthors{Mehta et al.}

\author[0000-0002-7351-6724]{Ajit Kumar Mehta}
\affiliation{Max Planck Institute for Gravitational Physics (Albert Einstein Institute), Am M\"uhlenberg 1, Potsdam 14476, Germany}

\author[0000-0002-5433-1409]{Alessandra Buonanno}
\affiliation{Max Planck Institute for Gravitational Physics (Albert Einstein Institute), Am M\"uhlenberg 1, Potsdam 14476, Germany}
\affiliation{Department of Physics, University of Maryland, College Park, MD 20742-2421, USA}

\author[0000-0002-1671-3668]{Jonathan Gair}
\affiliation{Max Planck Institute for Gravitational Physics (Albert Einstein Institute), Am M\"uhlenberg 1, Potsdam 14476, Germany}

\author[0000-0002-2666-728X]{M. Coleman Miller}
\affiliation{Department of Astronomy and Joint Space-Science Institute, University of Maryland, College Park, MD 20742-2421, USA}

\author[0000-0002-5794-4286]{Ebraheem Farag}
\affiliation{School of Earth and Space Exploration, Arizona State University, Tempe, AZ 85287, USA}
\affiliation{Joint Institute for Nuclear Astrophysics - Center for the Evolution of the Elements, USA}

\author[0000-0003-3784-6360]{R. J. deBoer}
\affiliation{Department of Physics, University of Notre Dame, Notre Dame, Indiana 46556, USA}
\affiliation{Joint Institute for Nuclear Astrophysics - Center for the Evolution of the Elements, USA}

\author[0000-0002-3409-3319]{M. Wiescher}
\affiliation{Department of Physics, University of Notre Dame, Notre Dame, Indiana 46556, USA}
\affiliation{Joint Institute for Nuclear Astrophysics - Center for the Evolution of the Elements, USA}

\author[0000-0002-0474-159X]{F.X.~Timmes}
\affiliation{School of Earth and Space Exploration, Arizona State University, Tempe, AZ 85287, USA}
\affiliation{Joint Institute for Nuclear Astrophysics - Center for the Evolution of the Elements, USA}

\correspondingauthor{Ajit {Kumar} Mehta}
\email{ajit.mehta@aei.mpg.de}

% abstract is 250 word max

\begin{abstract}

Using ground-based gravitational-wave detectors, we probe the mass function 
of intermediate-mass black holes (IMBHs) wherein we also include BHs in the 
upper mass gap $\sim 60-130~M_\odot$. Employing the projected sensitivity 
of the upcoming LIGO and Virgo fourth observing (O4) run, we perform Bayesian
analysis on quasi-circular non-precessing, spinning IMBH binaries
(IMBHBs) with total masses $50\mbox{--} 500\, M_\odot$, mass ratios
1.25, 4, and 10, and dimensionless spins up to 0.95, and
estimate the precision with which the source-frame parameters can be
measured.  We find that, at $2\sigma$, the mass of the
heavier component of IMBHBs can be constrained with an uncertainty
of $\sim 10-40\%$ at a signal-to-noise ratio of $20$. 
Focusing on the stellar-mass gap with new tabulations of the
$^{12}\text{C}(\alpha, \gamma)^{16} \text{O}$ reaction
rate and its uncertanties, we evolve massive helium core stars
using \MESA\, to establish the lower and upper edge of the mass gap
as $\simeq$\,59$^{+34}_{-13}$\,$M_{\odot}$ and
$\simeq$\,139$^{+30}_{-14}$\,$M_{\odot}$ respectively, where the error
bars give the mass range that follows from the $\pm 3\sigma$
uncertainty in the $^{12}\text{C}(\alpha, \gamma) ^{16} \text{O}$
nuclear reaction rate. We find that high resolution of the tabulated reaction 
rate and fine temporal resolution are necessary to resolve the peak of the BH mass spectrum.
We then study IMBHBs with components lying in the mass gap and show
that the O4 run will be able to robustly identify most such
systems. Finally, we re-analyse GW190521 with a
state-of-the-art aligned-spin waveform model, finding that the
primary mass lies in the mass gap with 90\% credibility.

\end{abstract}

% UAT concepts
         
\section{Introduction} \label{sec:intro}

The LIGO and Virgo detectors \citep{2015CQGra..32g4001L,2015CQGra..32b4001A}  have opened the gravitational-wave (GW) window onto the universe, 
reporting, as of today, 48 GW signals from binary black-hole (BH) mergers~\citep{O1-O2catalog,O3acatalog2020}. 
They have also opened the era of multi-messenger astronomy with GWs, shedding light on the origin 
of short-hard gamma-ray bursts with the observation of a coalescing binary neutron star~\citep{GW170817,GBM:2017lvd}. In addition, 
independent claims of GW observations have also been made~\citep{Nitz:2018imz,Nitz:2019hdf,Venumadhav:2019lyq,Zackay:2019btq}. Detections of these compact-object binaries have allowed us to probe various problems pertaining to astrophysics,
astronomy and cosmology, such as measuring the Hubble
parameter~\citep{Abbott:2017xzu}, establishing the rates and
population of compact binaries~\citep{Abbott:2020gyp}, and constraining the neutron-star radius and equation of
states~\citep{BNS2017,O3acatalog2020}. The detections of binary black-hole (BBH)
mergers with masses $\gtrsim 40~M_\odot$, in particular, can allow us to probe the
physical processes that are involved in the evolution of massive
stars --- for example the ${}^{12}\text{C}(\alpha, \gamma) {}^{16} \text{O}$
nuclear rate that significantly affects the mass of a BH formed through the collapse of a 
massive star~\citep{Brown:2001ua,woosley_2002_aa,farmer_2020_aa,Woosley:2021xba}.

\begin{figure*}[!htb] 
\centering
\includegraphics[width=1\apjcolwidth]{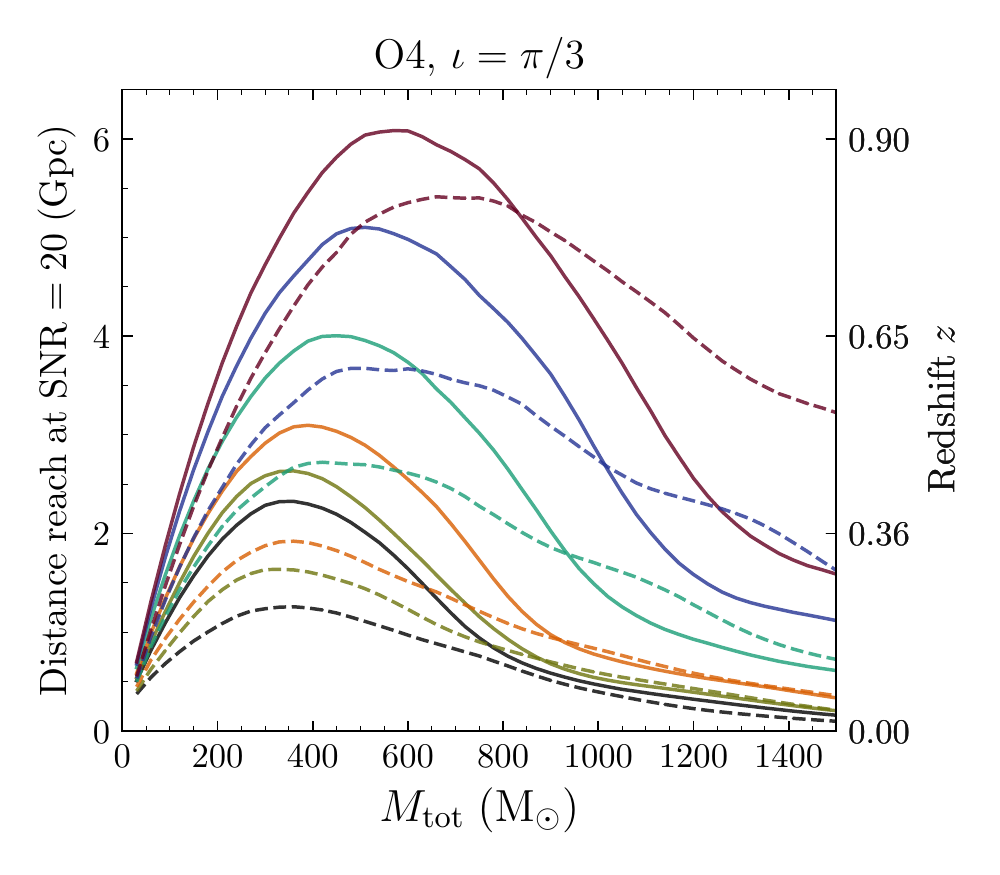}
\includegraphics[width=1\apjcolwidth]{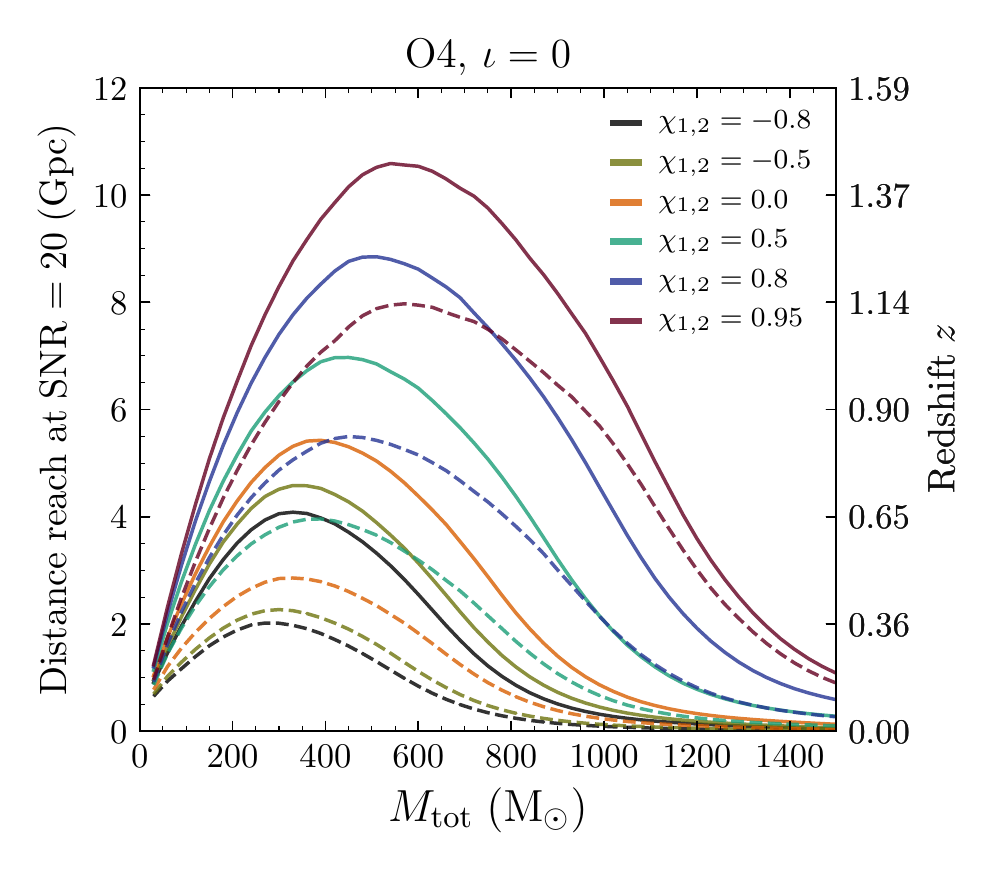}
\vspace*{-3mm}
\caption{The distance reach as a function of the total mass $M_{\rm{tot}}$ for spinning, non-precessing IMBHBs with mass ratio $q = 1.25$ (solid lines) and $q=4$ (dashed lines) at inclination angle $\iota = \pi/3$ (left panel) and $\iota = 0$ (right panel), and $\text{SNR}=20$. The different curves correspond to different values of the component spins $(\chi_{1}, \chi_{2})$, assumed to 
 be equal for the two BHs. We use the noise spectral densities expected for the upcoming O4 run~\citep{Abbott:2020qfu} and the spinning, non-precessing ${\tt SEOBNRHM}$ waveform model. The distance reach shown here is computed by averaging over the antenna pattern functions (see Equation \ref{eq:strain}), that is the angles that specify the location of the source in the sky and the polarization angle. The maximum redshift (distance reach) for IMBHBs with inclination $\iota=0$ (face-on) and $\pi/3$ (near edge-on) is $z\sim1.4$ ($11$ Gpc) and $\sim 0.9$ ($6$ Gpc), respectively.}
\label{fig:dL_reach_O4}
\end{figure*}

\begin{figure*}[!htb] 
\centering
\includegraphics[width=1\apjcolwidth]{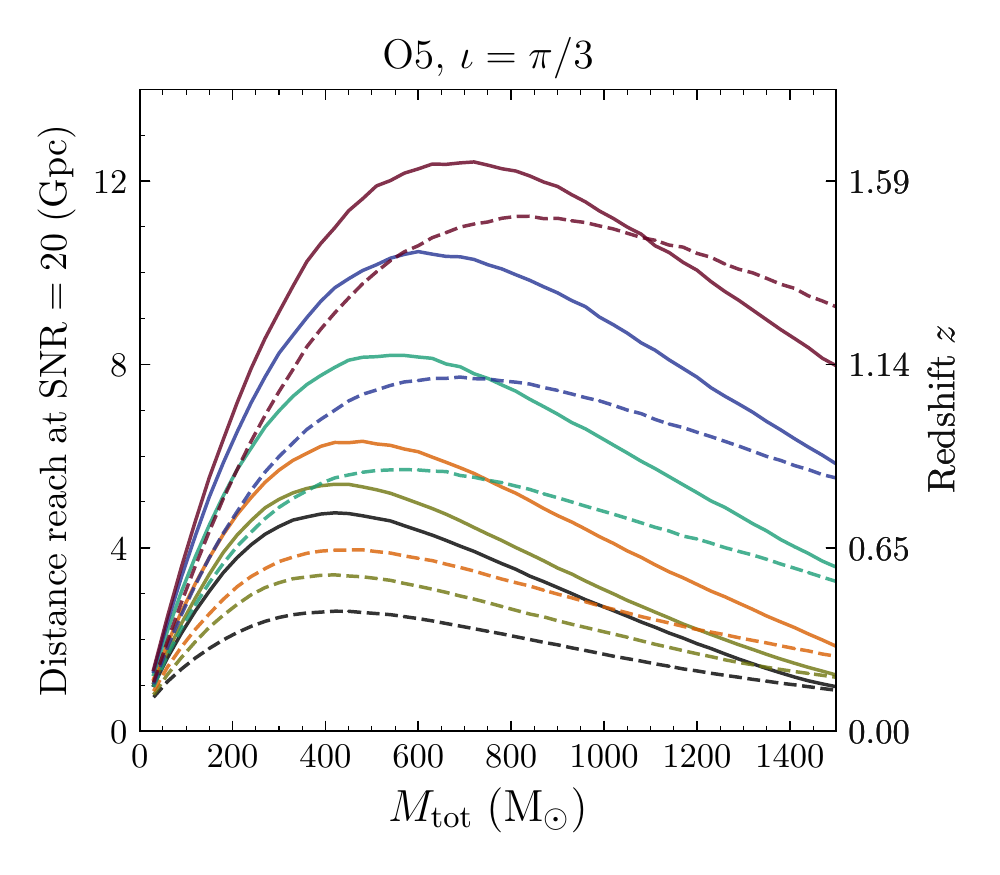}
\includegraphics[width=1\apjcolwidth]{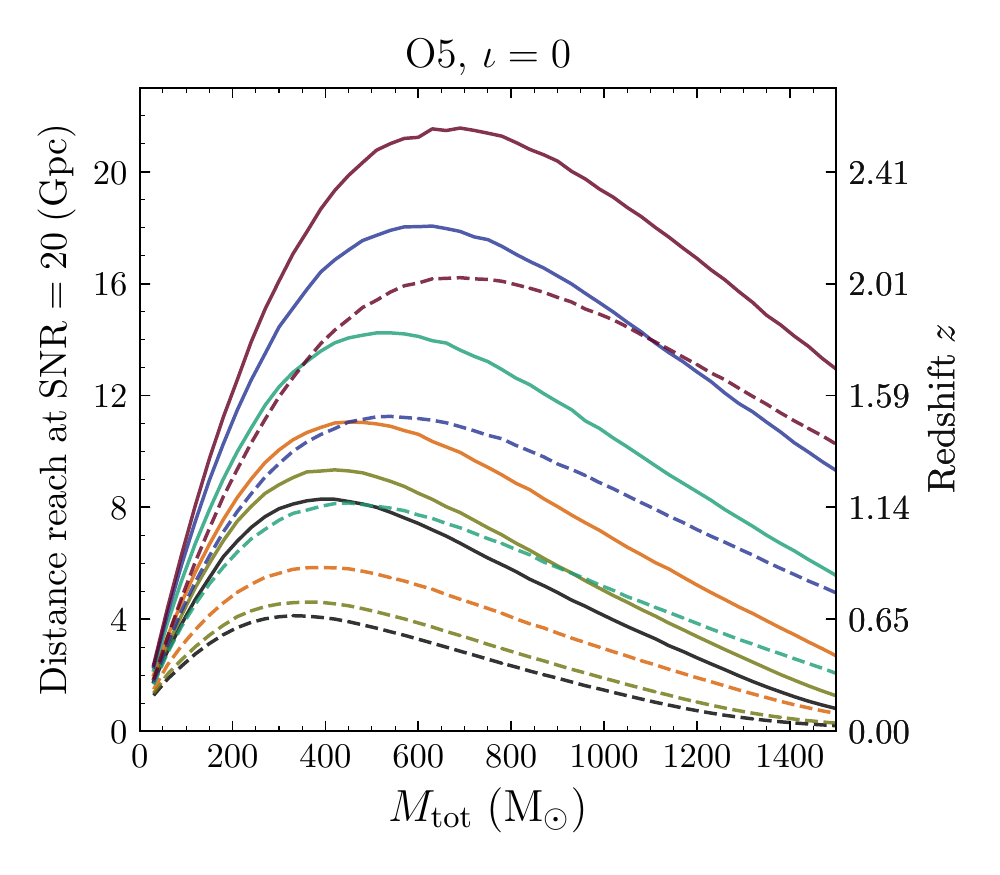}
\vspace*{-3mm}
\caption{Same as Figure~\ref{fig:dL_reach_O4} but for the noise spectral densities of O5 run~\citep{Abbott:2020qfu}. The maximum redshift (distance reach) for IMBHBs with inclination $\iota=0$ (face-on) and $\pi/3$ (near edge-on) is $z\sim2.5$ ($20.9$ Gpc) and $\sim 1.6$ ($12.1$ Gpc), respectively.}
\label{fig:dL_reach_O5}
\end{figure*}

The GW detections during the first and second observing (O1 and O2) runs~\citep{O1-O2catalog} revealed a
population of BBHs with component source masses $\lesssim 50 M_{\odot}$
and total source mass $\lesssim 84M_{\odot}$. These component masses are 
mostly consistent with the definition of stellar-mass BHs.  However, in the first half of the third (O3a) observing run, an event (GW190521)~\citep{Abbott:2020tfl,Abbott:2020mjq} 
was detected with a pre-merger binary total source mass of $\sim
150M_{\odot}$ and a remnant source mass of $\sim 140M_{\odot}$. The best
estimates of the component BH source masses, obtained with quasi-circular spinning, precessing 
waveforms, are $\sim 85 M_{\odot}$ and $\sim
65 M_{\odot}$~\citep{Abbott:2020tfl}. The remnant of this GW event falls in the category of 
intermediate-mass black holes (IMBHs), which are usually
defined as BHs with mass between $\sim 10^{2}\mbox{--}10^{5}
M_{\odot}$. The latter is not a strict definition.  Because we are also interested in studying BHs in the upper mass gap (i.e., the gap produced by pair instability supernovae), we define IMBHs as those with masses above the lower edge of this mass gap.  As we show in Section 4.3, that lower edge is at $\sim 60~M_\odot$, and thus for our purposes BHs with masses $M\gtrsim 60~M_\odot$ are IMBHs.
%The latter is not a strict definition, and since 
%here we are also interested in studying BHs in the upper--stellar-mass gap (i.e., the pair-instability--supernova mass gap), we adopt a 
%threshold of $M\gtrsim 60~M_\odot$ for IMBHs. 
Minimal-assumption LIGO-Virgo pipelines dedicated to searches for IMBHs did not report any detection except for GW190521, 
so far~\citep{Abbott:2017iws,Salemi:2019ovz}. 

We note that \cite{Nitz_2021} have recently found evidence, using a waveform model 
that was not employed in \cite{Abbott:2020tfl,Abbott:2020mjq}, that the source-mass 
posterior distributions of GW190521 are multi-modal, opening the possibility that GW190521 
was an intermediate mass-ratio binary (i.e., a binary with mass ratio around 10). We will comment on their analysis, and we re-analyse this GW event  as well in Sec.~\ref{GW90521_analysis} below. A similar conclusion was drawn by~\cite{Fishbach_2020}, in which the authors employed a different, namely, population-informed prior on the secondary mass rather than a uniform prior on the secondary mass. Given the very short signal, other analyses in the literature~\citep{Romero-Shaw:2020thy, CalderonBustillo:2020odh, Gayathri190521, 2021PhRvL.126h1101B}  pointed out the importance of re-analysing GW190521 
with waveform models of eccentric compact binaries, which, however, are not yet available.

IMBHs are difficult to observe. Indeed, there is no definitive electromagnetic evidence for their existence. Thus, their formation
channels and mass function are highly uncertain~\citep{2004IJMPD131M,  AmaroSeoane:2007aw, Gair_2011, Belczynski_2014}. GW observations 
have the potential to solve these mysteries, by providing
accurate measurements of their properties, such as their masses, spins, and location.

\cite{Graff:2015bba,Veitch:2015ela,Haster:2015cnn} studied the precision with which IMBH masses and 
spins could be measured with LIGO and Virgo detectors. \cite{Graff:2015bba} 
used multipolar waveform models that describe the entire coalescence process of non-spinning IMBHBs, 
and performed a Bayesian analysis to estimate the precision with which the 
parameters of IMBHBs can be estimated with LIGO detectors. \cite{Veitch:2015ela} employed spinning, non-precessing waveforms, 
but did not include subdominant modes; these are relevant for high total-mass binaries, 
because they break degeneracies between parameters and reduce the measurement uncertainties. 
Here we extend these analyses in several directions. We consider state-of-the-art multipolar spinning, non-precessing models for gravitational waves from IMBHBs, including the five strongest gravitational modes, and explore a larger region of the parameter space (e.g., mass ratios 
$1-10$ and dimensionless spin values up to $0.95$). We focus on masses in the source frame rather than the detector frame, since the former provides us with information about the upper--stellar-mass 
gap and, more generally, about the IMBH mass function. We employ for our study the projected 
noise-spectral densities~\citep{Abbott:2020qfu} of the upcoming fourth observing (O4) run (scheduled to start not earlier than the second half of 2022). We also comment on results that could be obtained during the fifth observing (O5) run~\citep{Abbott:2020qfu} (expected to start in 2025). 

Inference on the IMBH population requires not only accurate parameter measurements, but also a sufficiently high rate of observations of IMBHB mergers in upcoming LIGO and Virgo runs (see \cite{Ezquiaga:2020tns} for a study {when the BH masses} are above $120 M_{\odot}$). Based on the observation of one event, GW190521, \cite{Abbott:2020mjq} estimated an astrophysical merger 
rate of ${0.13}_{-0.11}^{+0.30}\,{\rm Gpc}^{-3}\,{\rm yr}^{-1}$. In Figure~\ref{fig:dL_reach_O4}, we show the distance reach
of the LIGO-Virgo--detector network expected during the O4 run to binaries with signal-to-noise (SNR) of $20$. We display results using multipolar spinning, non-precessing 
waveforms, for a variety of spin values and for mass ratios of 1.25 and 4, and binary inclinations $\iota=0$ (face-on) and  $\iota=\pi/3$ (close to edge-on). As we can see, the distance reach for face-on binaries even at SNR=20 could go up to a redshift $z\sim1.4$ ($11$ Gpc), while for near edge-on binaries this reduces a bit but sources at a redshift of $\sim 0.9$ ($6$ Gpc) can still be probed. {When combined with the measured astrophysical rate, this distance reach implies} that we could expect a detection rate as high as %$\sim 42.86_{-36.26}^{+ 184.63}\, {\rm yr}^{-1}$ 
$\sim 43_{-36}^{+ 185}\, {\rm yr}^{-1}$ for face-on binaries and %$\sim 21.32_{-18.04}^{+91.86}\, {\rm yr}^{-1}$ 
$\sim 21_{-18}^{+92}\, {\rm yr}^{-1}$ for near edge-on IMBH binaries with $\text{SNR}\sim 20$ at O4 sensitivity.
At O5 sensitivity at the same SNR, the maximum redshift reach for face-on binaries can go up to $\sim 2.5$, while for near edge-on binaries it is $\sim 1.6$ (Figure \ref{fig:dL_reach_O5}).  The previous numbers then increase to %$\sim 115.96_{-98.12}^{+499.54}\, {\rm yr}^{-1}$ 
$\sim 116_{-98}^{+500}\, {\rm yr}^{-1}$ and  %$\sim 54.56_{-46.16}^{+235.03}\, {\rm yr}^{-1}$
$\sim 55_{-46}^{+235}\, {\rm yr}^{-1}$. %, respectively at O5 sensitivity. 
We note that these numbers assume a fixed model for the mass distribution of IMBHBs and the evolution of the rate density with redshift, neither of which has been constrained by previous observations. The uncertainties are therefore underestimated.
% as they do not include all astrophysical uncertainties.}
%Note  that these numbers are just for illustrative purposes. A real computation of these numbers would require a knowledge of the mass distribution of the IMBHs in binaries as a function of redshift, in addition to the merger rate density of IMBHBs. These are not yet very accurately constrained. %known very accurately.

As said earlier, inferring IMBH parameters and estimating the corresponding measurement uncertainties will have important implications 
in understanding the formation of high-mass BHs and the evolution of massive stars. The theory of stellar evolution
predicts that stars with zero-age main-sequence (ZAMS) masses 100\,\Msun\,$\lesssim$\,\Mzams\,$\lesssim$\,130\,\Msun\ 
are subject to pair-instability
\citep{fowler_1964_aa, barkat_1967_aa, rakavy_1967_ab}, which causes the stars
to lose mass and leave behind a remnant with a typical mass smaller than
$\sim$\,65\,\Msun~ \citep{heger_2003_aa, blinnikov_2010_aa, chatzopoulos_2012_aa, yoshida_2016_aa, woosley_2017_aa, umeda_2020_aa}. 
These events set the lower edge of the BH mass gap.

Stars with masses 130\,\Msun\,$\lesssim$\,\Mzams\,$\lesssim$\,250\,\Msun \ 
are subject to the pair instability, which disrupts them completely, and hence no BH
forms. 
Stars with \Mzams\,$\gtrsim$\,250\,\Msun \ can collapse directly
to IMBHs with a mass $\gtrsim 135 M_{\odot}$. Thus, in the standard
picture, there should be an upper--stellar-mass BH gap in the range $[65, 135] M_\odot$, 
and any BH observed in this range (e.g., the primary BH of GW190521) have to 
form via other formation channels --- for example through hierarchical coalescence of smaller BHs or direct
collapse of a stellar merger between an evolved star and a main-sequence companion
\citep[e.g.,][]{1989ApJ...343..725Q, 2000ApJ...528L..17P,2001ApJ...562L..19E,2002MNRAS.330..232C, oleary_2006_aa, gerosa_2017_aa, 2019MNRAS.487.2947D, antonini_2019_aa, rodriguez_2019_aa, gayathri_2020_aa,2020arXiv201105332K,2020MNRAS.497.1043D, 2021arXiv210305016M}.
However, the exact mass boundaries of the gap
depend on parameters that are uncertain. 
For example, the ${}^{12}\text{C}(\alpha, \gamma) {}^{16} \text{O}$ nuclear reaction rate, which converts 
carbon to oxygen in the core, can affect the boundary significantly~\citep{takahashi_2018_aa,farmer_2020_aa,Woosley:2021xba,2021MNRAS.501.4514C}. 
Here, to better determine whether future GW observations will be able
to observe BHs in the mass gap, we re-compute the mass-gap boundaries
with updated ${}^{12}\text{C}(\alpha, \gamma) {}^{16} \text{O}$
reaction rates, and increased mass and temporal resolution.
The complexity of this system has made a reliable analysis of the reaction a decades old challenge~\footnote{The reaction rate for ${}^{12}\text{C}(\alpha, \gamma) {}^{16} \text{O}$ is determined by the quantum structure of the compound nucleus ${}^{16} \text{O}$ as an $\alpha$ cluster system. It is characterized by the interfering 
$\ell$=1 waves of the J$^{\pi}$=1$^-$ resonances and sub-threshold levels defining an E1 component for the reaction cross section as well as by the $\ell$=2 components and interference from broad J$^{\pi}$=2$^+$ resonances and the non-resonant E2 external capture to the ground state of  ${}^{16} \text{O}$. In addition to these two main E1 and E2 ground state components, transitions to higher lying excited states occur that also add to the total cross section (see, e.g., \cite{Buchmann2006254} and \cite{deboer_2017_aa}).}. The rapidly declining cross section at low energies has prohibited a direct measurement of the reaction at stellar temperatures and the reaction rate is entirely based on the theoretical analysis and extrapolation of the experimental data towards lower energies. The newly derived rate by \citet{deboer_2017_aa} using a multi-channel analysis approach, derives for the first time a reliable prediction for the interference patterns within the reaction components by taking into account all available experimental data sets that cover the near threshold energy range of the ${}^{12}\text{C}(\alpha, \gamma) {}^{16} \text{O}$ process.    

The paper is organized as follows. In Section~\ref{sec:setup} we
introduce the gravitational waveform models that we employ for our
parameter-estimation studies, and briefly review the Bayesian-analysis
method that we use to infer the source properties from the GW
signals. In Section~\ref{sec:inference} we first describe the
parameter space of the binary simulations that we investigate and the
choice of priors. Then, we present the results for the expected
measurement uncertainties that could be obtained with observations
made during the LIGO-Virgo O4 run and also comment on results that
could be obtained during the O5 run. We also discuss the bi-modality
that appears in the posterior distributions for some parameters in
some regions of the parameter space. In
Section~\ref{sec:implications}, after a brief review of the BH mass
gap and a discussion of the current estimate of the
${}^{12}\text{C}(\alpha, \gamma) {}^{16} \text{O}$ reaction rates, we
evolve massive He stars by incorporating new uncertainties in the
nuclear-reaction rates and we establish new bounds on the lower and
upper edges of the mass gap. Then, using these results, we estimate
the probability with which LIGO-Virgo O4 and O5 runs can identify
IMBHB systems whose primary and secondary masses lie in the BH mass
gap. We also re-analyse GW190521 with the spinning,
  non-precessing waveform models employed in this work, and find
that, although a bi-modality in the posterior distributions of the
detector-frame masses is present, it is absent in the posteriors of
the source-mass parameters. To contrast these findings with
precessing waveforms, we also analyze GW190521 with one precessing waveform 
model, whose former public version was employed in \cite{Nitz_2021}. Finally, in
Section~\ref{sec:conclusion}, we present our main conclusions and
discuss possible future research directions.

\section{Setup}\label{sec:setup}

\subsection{Waveform models}\label{sec:waveforms}

We focus our study on GW signals generated by BBHs with 
non-precessing spins, moving on quasi-circular orbits. Such signals are described 
by 11 parameters: $\boldsymbol{\theta} \equiv \{ m_{1}, m_{2}, \chi_{1}, \chi_{2}, d_{L},
t_{c}, \delta, \alpha, \iota, \psi, \phi_{c} \}$. The parameters $m_{1,2}$ are
the redshifted (i.e., detector-frame) component masses $m_{i} = (1+z)\,m_i^s$ for $i=1,2$, where $m_i^s$ is the source-frame
mass and $z$ the redshift. The quantities $\chi_{1,2}$ are the dimensionless component spins along the orbital
angular momentum $\mathbf{L}$ of the binary (i.e., $\chi_{i} =
\mathbf{S}_{i} \cdot \mathbf{L} /m_{i}^{2}$ for $i=1,2$). 
The parameter $d_{L}$ is the luminosity distance to the binary, which along with the declination
$\delta$ and the right ascension $\alpha$ define the location of the
binary in the sky. The parameter $t_{c}$ is the merger time or more
specifically, it is the peak time of the $\ell=2, m=2$ gravitational 
mode at the geocenter. The angle $\iota$ measures the inclination of the binary's
total angular momentum $\mathbf{J}$ (which, for non-precessing binary
systems, has the same direction as the orbital angular momentum
$\mathbf{L}$) with respect to the line of sight from the detector 
at the geocenter. The remaining parameters $\phi_{c}$ and $\psi$ are the merger phase and 
the gravitational wave polarization, respectively.

It is also useful to define the following binary parameters: the total mass $M_{\rm tot}=m_{1}
+ m_{2}$, the mass ratio $q=m_{1}/m_{2} \geq 1$, the symmetric mass ratio
$\nu=q/(1+q)^2$, the chirp mass $M_{c}=M_{\rm tot}\nu^{3/5}$ and the
effective spin of the binary $\chi_{\text{eff}} = (m_{1}\chi_{1} +
m_{2}\chi_{2})/M_{\rm tot}$.

In general relativity, GWs are described by the two polarizations $h_{+}(t)$ and $h_{\times}(t)$. The complex waveform defined 
by $h(t) \equiv h_{+}(t) -ih_{\times}(t)$ can be conveniently decomposed in a basis of -2 spin-weighted spherical harmonics~\citep{Ypan2011}:
\begin{equation}
	h(t;  \boldsymbol{\lambda}, \iota, \varphi_c) = \dfrac{1}{d_{L}} \sum_{\ell \geq 2}\sum_{|m| \leq \ell} {}_{-2} Y_{\ell m}(\iota, \varphi_c) \, h_{\ell m}(t,\boldsymbol{\lambda})\,,
	\label{eq:hoft_sphericalH}
\end{equation} 
where $\boldsymbol{\lambda}$ denotes a subset of the $\boldsymbol{\theta}$ parameters, namely, the  intrinsic parameters of the binary systems such as masses ($m_{1}, m_{2}$) and spins ($\chi_{1}, \chi_{2} $). 

The GW signal emitted throughout the coalescence of a BBH can be divided into three phases: 
inspiral, merger and ringdown (IMR). The inspiral phase describes the steady, adiabatic evolution of 
the system where the component BHs come closer and closer to each other, losing orbital energy because of GW emission. 
At the end of the inspiral, the BHs plunge into each other, form a common apparent horizon and merge. The ringdown 
phase describes the evolution of the system as the remnant object settles down to a stationary (Kerr) BH.

Here, we employ, as the main IMR waveform model, the one developed within the effective one-body formalism (EOB), 
which is a semi-analytical method that combines results from post-Newtonian (PN) theory for  the inspiral, 
BH perturbation theory for the ringdown, and numerical relativity (NR) for the merger stage.  
More specifically, since we are interested in studying high-mass BBHs with mass ratio as large as 10, 
we employ the quasi-circular, non-precessing spinning waveform models with gravitational modes beyond the 
dominant, quadrupolar one~\citep{Cotesta:2018fcv} (henceforth, ${\tt SEOBNRHM}$)~\footnote{In the LIGO Algorithm Library (LAL) 
the technical name of this waveform model is {\tt SEOBNRv4HM}$\_{\rm ROM}$. It is the reduced-order model (ROM) of 
the time-domain waveform model ${\tt SEOBNRv4HM}$~\citep{Cotesta:2018fcv,Cotesta:2020qhw}.}. The ${\tt SEOBNRHM}$ model contains the five strongest modes $(\ell,
m)=(2,\pm1), (2,\pm2), (3,\pm3), (4,\pm4), (5,\pm5)$ (see Equation~(\ref{eq:hoft_sphericalH})). We also use for 
synthetic (injection) signals a spinning, precessing IMR model built directly interpolating NR waveforms {\tt NRSurPHM}~\citep{VarmaPrecess}
~\footnote{In LAL this waveform model is denoted {\tt NRSur7dq4}~\citep{VarmaPrecess}.}. Such a model contains all 
modes with $\ell \leq 5$. However, the extra modes present in {\tt NRSurPHM} and absent in {\tt SEOBNRHM} are 
not expected to contribute significantly for our choices of
parameters. Finally, we also use a phenomenological IMR waveform model built in the frequency 
domain by combining EOB and NR waveforms, {\tt PhenomHM}~\citep{IMRPhenomXHM}
~\footnote{In LAL this waveform model is denoted ${\tt IMRPhenomXHM}$~\citep{IMRPhenomXHM}.}. 
We stress that, in general, the higher-order (or subdominant) modes become
important in the parameter estimation of the binaries when the inclination angle 
is large and the mass ratio is large ($q \equiv m_{1}/m_{2}\geq
1$) (see, e.g., \cite{Cotesta:2018fcv}). Moreover, higher total mass and higher spins also increase the
amplitude of subdominant modes, especially close to the merger of the
binary. We will see in Sections~\ref{sec:discussion} and
\ref{sec:bimodality} that higher modes can be important to precisely 
infer the parameters even when the mass ratio is as low as $q\sim
1.25$, but the total mass and/or the spins are high.

\subsection{Bayesian statistics}\label{sec:bayes}

{Bayes'} theorem allows us to construct the probability distribution of parameters $\boldsymbol{\theta}$ given a hypothesis (or a model) $\mathcal{H}$ and a data set $d$. It states:
\begin{equation}
	P(\boldsymbol{\theta}|d, \mathcal{H}) = \dfrac{ P(d|\boldsymbol{\theta}, \mathcal{H}) P(\boldsymbol{\theta}|\mathcal{H})}{P(d|\mathcal{H})}\,,
	\label{eq:Bayes_thm}
\end{equation}
where $P(\boldsymbol{\theta}|d, \mathcal{H})$ is the
posterior probability distribution of parameters $\boldsymbol{\theta}$, 
given a data set $d$, under the hypothesis
$\mathcal{H}$. The quantity $P(\boldsymbol{\theta}|\mathcal{H})$ in Equation~(\ref{eq:Bayes_thm}) is the prior
probability distribution of the parameters $\boldsymbol{\theta}$ under
the hypothesis $\mathcal{H}$. The function $P(d|\boldsymbol{\theta}, \mathcal{H})$ 
is the likelihood (very often denoted as $\mathcal{L}(\boldsymbol{\theta})$) for obtaining the data set $d$ with a specific parameter set
$\boldsymbol{\theta}$, under the hypothesis $\mathcal{H}$. Bayes' theorem updates our prior knowledge of the parameters using the likelihood of the data to finally provide us with the posterior probability distribution. Lastly, the quantity $P(d|\mathcal{H})$  in Equation~(\ref{eq:Bayes_thm}) is known as
the evidence for the data set $d$ under the hypothesis
$\mathcal{H}$. It is the normalization factor of the posterior
probability distribution in Equation~(\ref{eq:Bayes_thm}) and, thus, it 
does not matter in our parameter-estimation study. It is, however, widely used for comparing different hypotheses.

In this work, we assume that we have already detected a GW signal and our job is to extract the parameters that describe this signal most closely. The data output from the GW detectors can be written as follows (under the assumption of additive noise):
\begin{equation}
		{d} = {n} + {h}\,,
\end{equation}
where ${n}$ represents the noise realization from the GW detectors and ${h}$ is the  GW strains  measured at the different detectors.  Here
\begin{eqnarray}
	{h} = {F}_{+} (\alpha, \delta, \psi)\,  h_{+} + {F}_{\times}(\alpha, \delta, \psi)\,  h_{\times}\,,
	\label{eq:strain}
\end{eqnarray}
where ${F}_{+,\times} (\alpha, \delta, \psi)$ denote the antenna-pattern functions~\citep{Finn:1992xs} that account for the angular sensitivity of the GW detectors and thus depend on the source location $(\alpha, \delta)$, and on the polarization angle, $\psi$, that defines the relative orientation of the polarization axes with respect to which the polarization states $h_{+, \times}$ are defined. %which defines the plane where the polarization $h_{+, \times}$ are defined. 
To construct the posterior distribution of parameters $\boldsymbol{\theta}$, which describe the GW signal ${h}$, we must first write down the likelihood function.

\subsubsection{Likelihood function}

Let ${d}_{a}$ denote a data stream in a particular detector
$a$. Then the likelihood $\mathcal{L}_{a}(\boldsymbol{\theta})$ or
$P({d}_{a}|\boldsymbol{\theta}, \mathcal{H_{S}})$ is, by
definition, the probability of obtaining the data ${d}_{a}$
with a specified set of parameters $\boldsymbol{\theta}$. Thus, for a GW
signal ${h}_a(\boldsymbol{\theta})$, the likelihood should be
given by the probability of observing the noise realization,
${n}_a = {d}_a - {h}_a(\boldsymbol{\theta})$. The noise
realizations in GW detectors are modeled as independent Gaussian
distributions in each frequency bin with a zero mean and a variance
given by the detector's power spectral density (PSD). Thus, up to an additive constant, 
\begin{equation}
	\log \mathcal{L}_{a}(\boldsymbol{\theta}) \propto
	-\dfrac{1}{2}\sum_{i} \Bigg[  \dfrac{4}{T_a} \dfrac{|\tilde{{d}}_{a}(f_i) - \tilde{{h}}_{a}(f_i,\boldsymbol{\theta})|^{2}}{S^{(a)}_{n}(f_{i})} \Bigg]\,,
	\label{eq:lik}
\end{equation}
where $i$ runs over each frequency bin, $T_a$ is the duration of the GW signal in detector $a$, and a tilde represents the Fourier transform of the data series.  Here $S^{(a)}_{n}(f)$ is the (one-sided) noise PSD associated with the detector $a$. Equation~(\ref{eq:lik}) is the discrete approximation to the inner product between the data $d$ and the waveform model (or template) $h$, denoted by $(d,h)$, where
\begin{equation}
(d,h) \equiv  4\text{Re} \int_{f_\text{min}}^{f_\text{max}} df \dfrac{\tilde{d}(f)\tilde{h}^{*}(f) }{S_{n}(f)}\,,
\label{eq:inner_prodcut}
\end{equation} 
the asterisk denotes the complex conjugate, and $f_\text{min}$ and $f_{\text{max}}$ are the minimum and maximum frequency over which the integration is performed. Generally, the frequency limits are different for different detectors. The signal to noise ratio (SNR) of a signal $h$ is defined as $\sqrt{(h,h)}$.

The overlap between two signals $h_{1}$ and $h_{2}$ is
\begin{equation}
\mathcal{O}_{12} = \max_{t_{c}, \phi_{c}} \Bigg[ \dfrac{(h_{1}, h_{2})}{ \sqrt{(h_{1}, h_{1})(h_{2}, h_{2})} } \Bigg]\,.
\label{eq:overlap}
\end{equation}
By definition, the overlap varies between $0$ and $1$, with the latter representing the case when the two signals are scaled versions of each other.

Assuming that the data streams in the different detectors are independent, we can construct the total log likelihood function by summing the individual log likelihoods of the detectors, that is
\begin{equation}
	\log \mathcal{L}(\boldsymbol{\theta}) = \sum_{a} \log \mathcal{L}_{a}(\boldsymbol{\theta})\,.
	\label{eq:tot_lik}
\end{equation}
From Equation~(\ref{eq:tot_lik}), one can show that the total SNR from all the detectors is the sum in quadrature of the individual SNRs, 
\begin{equation}
\rho = \sqrt{\rho_{1}^{2} + \rho_{2}^{2} + \dotsb  + \rho_{N}^{2}}\,,
\label{eq:SNR}
\end{equation}
where $\rho_{a}$ represents the SNR of a signal in detector a, and $N$ is the number of detectors.

In this work, when we evaluate Equations~(\ref{eq:lik})--(\ref{eq:SNR}), we use
$f_{\textrm{min}}=11$ Hz, and we set $f_{\text{max}}$ to half the
sampling frequency, which is $4096$ Hz. For the results displayed in figures and tables, we use the projected
PSDs of the LIGO-Virgo detector network for the upcoming O4 run~\citep{Abbott:2020qfu}. 
We also regenerated most of the results with O5 PSDs~\citep{Abbott:2020qfu} at the same SNR. 
The duration of the signal $T_a$ in Equation~(\ref{eq:lik}) is chosen to vary between $4$ s to $128$ s depending
on the total mass of the injected signal.

\subsubsection{Priors}\label{sec:priors}

In order to construct the
posterior distribution of the parameters $\boldsymbol{\theta}$ using 
Equation~(\ref{eq:Bayes_thm}), we also need to specify our prior probability distribution
$P(\boldsymbol{\theta}|\mathcal{H})$.

For our analysis, we assume a flat prior in the component (detector-frame) masses $m_{1,2} \in [1, 600] M_{\odot}$ with $m_{1}\geq m_{2}$. 
For the location of the binary in the sky, we use flat priors in $\cos \delta$ and $\alpha$ with $\delta \in [0, \pi]$ rad and $\alpha \in [0, 2\pi)$ rad. 
We assume an isotropic distribution for the orientation of the binary with respect to the observer. This implies that the orientation-angle priors are flat 
in $\cos(\iota)$ for $\iota \in [0, \pi]$ rad, flat in $\psi$ and $\phi_{c}$ for $\psi \in [0, \pi]$ rad and $\phi_{c} \in [0, 2\pi]$ rad. 
The priors for the dimensionless spins, $\chi_{1,2}$,  are chosen to follow a uniform distribution between $-0.99$ and $0.99$. 

\begin{figure*}[!htb] 
	\includegraphics[width=2.15\apjcolwidth]{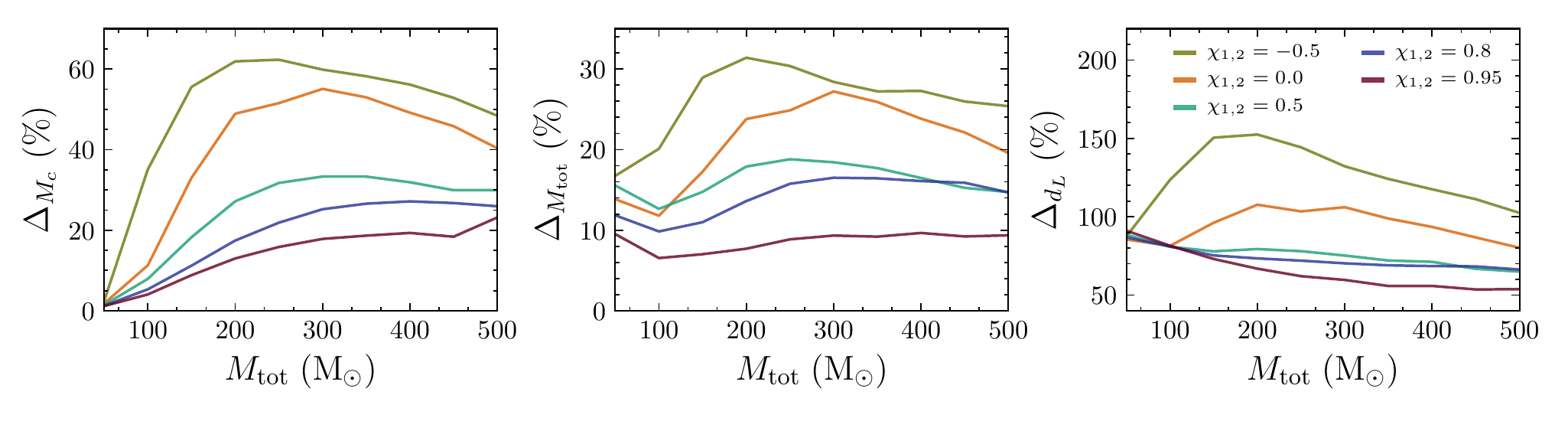}
	\vspace*{-3mm}
	\caption{The 95$\%$ relative width in the measurement of the parameters $\Theta=\{M_{c}, M_{\rm{tot}}, d_{L} \}$ defined by $\mathrm{\Delta}_{\Theta} = {\Delta\Theta ^{95\%}}/{\Theta^{inj}} \times100$, where  $\Delta\Theta ^{95\%}$ is the $95\%$ absolute width and $\Theta^{inj}$ is the true value of the parameter $\Theta$. These results correspond to spinning, non-precessing BBH systems with $q=4$, $\iota=\pi/3$ at $\text{SNR}=20$. The different color lines represent the different values of component spins (with $\chi_{1}=\chi_{2}$), as illustrated in the legend in the rightmost plot.  We use ${\tt SEOBNRHM} $ waveforms for the injection and recovery. For $\chi_{1}=\chi_{2}=0.95$ systems, the worsening in the precision of the chirp mass (left panel) for $M_{\rm{tot}}\geq 450 M_{\odot}$ is due to the occurance of bi-modality in the posterior of $M_{c}$.}
	\label{fig:q4_width}
\end{figure*}

We make use of two distinct distance priors: i)  uniform in Euclidean volume (i.e., flat in $d_{L}^{2}$); and ii) uniform in comoving volume ($V_{c}$), i.e., 
\begin{equation}
	P(d_{L}) \propto \dfrac{1}{1+z} \dfrac{dV_{c}}{dz}  \Bigg(\dfrac{dd_{L}}{dz}\Bigg)^{-1}\,,
	\label{eq:comov_vol}
\end{equation}         
where the conversion from redshift, $z$, to luminosity distance,
$d_{L}$, depends on the cosmology under consideration. In this work, we 
use the standard $\Lambda$CDM model of the universe~\citep{PlanckCosmo2015}.  For a spatially flat universe, we have 
\begin{equation}
	\dfrac{d d_{L}}{dz} = d_{C} + (1+z)\dfrac{d_{H}}{E(z)}\,,
	\label{eq:dLtoz}
\end{equation}
where $d_{C}$ is the comoving distance, $d_{H}=c/H_{0}$ is the Hubble
distance and $E(z)$ is the normalized Hubble parameter at the redshift
$z$. For both distance
priors, we have $d_{L} \in [100 \,\text{Mpc}, 12 \,\text{Gpc}]$. 
By inverting Equation~(\ref{eq:dLtoz}), we can obtain the redshift of the 
source for a given luminosity distance $d_{L}$. The source-frame 
masses are obtained from the detector-frame masses via $m^{s}=m/(1+z)$.

To sample the posterior distribution we use the 
\textsc{LALInferenceNest} code~\citep{VeitchLALInf}. This is a software
package for sampling posterior distributions of the parameters of
compact-binary GW sources that is part of LAL. It uses nested sampling to explore the
posterior distribution. Nested sampling~\citep{10.1214/06-BA127} was
originally introduced as an efficient way to compute the Bayesian
evidence, as a tool for model selection, but also returns
independent samples from the posterior distribution on the
parameters. The algorithm evolves a set of \emph{live points},
replacing the lowest likelihood point at each step by another point of
higher likelihood chosen uniformly from the prior
distribution. The \textsc{LALInferenceNest} code achieves these updates using
short Markov-chain Monte-Carlo evolutions. We refer the reader
to ~\cite{VeitchLALInf} for further details of the implementation.

\section{Measurement of IMBH properties}\label{sec:inference}
\subsection{Parameter space of simulations}\label{sec:simulations}

To understand the uncertainty with which  the parameters of GW signals from IMBHBs could be constrained with upcoming observations~\citep{Abbott:2020qfu}, we simulate a set of (synthetic) GW events and
analyze them using Equation~(\ref{eq:Bayes_thm}). For simplicity,
the signals are simulated (injected) in a zero-noise background. The addition of
noise is expected mainly to change the peak of the 
posteriors not the widths (or the uncertainties), which are our primary 
interest. %However, at our assumed SNRs, the change in the peak should also not be significant. 
Unless otherwise stated, we choose simulated signals that have LIGO-Virgo network SNR of $20$. 
The event GW190521 was observed with SNR $\approx 15$~\citep{Abbott:2020tfl,Abbott:2020mjq}, so with the improved sensitivity expected in O4, seeing similar events at SNR$\sim 20$ is not unreasonable. Higher SNR events will provide the best parameter estimates and hence are those that are most likely to be 
%We restrict our analysis to high SNRs, mainly because we are looking for the best-constrained events given that these are the ones that will be most 
confidently identified as IMBHs.

We work with IMBHBs with mass ratios $q=1.25$, $q=4$ and 
$q=10$. Based on the trend that we observe in the results, we expect that for any other mass ratio between them, the
associated uncertainty is contained within the uncertainties of
these three mass ratios.  We fix the inclination angle to $60^{\circ}$ (i.e, $\iota=\pi/3$) and $0^{\circ}$ (face-on). 
We vary the (detector-frame) total mass in the range, $M_{\rm tot} \in [50 , 500]M_{\odot}$ with steps of $50
M_{\odot}$. Given that we still do not know very accurately the spin 
distribution for IMBHs, we choose a very wide range of values for the component spins,
namely, $\chi_{1}=\chi_{2}=\{-0.8, -0.5, 0., 0.5, 0.8,
0.95\}$. For the same reason, we
also explore binaries with opposite spins --- for example, $\chi_{1}=0.5,
\chi_{2}=-0.5$. We find that the results are contained
within the range of results set by the equally spinning binaries. %We, however, caution the reader that this conclusion might fail for some points at the corners of the parameter space.

\subsection{Results using Bayesian analysis}\label{sec:discussion}

\begin{figure*}[!htb] 
	\centering
	\includegraphics[width=2.15\apjcolwidth]{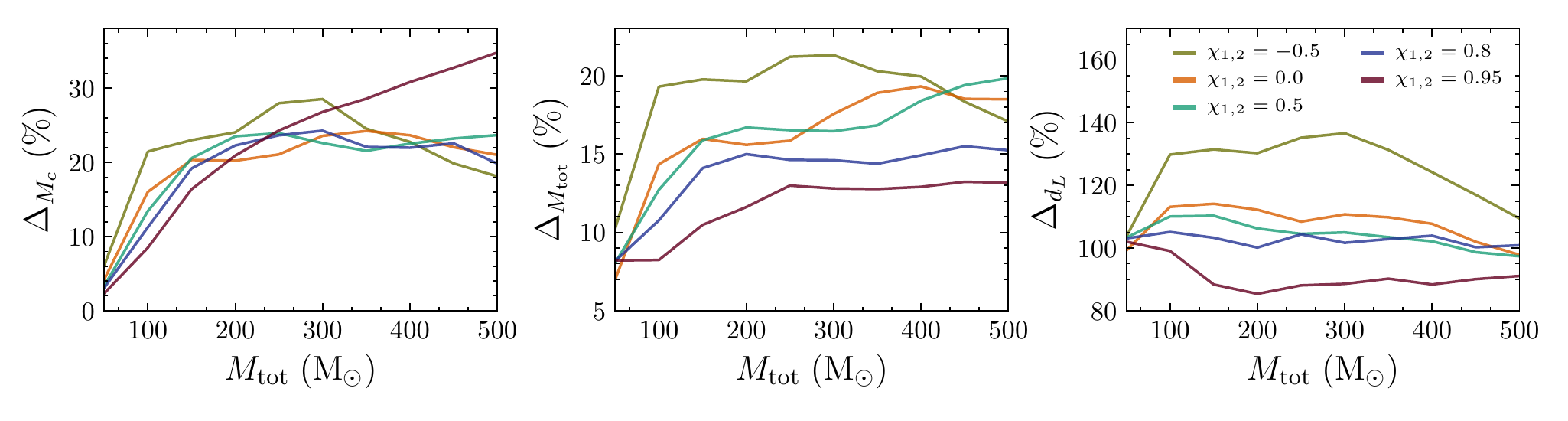}
	\vspace*{-3mm}
	\caption{The $95\%$ relative widths in the measurement of the parameters for the binaries with mass ratio, $q=1.25$. 
			The definitions are same as in Figure~\ref{fig:q4_width}. For $\chi_{1}=\chi_{2}=0.95$ systems, the worsening in the precision of the chirp mass, due to bi-modality, starts at 
$M_{\rm{tot}}\sim 250 M_{\odot}$.}
	\label{fig:q1_25_width}
\end{figure*}

As a cross-check of our analysis, we start our study by reproducing the 
results of \cite{Graff:2015bba}. They focused on IMBHBs with (detector-frame) total masses  
in the range $[50, 500]M_{\odot}$, mass ratios $q = 1.25$ and $q= 4$, and $\text{SNR}=12$ and used 
a version of the LIGO noise-spectral density at design sensitivity available at that time. They 
employed for the Bayesian analysis the multipolar non-spinning waveforms, {\tt EOBNRHM}~\citep{Ypan2011}~\footnote{In LAL this waveform model is denoted 
${\tt EOBNRv2HM}$~\citep{Ypan2011}.}. Using our waveform model ${\tt SEOBNRHM}$ in the non-spinning limit, 
we could recover the results of \citealp{Graff:2015bba} with some small differences 
--- for example, we find that the maximum discrepancy (i.e., the absolute difference between the estimated precisions) is $8\%$ in $M_{c}$ and $\nu$ for high total masses
($M_{\rm tot}\gtrsim 300 M_{\odot}$), where the merger and ringdown
phase of the signal dominates in the most sensitive frequency band of
the detectors. These discrepancies are mainly due to differences between the 
waveform models. The waveform model~\citep{Cotesta:2018fcv, Cotesta:2020qhw} 
used in this work is more accurate than the one employed in \cite{Graff:2015bba}, since 
it was calibrated to a much larger set of NR simulations and contains more information from PN theory.

\begin{figure*}[!htb] 
	\centering
	\includegraphics[width=2.15\apjcolwidth]{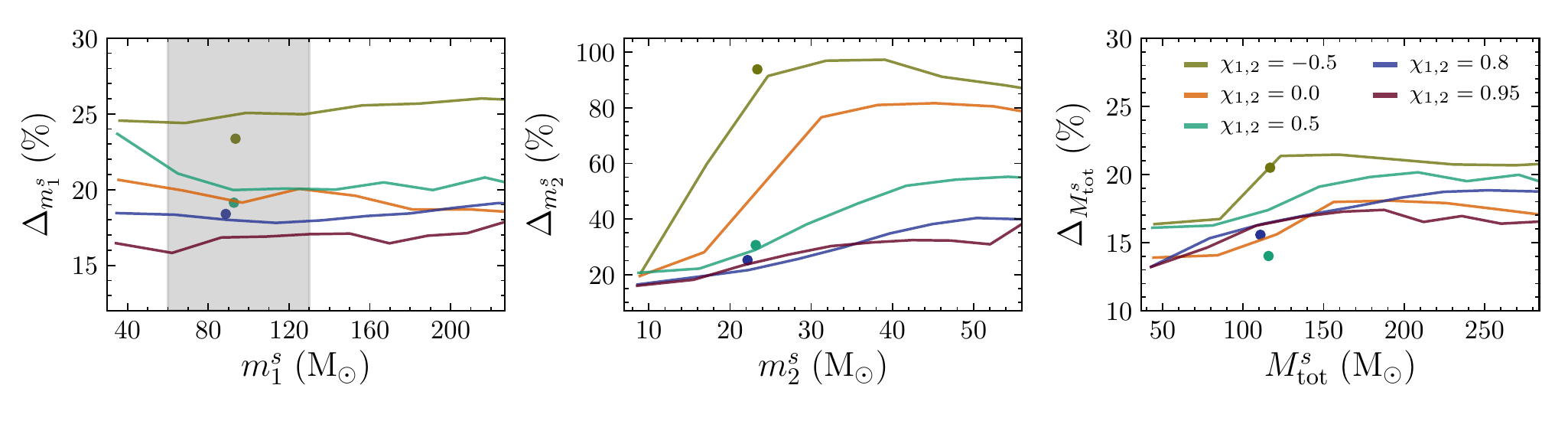}
	\vspace*{-3mm}
	\caption{The $95\%$ relative width in the measurement of the source-frame masses $\{m_{1}^{s}, m_{2}^{s}, M_{\rm{tot}}^{s} \}$ of the injected signals from Figure~\ref{fig:q4_width} (i.e., for $q=4$). The upper limits of the $x$-axes are restricted to the final source-frame masses associated with $\chi_{1}=\chi_{2}=0.95$ IMBHB systems which, because of higher amplitude and hence higher redshift reach, provide the smallest source-frame mass for a given detector-frame mass (e.g., at $M_{\rm{tot}}=500M_{\odot}$). The shaded region represents the BH's mass gap $[60, 130]M_{\odot}$ derived in Section~\ref{sec:c12ag} and computed at the median ($\sigma =0 $) of the \textsuperscript{12}C($\alpha$,$\gamma$)\textsuperscript{16}O reaction rate (see Figure~\ref{fig:bh_mass_gap}). The dots represent the uncertainties when we inject spinning, precessing signals {\tt NRSurPHM}, and recover them with spinning, non-precessing ${\tt SEOBNRHM}$ waveforms. The $95\%$ uncertainty of each dot should be compared to the one of the curve with the same color at the same value of the source-frame parameter. The primary mass can be estimated with the precision  $\sim 15 \mbox{--} 25\%$ while the total mass can be estimated with a slightly better precision $\sim 12 \mbox{--} 22\%$.}
\label{fig:q4_width_2}
\end{figure*}

\begin{figure*}[!htb] 
	\centering
	\includegraphics[width=2.15\apjcolwidth]{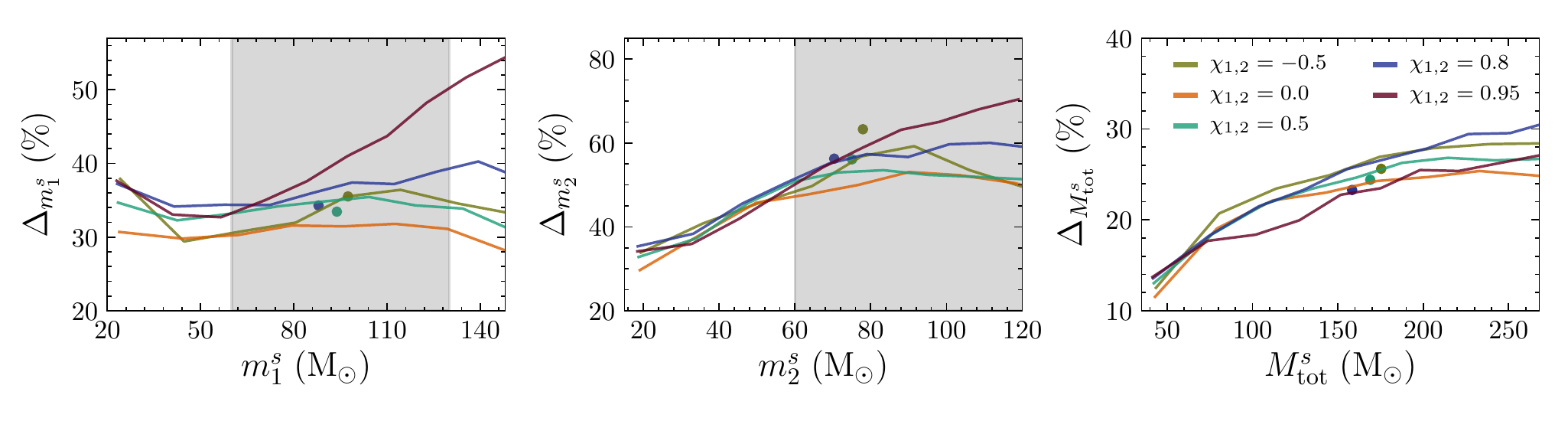}
	\vspace*{-3mm}
	\caption{The $95\%$ relative width in the measurement of the source-frame masses $\{m_{1}^{s}, m_{2}^{s}, M_{\rm{tot}}^{s} \}$ of the injected signals from Figure~\ref{fig:q1_25_width} (i.e., for $q=1.25$). The other definitions are same as Figure~\ref{fig:q4_width_2}. Even for symmetric IMBHBs, except when spins are high ($\chi_{1}=\chi_{2}> 0.8$), for which bi-modality occurs, the primary mass (total mass) can still be estimated with a precision better than $40\%$ ($30\%$).}
	\label{fig:q1_25_width_2}
\end{figure*}

As described in Section~\ref{sec:simulations}, here we extend the study of \cite{Graff:2015bba} in several 
directions. We consider multipolar spinning, non-precessing IMBHB systems with (detector-frame) total masses in the range $[50, 500]M_{\odot}$, 
but mass ratios up to $10$ (i.e., $q \in [1.25, 10]$) and $\text{SNR}=20$. We also use 
updated LIGO and Virgo PSDs, notably the ones for the upcoming O4 and O5 runs~\citep{Abbott:2020qfu}.
In Figures~\ref{fig:q4_width}, \ref{fig:q1_25_width}, \ref{fig:q4_width_2} and \ref{fig:q1_25_width_2} we summarize our results for mass 
ratios $q=1.25$ and $q=4$, inclination $\pi/3$, and a variety of 
spin values, $\chi_{1}=\chi_{2}=-0.8, -0.5, 0, 0.5, 0.8, 0.95$, while 
in Tables~\ref{table:q125_table}, \ref{table:q4_table} and \ref{table:q10_table}, we provide results for 
zero inclination (i.e., face-on configuration) and for larger mass ratio (i.e., $q=10$). Furthermore, 
we do not show the $\chi_{1}=\chi_{2}=-0.8$ results in our figures because, sometimes, they have much higher 
posterior widths compared to the other cases owing to their very small number of GW cycles in the 
detectors' bandwidth. The results also depend on the specific prior choices on the luminosity distance 
--- for example priors flat in $d_{L}^{2}$ or flat in the comoving volume produce noticeably different
  results for higher total masses. For all other spin configurations, we do not see any significant 
differences between the results of the two distance priors. In all figures and Tables we display results with a prior  flat 
in comoving volume.

Figure~\ref{fig:q4_width} shows the $95\%$ relative widths of the posteriors of different parameters of the binary systems with mass ratio $q=4$ and several choices for the BH spins. It can be seen that all the systems at the lowest total
mass (i.e., $M_{\rm tot}=50M_{\odot}$) provide measurement of the chirp mass ($M_{c}$)
better than the total mass. This is because the waveform from the
systems at such a low total mass is dominated by its inspiral phase in
the most sensitive frequency band of the detectors. At leading PN order, 
the phase of inspiral gravitational waveforms depends only on the chirp mass 
$M_{c}$~\citep{Sathyaprakash:1991mt}. At higher
total mass, the waveform is dominated by its post-inspiral phase (i.e., by  the merger),
which is better described by $M_{\rm tot}$, and thus measures the
total mass better \citep{Graff:2015bba}.  We find that the total mass at which $M_{\rm tot}$ starts to be
measured better than $M_{c}$ depends also on the spins of the
components. We note that mergers of binaries with component spins
aligned (anti-aligned) in the direction of the orbital angular momentum $\mathbf{L}$
are delayed (accelerated) compared to their non-spinning counterparts~\citep{Campanelli:2006uy}, 
and thus, the binary remains in the inspiral
phase longer (shorter) than the non-spinning
systems. For example, at $M_{\rm{tot}}\sim 100M_{\odot}$, the aligned-spin systems provide more precise measurements of the chirp mass than the total mass, whereas the anti-aligned--spin systems (see, e.g., the magenta curve) provide more precise measurements of $M_{\rm{tot}}$. For $M_{\rm{tot}} \gtrsim 150 M_{\odot}$ onward, 
regardless of the spin magnitude and orientation, all systems provide more precise measurements of total mass than chirp mass.

As can be seen from Figure~\ref{fig:q4_width}, the precision of the chirp mass ($M_{c}$) measurement initially degrades as the total mass is
increased, before starting to improve for sufficiently high masses.  The initial increase in the uncertainties is because of the decrease
in the number of GW cycles, as we increase the total mass of the
systems. However, after a certain total mass (depending on the spins), the uncertainty starts to decrease as the merger-ringdown phase of the waveform starts to match well with the minimum of the PSDs, and also the subleading modes reach the most sensitive frequency range of the detector. The
$\chi_{1}=\chi_{2}=0.95$ systems, in particular, behave somewhat
unexpectedly above $M_{\rm tot}\gtrsim 450M_{\odot}$. We find that
these systems have a bi-modal distribution in the parameters $M_{c}$,
$\nu$, $m_{1,2}$, and $\chi_{2}$, which cause the width of the
posteriors to increase significantly. We shall discuss these features in more
detail in Section~\ref{sec:bimodality} below.

Figure~\ref{fig:q1_25_width} shows the posterior relative widths for the parameters when
the mass ratio of the binaries is $q=1.25$. As with the $q=4$
binaries, there is a trend for each spin, with a few exceptions. 
First, we find that systems with $\chi_{1}=\chi_{2}=0.95$ show bi-modality even when the
total mass is as low as $200M_{\odot}$, and continue doing so for
higher total masses. The bi-modality in $\chi_{2}$ appears throughout (i.e.,
even when the total mass is $\sim 100M_{\odot}$). Second, some systems show
an unusual trend with total mass (e.g., see  the total mass uncertainty plot in Figure~\ref{fig:q1_25_width}). 
The uncertainty increases again after $350 M_{\odot}$,
where we would expect it to decrease. We find that for such
parameter choices, there is a strong correlation between the total
mass $M_{\rm tot}$ and the primary spin $\chi_{1}$. Additionally, the posterior for
$\chi_{1}$ also develops a mild bi-modality.

As stressed before, we are mostly interested in characterizing the
uncertainty in the measurements of source-frame masses, as they could
help us understand the precision with which, e.g., the mass function
of IMBHs can be constructed from GW measurements in upcoming
observations, and also determine the probability that the observed
BH's mass is in the mass gap. This information can be extracted from
Figures~\ref{fig:q4_width_2} and \ref{fig:q1_25_width_2}. The primary
mass for all the systems with $q=4$ can be measured with an
uncertainty of $\sim 17-25\%$~\footnote{Including the
  $\chi_{1}=\chi_{2}=-0.8$ results from
  Table~\ref{table:q4_table}.} while for binaries with $q=1.25$ the
uncertainties are $\sim 30-40\%$ except when $\chi_{1}=\chi_{2}>0.8$
where bi-modality further worsens the precision, pushing the
uncertainties to $\sim 60\%$. The improvement in the precision for
$q=4$ binaries is due to the presence of higher modes in the
gravitational signal, which break the degeneracy among parameters and
lead to a better measurement of $m_{1}$ and $d_{L}$ measurement.  For
high mass-ratio ($q=10$), the precision improves further, bringing the
uncertainties down to $\sim 11-20\%$ (see Table~\ref{table:q10_table})
except when $M_{\rm{tot}}<100 M_{\odot}$ where we see a very strong
correlation among the parameters $M_{\rm{tot}}$, $m_{1}$, $m_{2}$ and
$q$. This is also true for even smaller mass ratios. We can see this
from Figure~\ref{fig:q4_width} for $q=4$, where, contrary to our expectation, 
the precision of the total-mass measurement at $50 M_{\odot}$ is 
poorer than the one at $100 M_{\odot}$. We observe a similar trend in the
measurement of the primary mass $m_{1}$. The secondary mass, on the
other hand, is measured with relatively poor precision: for symmetric
systems ($q\sim 1$), the uncertainties lie between $30-60\%$ except
for binaries with $\chi_{1}=\chi_{2} \geq 0.8$, for which the uncertainty
can reach $80\%$. For higher mass-ratio signals, the uncertainty can
exceed $100\%$, except for a few highly spinning aligned systems
(e.g., $\chi_{1}=\chi_{2}\geq 0.5$) where the uncertainty can reduce
to $\sim 10-30\%$ (see Table~\ref{table:q10_table}). The total source
mass ($M_{\rm{tot}}^{s}$), however, is measured with a much better
precision, with uncertainties in the range $\sim 10-30\%$,
regardless of the details of the signals.

From Figures~\ref{fig:q4_width} and \ref{fig:q1_25_width}, we also see that component spins can only be measured poorly.   The primary spin ($\chi_{1}$) can be measured better than
$\sim 50\%$ only for asymmetric binaries which have high component
spins, $\chi_{1}=\chi_{2} > 0.5$. Figure~\ref{fig:q1_25_width} shows that for nearly symmetric systems, unless the spins are $\chi_{1}=\chi_{2} > 0.8$, we might not be able to measure the primary component spin better than $50\%$. But, as
expected, for very high mass-ratio signals, we can measure the primary
spin with an uncertainty lower than $\sim30 \%$ if the systems are
aligned (see Table~\ref{table:q10_table}).  Measuring the secondary
spin, however, seems to be difficult for almost all of the IMBH binaries.

To understand how much the neglect of spin-precession in our waveform model 
  affects these precisions, we simulate a few moderately spinning,
  precessing GW signals with the ${\tt NRSurPHM}$ waveform model
  \citep{Varma2019}, and analyze them with the spinning,
  non-precessing ${\tt SEOBNRPHM}$ model.  Mild spin-precession is
  motivated from the LIGO-Virgo
  observations~\citep{O1-O2catalog,O3acatalog2020}, so far. More specifically, we 
  fix the angle between the primary spin $\mathbf{\chi_{1}}$ and the total angular
  momentum $\mathbf{J}$ (i.e., the tilt angle) to be 
  $30^{\circ}$ except for anti-aligned binaries for which the tilt
  angle is taken to be $210^{\circ}$, while the magnitude of the
  component spin vectors are taken to be the same as their non-precessing
  counterparts. The other angles required to define the components
  of the spins on the orbital plane are taken to be zero. All of these quantities are defined at a 
  reference frequency, that we choose to be $f_{\text{ref}}=11$ Hz. 
 These results are indicated by the small dots in Figures~\ref{fig:q4_width_2} and
\ref{fig:q1_25_width_2}. We can see that the uncertainties in the
component-mass measurement change only by $\lesssim 5\%$. The additional 
systematic bias introduced by the neglect of precession in the recovery model is also $\lesssim 5\%$. 
Thus, a mild precession in the signals is not expected to change the results
established here significantly, as long as they are recovered with
spinning, non-precessing waveforms. We plan in the future to carry out a comprehensive 
study that will analyze spinning, precessing GW signals with precessing
waveforms.

\begin{figure*}[!htb] 
	\centering
	\includegraphics[width=2.15\apjcolwidth]{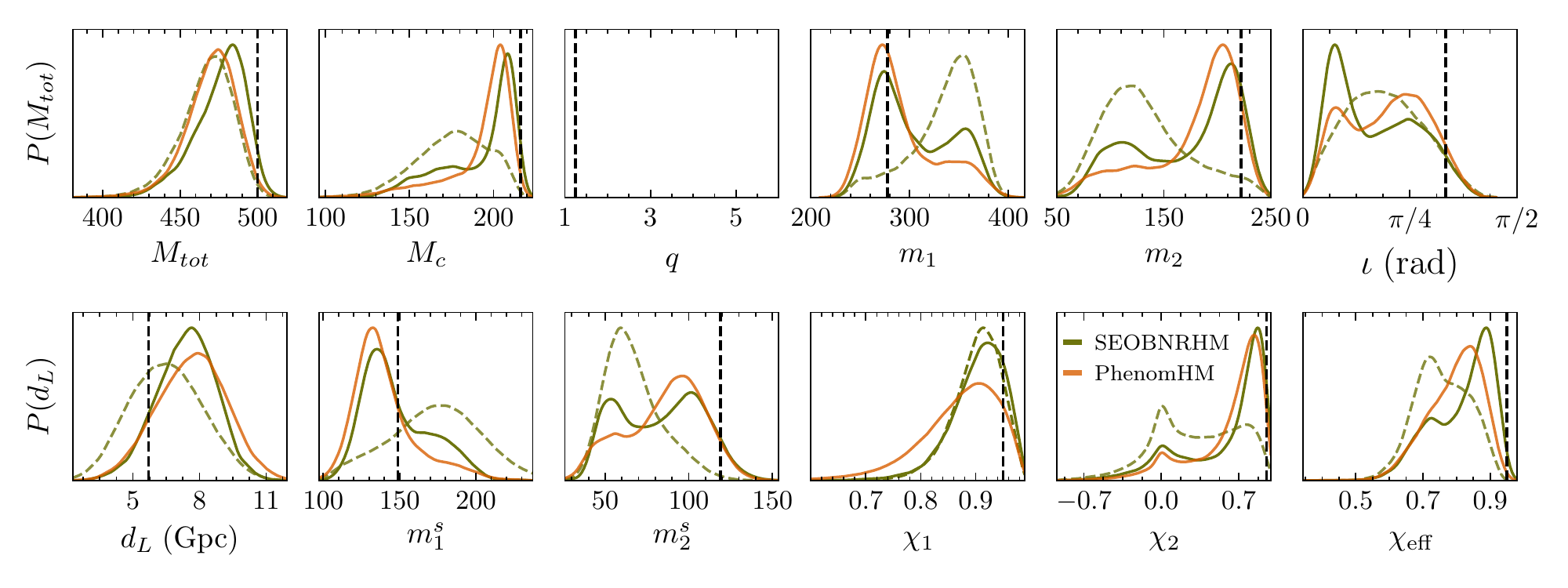}
	\vspace*{-3mm}
	\caption{Posterior distributions of the parameters for an
          injected signal with $M_{\rm tot}=500 M_{\odot}$, $q=1.25$,
          $\chi_{1}=\chi_{2}=0.95$, $\iota=\pi/3$ at
          $\text{SNR}=20$. The dashed lines show the posteriors when
          ${\tt SEOBNRHM}$ injected signal is recovered with 22 mode
          waveform model ${\tt SEOBNR}$. In each panel the vertical
          dashed lines indicate the true (injected) value of the
          parameter. Both ${\tt SEOBNRHM}$ and ${\tt PhenomHM}$ models 
          show bi-modality in various parameters (e.g., the component masses
            $m_{1,2}$). The ${\tt SEOBNR}$ model, which only contain the dominant (2,2) mode, 
           hardly shows bi-modality in most of the parameters, 
            but the posteriors peak away from the true
            (injected) values. Higher modes can, thus, be
            important even when the mass ratio is close to 1, but the
            spins and total mass are high.}
	\label{fig:bimodality}
\end{figure*}

\begin{figure*}[htb] 
	\begin{minipage}[b]{0.5\textwidth}
		\includegraphics[width=1.1\apjcolwidth]{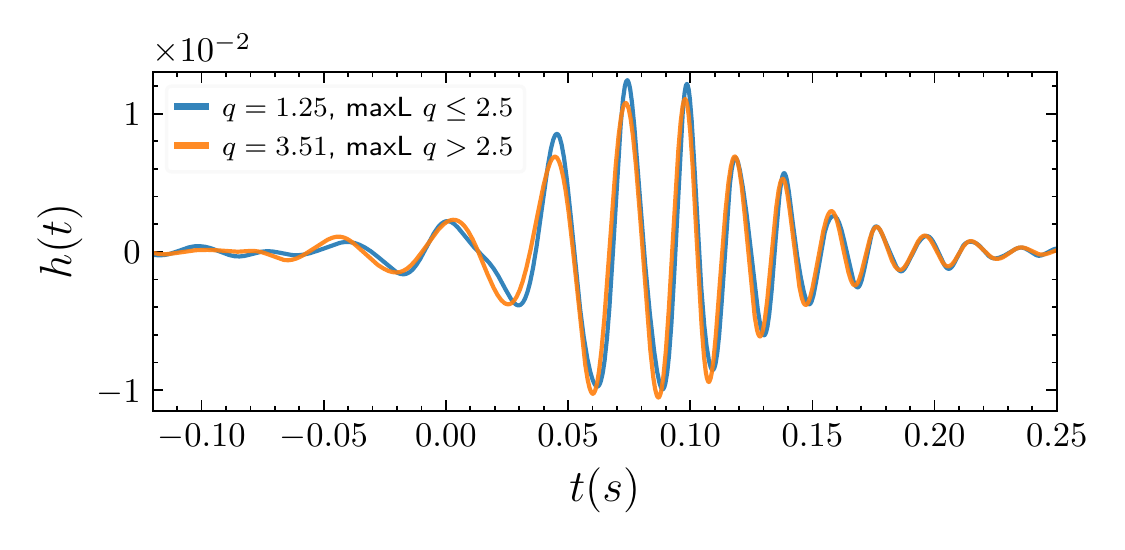}
	\end{minipage}
	\begin{minipage}[b]{0.5\textwidth}
		\includegraphics[width=1.1\apjcolwidth]{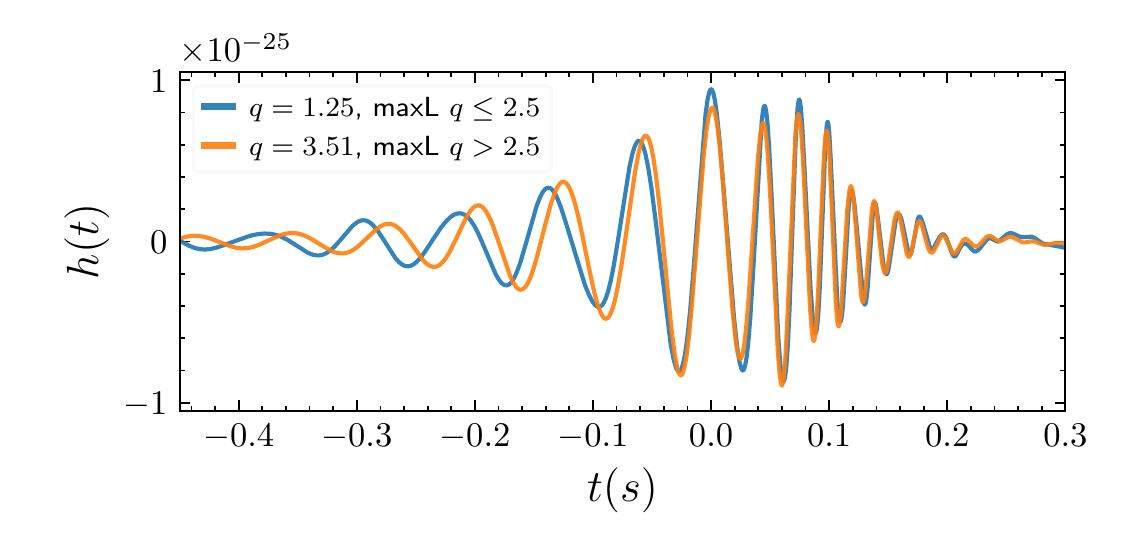}
	\end{minipage}
	
	\caption{Whitened (left panel) and unwhitened (right panel) strains, at the LIGO-Livingston detector, for the parameters of the two 
maximum-likelihood points of the injection study shown in Figure~\ref{fig:bimodality} (for ${\tt SEOBNRHM}$-${\tt SEOBNRHM}$ model),  that is from the region $q\leq 2.5$  and $q>2.5$. The change in amplitude between the two plots reflects the fact that the PSD is $\sim10^{-46}$ in the band of interest, but is not relevant. What is relevant is the change in the shape of the waveform --- the two whitened waveforms match very well (left panel) even though the unwhitened waveforms differ more significantly (right panel) at earlier times.
}
	\label{fig:overlap}
\end{figure*}

We also produced results with the noise-spectral density projected for the 
O5 run~\citep{Abbott:2020qfu}, but at the same SNR=20, since  we a priori do not know what the distribution of the observed SNRs is going to be. However, we do expect that during O5 a larger number of IMBHBs should be observed at a given SNR compared to O4, as also shown in Section~\ref{sec:intro}. We find that, at O5 sensitivity and SNR = 20, the precision of detector-frame masses improves only by a few percent (e.g., for the chirp mass $M_{c}$ it improves by $\sim 6\%$), 
while for the source-frame component masses, which we are mainly interested in, it remains mostly unchanged.

When considering also the results in Tables~\ref{table:q125_table}, \ref{table:q4_table} and
\ref{table:q10_table}, obtained at inclination angle of $\pi/3$ and mass ratio $q=10$, 
and at zero inclination angle for mass ratios $q = 1, 4, 10$, we can summarize the main findings as 
follows. At inclination $\pi/3$, for quite asymmetric IMBHB systems (i.e., $q \sim 10$), 
we could measure the primary mass with uncertainties $11-25\%$, whereas for symmetric binaries
($q\sim1$), the uncertainties are expected to be $\sim 30-40\%$ except
for systems with component spins $\chi_{1}=\chi_{2}\gtrsim 0.8$, where
the uncertainties can reach $\sim 60\% $ due to the presence of
bi-modality. At zero inclination, however, the uncertainty in the primary mass can also go up to $\sim 60\%$ independently of the mass ratio (except for $\chi_{1,2}\gtrsim 0.80$ systems where the 
precision becomes even worse due to bi-modality). However,  for the bulk of the parameter space, even at zero inclination, the uncertainty  is $\lesssim 40\%$ for the primary mass, where this
upper limit is set by the high-mass IMBHBs (e.g., $M_\text{tot}\sim 500
M_{\odot}$). An important question in astrophysics concerns the nature of the IMBH mass function. Although upcoming LIGO and Virgo observations may not be capable of 
observing enough IMBHs to reconstruct their mass function, they hold 
the potential to provide us with much better mass measurements than what might be  possible with electromagnetic observations. %, e.g.,  for high mass-ratio systems, where higher modes come into action and thus improve the mass measurement by breaking the degeneracy with the luminosity distance, as we discussed above, we can extract the primary BHs from IMBHBs with an uncertainty of $\sim 11-25\%$ (see Figure~\ref{fig:q4_width_2} and Table~\ref{table:q10_table}), and also the total source mass with an uncertainty of $\sim 10-25\%$. 
In Section~\ref{sec:mass_gap_measurement}, we shall study in more detail the implications of the  
component-mass measurements in assessing the BH upper--stellar--mass gap.

From Figures~\ref{fig:q4_width} and \ref{fig:q1_25_width}, we can see that measuring the luminosity distance ($d_{L}$) with precision better than $50\%$ may not be possible at high inclination ($\iota=\pi/3$). On the other hand,  Table~\ref{table:q125_table} shows that symmetric face-on IMBH binaries, which emit GW signals with the highest amplitudes, can allow us to constrain $d_{L}$ with uncertainty $< 50\%$. When the total mass and spins are high (e.g., $M_{\rm{tot}}\gtrsim 400 M_{\odot}$, $\chi_{1}=\chi_{2}\gtrsim 0.5$), we can even constrain the luminosity distance with uncertainties less than $40\%$ ($20\%$) at O4 (O5) sensitivity. Given the large detection horizon distance for IMBH binaries, these luminosity distance measurements could be valuable for statistical constraints on cosmological parameters~\citep{Schutz:1986gp,Abbott_2021}, but it will depend on the rate of observed events.

\subsection{Bi-modality in component masses and spins}\label{sec:bimodality}

As we discussed in the previous section, for sufficiently high total
masses, highly spinning IMBHB systems (i.e., $\chi_{1}=\chi_{2}\gtrsim 0.8$)
can exhibit bi-modality in the posterior distributions of some parameters. As 
an example, we show in Figure~\ref{fig:bimodality} the results obtained for an IMBHB system with $M_{\rm
  tot}=500 M_{\odot}$, $q=1.25$, $\chi_{1}=\chi_{2}=0.95$,
$\iota=\pi/3$ at $\text{SNR}=20$. To better understand the bimodality we inject and run 
the Bayesian analysis with two spinning, non-precessing waveform models: ${\tt SEOBNRHM}$ and ${\tt PhenomHM}$~\citep{IMRPhenomXHM}.  
For both models we observe bi-modal posterior distributions in the component masses $m_{1,2}^{s}$, 
the secondary component spin $\chi_{2}$, and the inclination angle $\iota$.

To understand the results, we compare in the (left) right panel of Figure~\ref{fig:overlap} the (whitened) waveforms for the 
${\tt SEOBNRHM}$ model corresponding to the two maximum-likelihood points defined in the
region $q\leq 2.5$ (low $q$) and $q>2.5$ (high $q$). To obtain whitened waveforms in the time domain we first 
divide the waveform in the frequency domain by $\sqrt{S_{n}(f)}$, and then we inverse Fourier transform them into the time domain. 
Whitening of the waveforms helps to better understand the matching of the signal with the waveform, because we can see from 
Equation~(\ref{eq:lik}) that the power spectral density $S_{n}(f)$ appears inverse weighted in the likelihood function. 

From the left panel of Figure~\ref{fig:overlap}, we can see that there
is a very good agreement between the two whitened waveforms even
though the unwhitened waveforms, shown in the right panel of
Figure~\ref{fig:overlap}, have differences at earlier and earlier
  times before coalescence. Thus the bi-modality appears to stem from
a conspiracy: the total mass is very high and hence the number of GW
cycles is already just a few within the detectors' bandwidth, and the
early cycles of the signals, where they differ significantly, are
being suppressed by the worsening of the PSDs~\footnote{However, an
  unresolved question is why bi-modality does not seem to occur when
  spins $\chi_{1,2} < 0.8$ at high total masses.}. We also compute
overlaps using Equation~(\ref{eq:overlap}) between the maximum
likelihood signal and other points from the posterior samples. We find
 similar bi-modal behaviour in the distribution of the overlaps.
This indicates that there are two points in the parameter space
  that have larger match, as we also see
  visually in the left panel of
  Figure~\ref{fig:overlap}. Moreover, when we recover the signal with
  just the ${\tt SEOBNR}$ model, which only contains the (2,2) mode, we find that it is the
  high $q$ region which has the point with the highest posterior probability
  rather than the low $q$ region, where actually the injection lies. The 
  SNRs recovered with the ${\tt SEOBNR}$ model at the maximum likelihood points in these
  two regions are pretty close, 19.49 and 19.63, respectively. Note that the 
  injected SNR is 20. Thus,  almost all
  of the SNR is being recovered by the 22 mode waveform, yet the
  recovered posteriors are significantly different. This suggests that
  the inclusion of higher modes can matter for
  inferring the properties of the source (IMBHBs) even when the mass
  ratio is close to 1, but the spins and the total mass are high. In particular, the 
  global maximum likelihood, is still recovered in the region
  around the injection (low $q$).

\section{Upper--stellar-mass gap}
\label{sec:implications}

Before discussing the implications of our parameter-estimation study of IMBHB systems 
on the measurement of the BH's upper--stellar-mass gap, we first review the main 
results in the literature on this topic, and then perform, using updated 
$^{12}$C($\alpha$,$\gamma$)$^{16}$O reaction rates, a new study aimed at establishing 
more robustly the uncertainties in the upper and lower edges of the mass gap.

\subsection{What is the black hole mass gap?}\label{sec:mass_gap}

% what is the mass gap?

{As mentioned in Section~\ref{sec:intro}, single stars with masses of 20\,\Msun\,$\lesssim$\,\Mzams\,$\lesssim$\,100\,\Msun \ 
end their lives in core-collapse supernovae and are thought to exclusively form BHs
\citep{timmes_1996_ac,fryer_2001_ab,zhang_2008_ab,sukhbold_2014_aa,sukhbold_2018_aa}.
Stars with \Mzams\,$\gtrsim$\,100\,\Msun\
reach core temperatures of $\gtrsim$\,7$\times$10$^{8}$ K that allows 
for the production of electron–positron pairs from photons, $\gamma$+$\gamma$\,$\rightarrow$\,$e^{-}+e^{+}$
\citep{fowler_1964_aa, barkat_1967_aa, rakavy_1967_ab}.
The production of $e^{-}e^{+}$ pairs removes photons, softening the equation of state.
These stars are expected to become dynamically
unstable before core-oxygen depletion, as the pair production leads to
regions where the adiabatic index $\Gamma_1$\,=\,$d\ln P/d\ln \rho|_S$\,$\le$\,4/3
\citep{fraley_1968_aa, ober_1983_aa, bond_1984_aa, woosley_2002_aa, heger_2003_aa, takahashi_2018_aa, farmer_2019_aa, marchant_2020_aa}. 
The ensuing dynamical collapse results in vigorous oxygen burning whose outcome depends, in part, 
on the mass of the star and the adopted $^{12}$C($\alpha$,$\gamma$)$^{16}$O reaction rate.

Stars with 100\,\Msun\,$\lesssim$\,\Mzams\,$\lesssim$\,130\,\Msun \ 
can undergo a cyclic pattern of entering the pair instability region; contracting,
undergoing oxygen burning, and expanding
\citep{heger_2003_aa, blinnikov_2010_aa, chatzopoulos_2012_aa, yoshida_2016_aa, woosley_2017_aa, umeda_2020_aa}. This process yields a series of pulsations
that removes large amounts of mass from the star, leading to a
pulsational pair-instability supernova (PPISN) whose core collapse leaves
significantly lower mass BHs. PPISN set the lower edge of the BH mass gap.
{The importance of impact of angular momentum transport 
was investigated by \citet{marchant_2020_aa} studied, and the influence of 
metallicity, wind mass loss prescription, and treatments of chemical
mixing on the lower edge of the BH mass gap explored by \citet{farmer_2019_aa}.}

Stars with 130\,\Msun\,$\lesssim$\,\Mzams\,$\lesssim$\,250\,\Msun \ 
can produce a pair instability supernova (PISN) where the energy injected from
the first explosive oxygen-burning event completely unbinds the star
\citep{joggerst_2011_aa,chatzopoulos_2013_ab,kozyreva_2014_ab,gilmer_2017_aa,marchant_2020_aa,renzo_2020_aa}. 
PISN leave no compact object, making them responsible for the existence of the BH mass gap. 
Stars with \Mzams\,$\gtrsim$\,250\,\Msun \ 
reach core temperatures of log(T\textsubscript{c}/K) $\approx$ 9.8, 
where the rate of endothermic photodisintegration reactions absorbs enough energy to prevent the star from unbinding
\citep{heger_2003_aa}.
The star, once again, can reach core collapse. This sets the upper boundary of the mass gap.

In addition to the ZAMS mass, the $^{12}$C($\alpha$,$\gamma$)$^{16}$O
reaction rate plays a central role in determining the final outcome by
setting the C/O ratio in the core after helium burning.
\citet{takahashi_2018_aa} found that cores with reduced
$^{12}$C($\alpha$,$\gamma$)$^{16}$O rates have larger
C/O ratios, develop shell convection during central carbon burning, 
and sufficiently avoid the pulsational-instability regime to collapse as BHs.
\citet{farmer_2020_aa} found that cores with reduced
$^{12}$C($\alpha$,$\gamma$)$^{16}$O rates can have C/O ratios
$\simeq$\,0.4. These cores undergo a sequence of central carbon
burning, off-center carbon burning, central oxygen burning, and
core collapse to produce a BH.  Cores with median
$^{12}$C($\alpha$,$\gamma$)$^{16}$O rates can have C/O ratios
$\simeq$\,0.1.  These cores effectively skip central carbon burning to yield PPISN
with smaller BH masses.  Cores with large
$^{12}$C($\alpha$,$\gamma$)$^{16}$O rates can have C/O ratios
$\ll$\,0.1.  These cores effectively skip central and shell carbon burning. They proceed directly to
explosive oxygen burning and result in a PISN.  Given this sensitivity, we thus
undertake a new exploration of the $^{12}$C($\alpha$,$\gamma$)$^{16}$O reaction rate
and its uncertainties.

\subsection{Updated \texorpdfstring{$^{12}C(\alpha$,$\gamma$)$^{16}$O}{} reaction rates} \label{sec:c12ag}

The C/O content of stellar cores is determined by the competition
between triple-$\alpha$ to \textsuperscript{12}C and
\textsuperscript{12}C($\alpha$,$\gamma$)\textsuperscript{16}O nuclear-
reaction rates during helium burning \citep{deboer_2017_aa}. Because of the complexity of the calculations, most analyses have only considered a subset of the reaction channels and a few representative data sets. In order to obtain a comprehensive evaluation, \citet{deboer_2017_aa} considered the entirety of existing experimental data related to the
determination of the low energy $^{12}$C$(\alpha,\gamma)^{16}$O cross
sections, aggregating 60 year of experimental data consisting of more than 50 independent experimental studies. The more than 10,000 data points were then incorporated into a complete multi-channel phenomenological $R$-matrix
analysis~\citep{RevModPhys.30.257, 0034-4885-73-3-036301, azure} using
the code \texttt{AZURE2}~\citep{azure,azure2}. A main result was the
characterization of the uncertainty in the reaction rate, which was
accomplished through a Monte-Carlo--uncertainty analysis of the data
and the extrapolation to low energy using the $R$-matrix model. This
resulted in a rate uncertainty that had statistical significance,
which had only been accomplished in a few other previous works
\citep[e.g.,][]{Gialanella2001, Schurmann201235}, and there with much
more limited data sets. In \citet{deboer_2017_aa}, the 1$\sigma$
uncertainty of the reaction rate was given after finding an
approximately Gaussian underlying probability distribution for the
rate.

In the recent work by \citet{farmer_2020_aa}, the main rate that was
used was that of \citet{0004-637X-567-1-643}, but with the central
value adjusted to be the geometric mean of the upper and lower
1$\sigma$ uncertainty estimates, with the assumption that these
1$\sigma$ values reflect an underlying probability distribution that
is approximately Gaussian. This modified rate was implemented in the
\texttt{STARLIB} rate library~\citep{0067-0049-207-1-18}. The work of
\citet{deboer_2017_aa} was very much in the spirit of that of
\citet{0004-637X-567-1-643}, but used a Monte-Carlo method to estimate
the uncertainties (instead of $\chi^2$) and had at its disposal the
significantly increased amount of experimental measurements that had
accrued in the intervening time. In particular, the significantly more
stringent constraints for the values of the sub-threshold asymptotic
normalization coefficients determined through sub-Coulomb transfer
measurements~\citep{PhysRevLett.83.4025, PhysRevLett.114.071101} and
the facilitation of them using the alternative $R$-matrix
parameterization of \citet{PhysRevC.66.044611}.

To facilitate the present calculations, and future ones, we expand 
the tabulated reaction rate to a much finer temperature grid. This
is done to ensure that no temperature step results in variations
in the rate of more than an order of magnitude. The expanded reaction
rates of \citet{deboer_2017_aa} are shown in Figure~\ref{fig:c12ag},
over a region of $\pm$3$\sigma$. As shown previously by
\citet{farmer_2020_aa}, the uncertainty present in this nuclear-reaction rate 
translate into one of the primary sources of uncertainty
in the location of the BH mass gap boundaries.

\begin{figure}[!htb]  
\centering
\includegraphics[width=1.0\apjcolwidth]{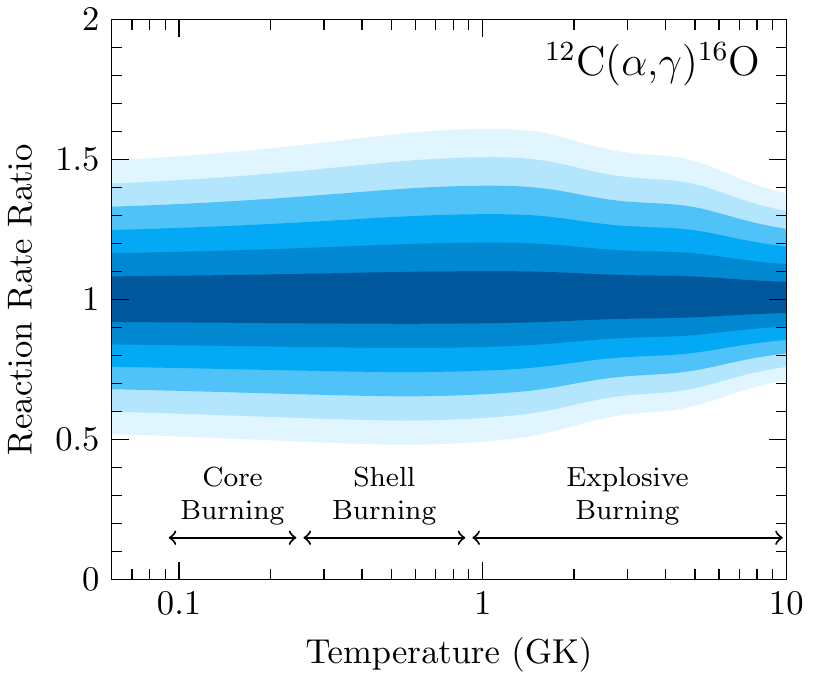}
\caption{Relative uncertainties in the
  $^{12}$C$(\alpha,\gamma)^{16}$O reaction rate of this work, expanded
  from those presented in \citep{deboer_2017_aa}. The uncertainties
 are normalized to the central value for clearer
  presentation. The regions of fading blue color represent 0.5$\sigma$
  steps in the Gaussian uncertainty distribution.  }
\label{fig:c12ag}
\end{figure}

\subsection{The black hole mass gap with updated \texorpdfstring{$^{12}C(\alpha$,$\gamma$)$^{16}$O}{} reaction rates} \label{sec:mesa_models}

We use \MESA\ version r11701
\citep{paxton_2011_aa,paxton_2013_aa,paxton_2015_aa,paxton_2018_aa,paxton_2019_aa}
to evolve massive helium cores with a metallicity of Z\,=\,10$^{-5}$ until 
they either collapse to form a
BH or explode as a PISN without leaving a compact remnant.  
We use the same inlists and \texttt{run{\_}star\_extras.f} as in
\citet{farmer_2020_aa} to calculate the boundaries of the BH mass gap with respect to the updated 
\textsuperscript{12}C($\alpha$,$\gamma$)\textsuperscript{16}O reaction-rate uncertainties.

Figure \ref{fig:bh_mass_gap} shows the location of the PISN
BH mass gap as a function of the uncertainty in
the \textsuperscript{12}C($\alpha$,$\gamma$)\textsuperscript{16}O
rate. As the reaction rate increases, through increasing
$\sigma$[\textsuperscript{12}C($\alpha$,$\gamma$)\textsuperscript{16}O], both the lower and upper
edges of the BH mass gap shift to lower masses while
maintaining a roughly constant width, of $\simeq$\,80$^{+9}_{-5}$\,\Msun. 
For the updated $\sigma$[\textsuperscript{12}C($\alpha$,$\gamma$)\textsuperscript{16}O] rates adopted in this work, 
the location of the lower and upper edge over the $\pm 3\sigma$ range 
is $\simeq$\,59$^{+34}_{-13}$\,$M_{\odot}$ and $\simeq$\,139$^{+30}_{-14}$\,$M_{\odot}$ respectively. 
These results are commensurate with \citet{farmer_2020_aa} at the $\simeq$\,20\% level for the 
lower edge of the BH mass gap and at the $\simeq$\,5\% for the upper edge of the BH mass gap.
We next discuss the main reasons why our results slightly differ and put them in context
with previous studies.

\begin{figure}[!htb]  
\centering
\centerline{\includegraphics[width=1.0\apjcolwidth]{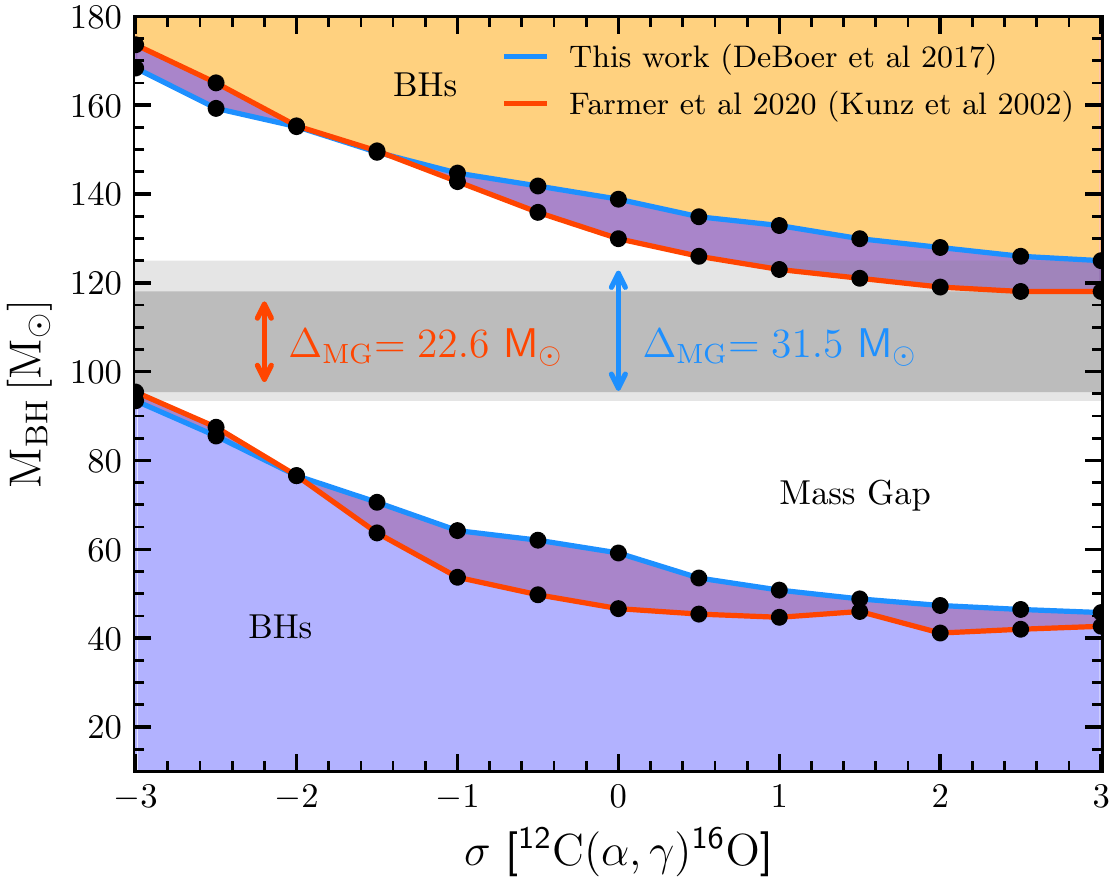}}
\caption{The location of the BH mass gap as a function of the
temperature-dependent uncertainty in
the \textsuperscript{12}C($\alpha$,$\gamma$)\textsuperscript{16}O
reaction rate. The blue lines mark the mass-gap boundaries predicted by
our updated \textsuperscript{12}C($\alpha$,$\gamma$)\textsuperscript{16}O rate
uncertainties. The orange lines mark the mass-gap boundaries, 
as found in Figure~5 of \citet{farmer_2020_aa}, predicted by the 
\citet{2002ApJ...567..643K} rate as expressed in the \texttt{STARLIB} 
reation-rate library \citep{2013ApJS..207...18S}. 
The white region denotes the mass gap, the
purple regions highlight differences of the 
adopted \textsuperscript{12}C($\alpha$,$\gamma$)\textsuperscript{16}O
rates, and the labeled grey horizontal bars denote the mass range
where a BH does not form for any value of
the adopted \textsuperscript{12}C($\alpha$,$\gamma$)\textsuperscript{16}O
rate.}
\label{fig:bh_mass_gap}
\end{figure}

\citet{marchant_2020_aa} found that the efficiency of angular-momentum transport
changes the lower edge of the BH's mass gap at the $\simeq$\,10\% level.
\citet{farmer_2019_aa} found that the lower edge of mass gap was robust at the $\simeq$\,10\% level
to changes in the metallicity, wind mass loss prescription, and treatment of chemical mixing.
{Our models use 2 times the mass resolution and about 2.5 times the temporal resolution 
as those used in \citet{farmer_2020_aa}, which we estimate means the results shown in 
Figure~\ref{fig:bh_mass_gap} should be robust with respect to mass and temporal resolution at the $\simeq$\,10\% level.}
For each $\sigma$[\textsuperscript{12}C($\alpha$,$\gamma$)\textsuperscript{16}O], our $\Delta M$\,=\,1~\Msun\ mass grid of \MESA\ models consumed $\simeq$\,60,000 core-hours,
with Figure~\ref{fig:bh_mass_gap} thus costing $\simeq$\,780,000 core-hours.

\subsection{Sensitivity to the resolution of the tabulated \texorpdfstring{$^{12}C(\alpha$,$\gamma$)$^{16}$O}{} reaction rates} \label{sec:sensitivity_c12ag}

Within the context of these specific \MESA\ models,
Figure~\ref{fig:converge01} shows the dependence of the BH mass spectrum
on the tabulated temperature resolution of the $\sigma$\,=0 $^{12}$C($\alpha$,$\gamma$)$^{16}$O reaction rate
at the baseline mass and temporal resolution.
When the reaction rate is defined by 52 temperature points, the BH mass spectrum reaches a 
maximum BH mass of 49.6 M$_{\odot}$ at an initial helium core mass of 55.0 M$_{\odot}$.
When the reaction rate is defined by 2015 temperature points, the BH mass spectrum reaches a 
maximum BH mass of 59.1 M$_{\odot}$ at an initial helium core mass of 60.0 M$_{\odot}$.
The 52 point rate produces a flatter BH mass spectrum, while the 2015 point rate 
sustains a linear trend of larger BH masses with larger initial helium core masses
until the peak at an initial helium core mass of 60.0 M$_{\odot}$.
Overall, the 52 temperature point rate produces smaller BH masses than the 2015 temperature point rate.

Figure~\ref{fig:r2000vs52} shows why the 52 temperature point reaction rate 
produces a different BH mass spectrum than the 2015 temperature point reaction rate:
the errors from interpolating the 52 temperature point reaction rate are larger 
than the formal uncertainties in the 2015 temperature point reaction rate.
Fundamentally, the 52 temperature point reaction rate is ``bad'' because the 
reaction rate changes by nearly an order of magnitude between tabulated temperature points.
When a reaction rate varies by this much between tabulated temperature points,
there is a limit to what interpolation can provide.

Delving deeper, Figure~\ref{fig:c12_center_compare} shows the evolution of the 
central $^{12}$C mass fraction from near the onset helium ignition to 
central carbon ignition for the M$_{{\rm He,init}}$\,=\,60M$_{\odot}$ stellar model 
as a function of the central temperature (a proxy for time).
The model computed with the 52 temperature point $\sigma$\,=\,0 
reaction rate achieves a central carbon mass fraction of $\sim 0.125$.
Stellar models computed using the 2015 temperature point $\sigma$\,=\,0 
reaction rate achieve a central carbon mass fraction of $\sim 0.17$.
This difference in the carbon mass fraction
of the core is the primary reason why the BH mass spectra shown in 
Figure~\ref{fig:converge01} differ.

We also calculate new $\pm3\sigma$ rates for the 
$^{12}$C\,+$^{12}$C, $^{12}$C\,+$^{16}$O, $^{16}$O\,+$^{16}$O
reactions. Consistent with \citet{farmer_2020_aa}, we find that these 
reaction rates move the the BH's mass-gap boundary by $\lesssim$\,1\,\Msun. 
Evidently, in this case, the total energy liberated by C-burning is more important 
than how quickly or slowly the C-burning energy is liberated.
A large carbon fuel reservoir from a small \textsuperscript{12}C($\alpha$,$\gamma$)\textsuperscript{16}O rate
leads to a more massive BH, an intermediate carbon fuel reservoir
from the recommended (i.e., median) \textsuperscript{12}C($\alpha$,$\gamma$)\textsuperscript{16}O rate
leads to less massive BHs, and a small carbon-mass fraction
from a large \textsuperscript{12}C($\alpha$,$\gamma$)\textsuperscript{16}O rate
leads to no compact object being formed (i.e., a PISN).

\subsection{Sensitivity to mass and time resolution} \label{sec:sensitivity_spacetime}

Figure~\ref{fig:converge02} shows the impact of enhanced time and mass
resolution on the BH mass spectrum.  The first spectrum, labeled (a),
is calculated as in \citet{farmer_2020_aa} with the \MESA\ controls:
\code{max\_dq}\,=\,1d-3 and \code{delta\_lgRho\_cntr\_limit}\,=\,2.5d-3. 
The \code{max\_dq} control limits the mass of any given cell to
contain no more than the specified fraction of the total mass. That
is, the minimum number of cells in a model is 
1/\code{max\_dq}. The \code{delta\_lgRho\_cntr\_limit} limits the size
of time-steps such that the central density does not change by more
than a specified fraction. The second BH mass spectrum, labeled (b),
doubles the mass resolution by halving \code{max\_dq}, implying a
{minimum} of 2000 cells.  The third BH mass spectrum, labeled (c),
increases the temporal resolution by a factor of 2.5 by decreasing
\code{delta\_lgRho\_cntr\_limit} by the same factor.

\begin{figure}[!htb]  
\centering
\centerline{\includegraphics[width=1.\apjcolwidth]{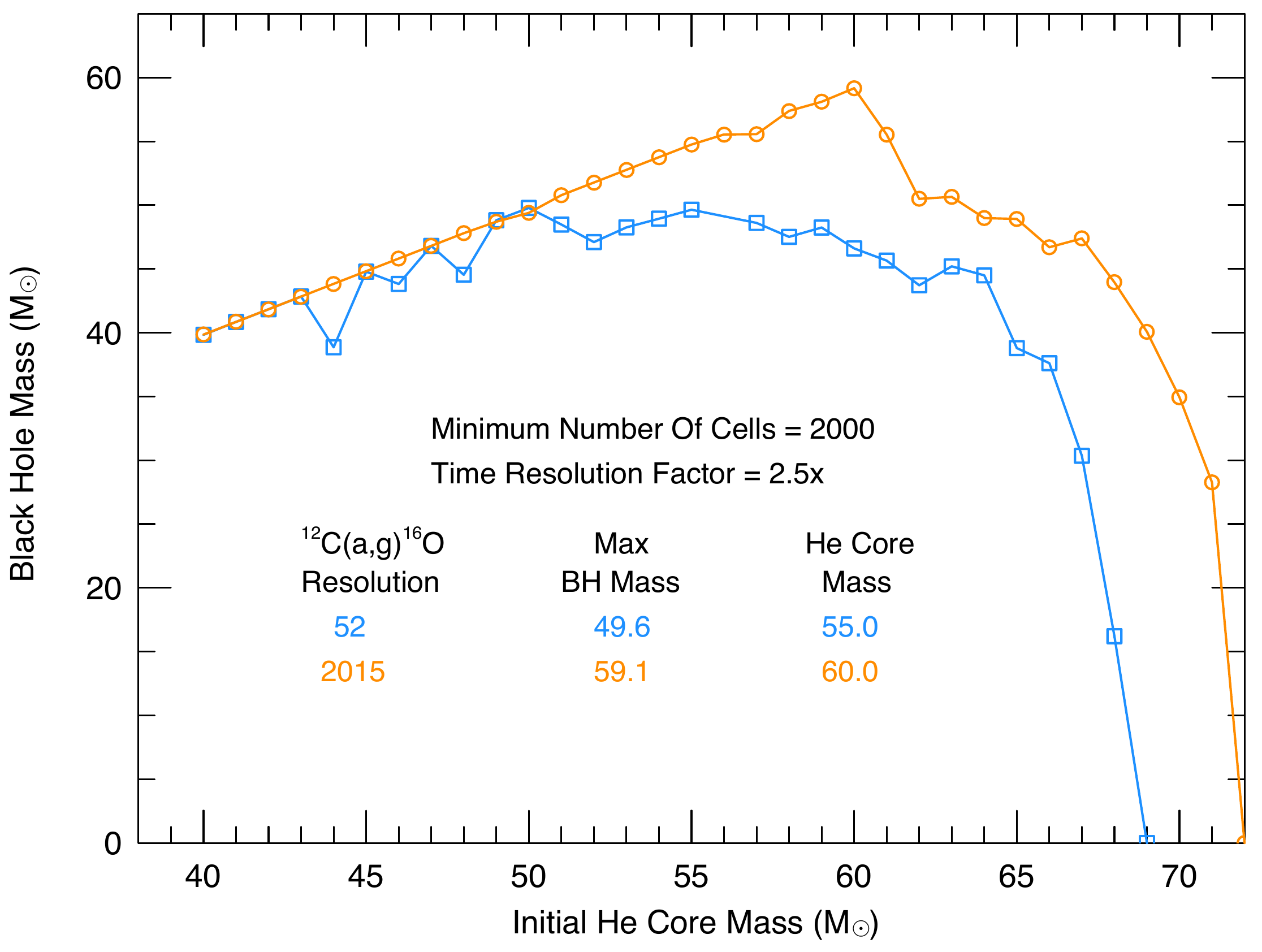}}
\caption{BH mass spectra for different
$^{12}$C($\alpha$,$\gamma$)$^{16}$O reaction rate resolutions at the adopted baseline mass and temporal resolution.
Commonly used reaction rate resolutions of 52 temperature points produce smaller BH masses.}
\label{fig:converge01}
\end{figure}

\begin{figure}[!htb]  
\centering
\centerline{\includegraphics[width=1.\apjcolwidth]{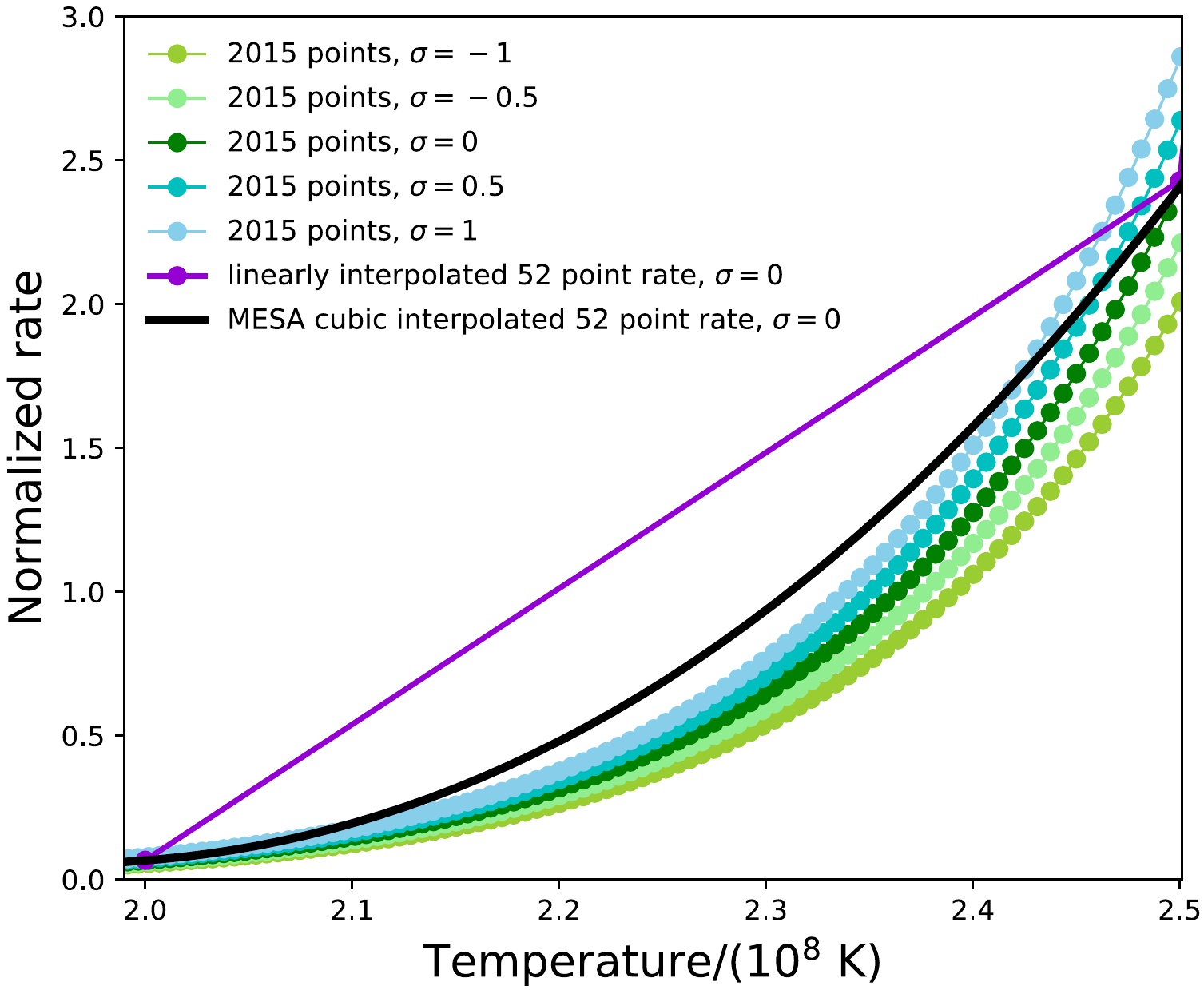}}
\caption{The 2015 temperature point $\sigma$\,=\, 0, $\pm$0.5, and $\pm$1.0 
normalized $^{12}$C($\alpha$,$\gamma$)$^{16}$ reaction rate (green/blue curves) over the relevant
helium burning temperature range. Also shown is the reaction rate that results from
linearly interpolating the $\sigma$\,=0 rate defined by 52 temperature points (purple curve) and the 
reaction rate from \MESA's cubic interpolation of the $\sigma$\,=0 rate defined
by 52 temperature points (black curve). The error from interpolating the 
52 temperature point $\sigma$\,=0  rate 
is larger than the $\sigma$\,= +1.0 rate defined by 2015 temperature points. 
}
\label{fig:r2000vs52}
\end{figure}

\begin{figure}[!htb]  
\centering
\centerline{\includegraphics[width=1.\apjcolwidth]{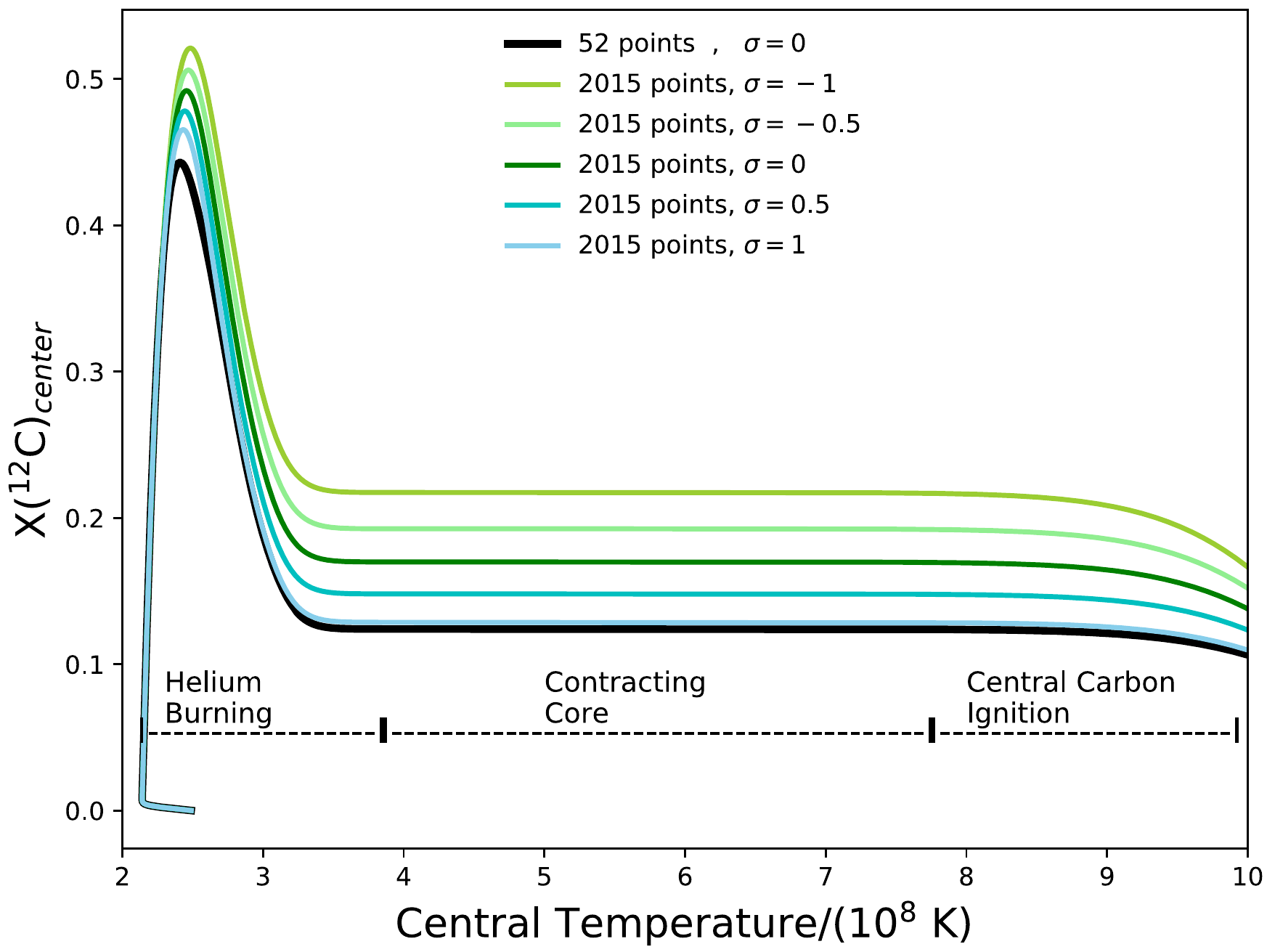}}
\caption{Evolution of the central $^{12}$C mass fraction with core temperature for  
the M$_{{\rm He,init}}$\,=\,60M$_{\odot}$ stellar models. The color scheme is the same as in
Figure~\ref{fig:r2000vs52}. The 52 temperature point $\sigma$\,=\,0 reaction rate yields
a smaller central carbon mass mass fraction than the 2015 temperature point $\sigma$\,=\,0 reaction rate.
}
\label{fig:c12_center_compare}
\end{figure}

\begin{figure}[!htb]  
\centering
\centerline{\includegraphics[width=1.\apjcolwidth]{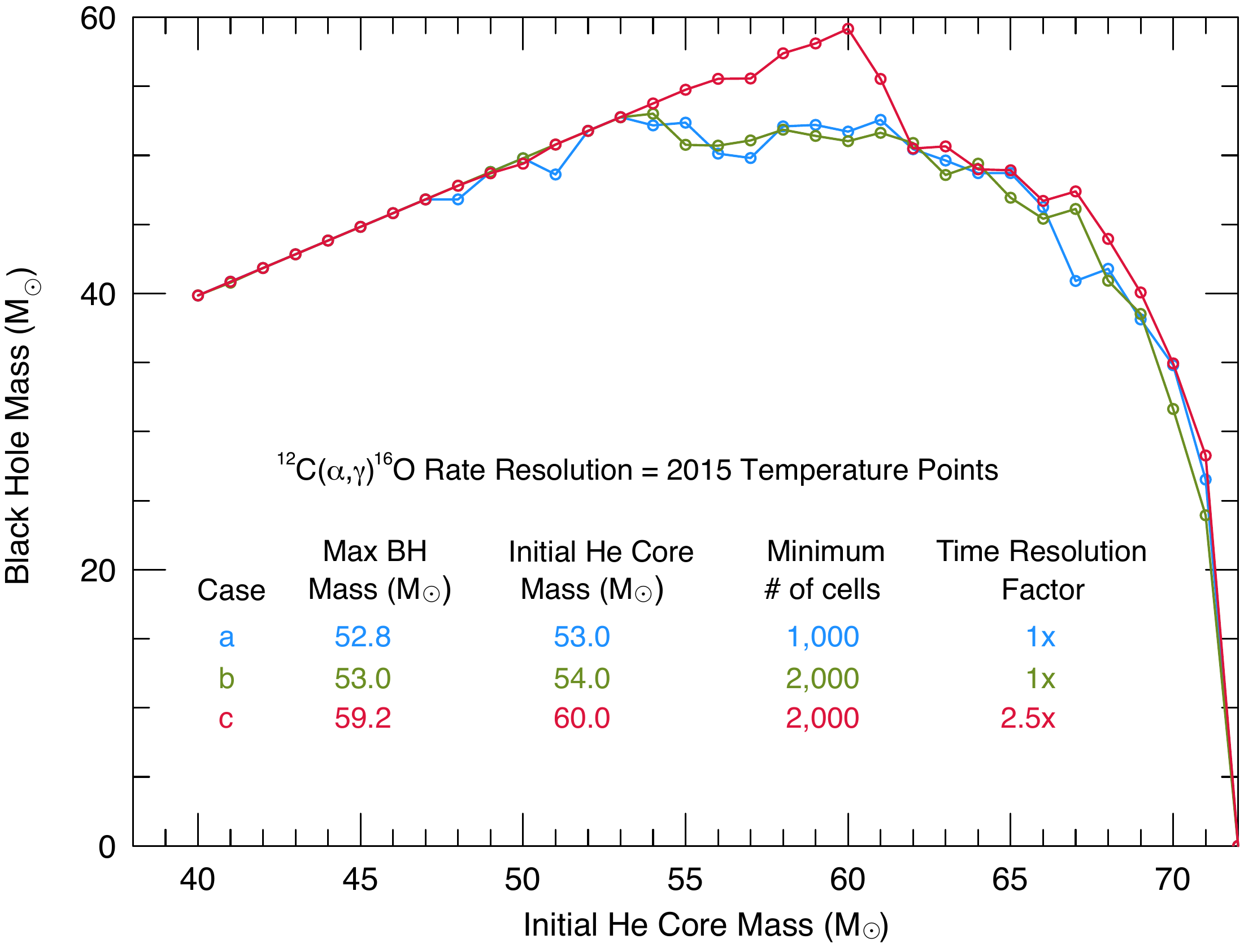}}
\caption{Black hole mass spectrum for different mass and time resolutions at the highest
$^{12}$C($\alpha$,$\gamma$)$^{16}$O reaction rate resolution. 
Increasing the mass resolution (olive curve) does not appreciably change the spectrum, 
while increasing the time resolution (red curve) allows the linear trend 
at low initial helium core masses to continue to $\simeq$\,60 \Msun.
}
\label{fig:converge02}
\end{figure}

Increasing the mass resolution of our models yields little discernible
difference in the BH mass spectrum, compare case (a) versus case (b)
in Figure~\ref{fig:converge02}. Increasing the temporal resolution,
case (c), increases the maximum BH mass in the BH mass spectrum from
53.0 M$_{\odot}$ in case (b) to 59.2 M$_{\odot}$ in case (c). This
difference is due to the smaller time-steps allowing the 
{pressure-weighted volume} average
$\langle\Gamma_{1}\rangle$ to get closer to 4/3 without dipping below it. See
{\citet{stothers_1999_aa} and} \citet{farmer_2020_aa} for a discussion of $\langle\Gamma_{1}\rangle$.  This
allows convective carbon shell burning to take place before core
oxygen ignition. For carbon mass fractions $\gtrsim 0.17$, off-center
carbon burning is strong enough to trigger convective mixing and
burning of the entire carbon shell, preventing the model from reaching
the $\langle\Gamma_{1}\rangle$ instability.  The model stabilizes long enough
for the core to burn a significant fraction of its oxygen before
$\langle\Gamma_{1}\rangle$ dips below 4/3.  In turn, this allows the stellar
interior to progress closer to core-collapse before the $\langle\Gamma_{1}\rangle$
instability coupled with oxygen burning triggers a pulse of
mass-loss. We conclude that for models with carbon mass fractions
$\gtrsim 0.17$, smaller timesteps are necessary to resolve the peak of
the BH mass spectrum, although additional time resolution could be necessary at lower mass fractions.

Figure~\ref{fig:60msol_panel} shows the evolution of the internal structure
of a M$_{{\rm He,init}}$\,=\,60\,M$_{\odot}$ model 
for the case (b) and case (c) time resolutions explored in Figure~\ref{fig:converge02}. 
Both models {highlight} an episode of radiative carbon burning in the core,
followed by convective carbon burning in a shell. However, the two time
resolutions show different carbon shell burning evolutions.

In the case (b) model, {carbon ignition occurs near model 3000, $\tau_{\rm ci}$\,=\,0 hr where $\tau_{\rm ci}$ is
the time after carbon ignition}.
Carbon shell burning generates a small convective region {starting at about model 3400, $\tau_{\rm ci}$\,$\simeq$\,470 hr}. 
Only a portion of the carbon shell is burned before 
$\langle\Gamma_{1}\rangle$\,$<$\,4/3 {which occurs near model 3430, $\tau_{\rm ci}$\,$\simeq$\,477 hr}.
Oxygen ignites radiatively in the core {at about model 3500, about 0.06 hr later.}
{The energy release from helium burning 
as a result of helium mixing deep into the structure
in the $\simeq$\,60\,s between models 4500 and 5000
is $\simeq$\,10$^{41}$ erg. This is  $\simeq$\,7 orders of magnitude smaller than the change 
in the total energy over the same model numbers, suggesting the integrated energy release is dynamically small.}
Oxygen burning causes a pulse of mass-loss that removes $\simeq$\,5\,M$_{\odot}$ of material from the surface
layers {by about model 8000, $\tau_{\rm ci}$\,$\simeq$\,757 hr (not shown in Figure~\ref{fig:60msol_panel}).}
A second pulse then removes an additional $\simeq$\,3\,$M_{\odot}$ of material from the surface layers
{at about model 16000, about 1940 yr later(not shown in Figure~\ref{fig:60msol_panel}).}

In the case (c) model, 
carbon ignition occurs {near model 4400, $\tau_{\rm ci}$\,=\,0 hr}.
Carbon shell burning becomes strong enough to grow the convective region, mixing the entire shell 
{starting at model 4800, $\tau_{\rm ci}$\,$\simeq$\,240 hr and ending at model 5300, $\tau_{\rm ci}$\,$\simeq$\,351 hr}.
This allows the model to stave off the $\langle\Gamma_{1}\rangle$
instability until the carbon mass fraction in the shell drops to $\simeq$\,10\% its initial value
{near model 5780, $\tau_{\rm ci}$\,$\simeq$\,388 hr.}
Only then does the model become dynamically unstable to
$\langle\Gamma_{1}\rangle$ $<$ 4/3. A weak pulse then removes
$\simeq$\,$0.3$\,M$_{\odot}$ of material from the surface layers
{by about model 11500, $\tau_{\rm ci}$\,$\simeq$\,433 hr (not shown in Figure~\ref{fig:60msol_panel})}.

%\revision{
%Convection in the M$_{{\rm He,init}}$\,=\,60\,M$_{\odot}$ 
%model allows a burning carbon shell to tap fresh fuel from the layers above.
%A larger convective burning shell can mix more fresh carbon fuel from the
%layers above than a smaller convective burning shell, 
%keeping the carbon mass fraction at the base of the convective shell larger for a longer period of time.
%Therefore a larger convective shell staves
%off the $\langle\Gamma_{1}\rangle$ instability longer than a smaller
%convective carbon burning shell.
%}

The difference in the final BH mass for the two different time resolutions
is the result of a tight coupling between the nuclear burning and convection.
Smaller time-steps better resolve the coupling. 
These results suggest a more extensive convergence study may be needed
to accurately resolve the peak of the BH mass spectrum.

\begin{figure*}[!htb]  
\centering
\centerline{\includegraphics[width=2.\apjcolwidth]{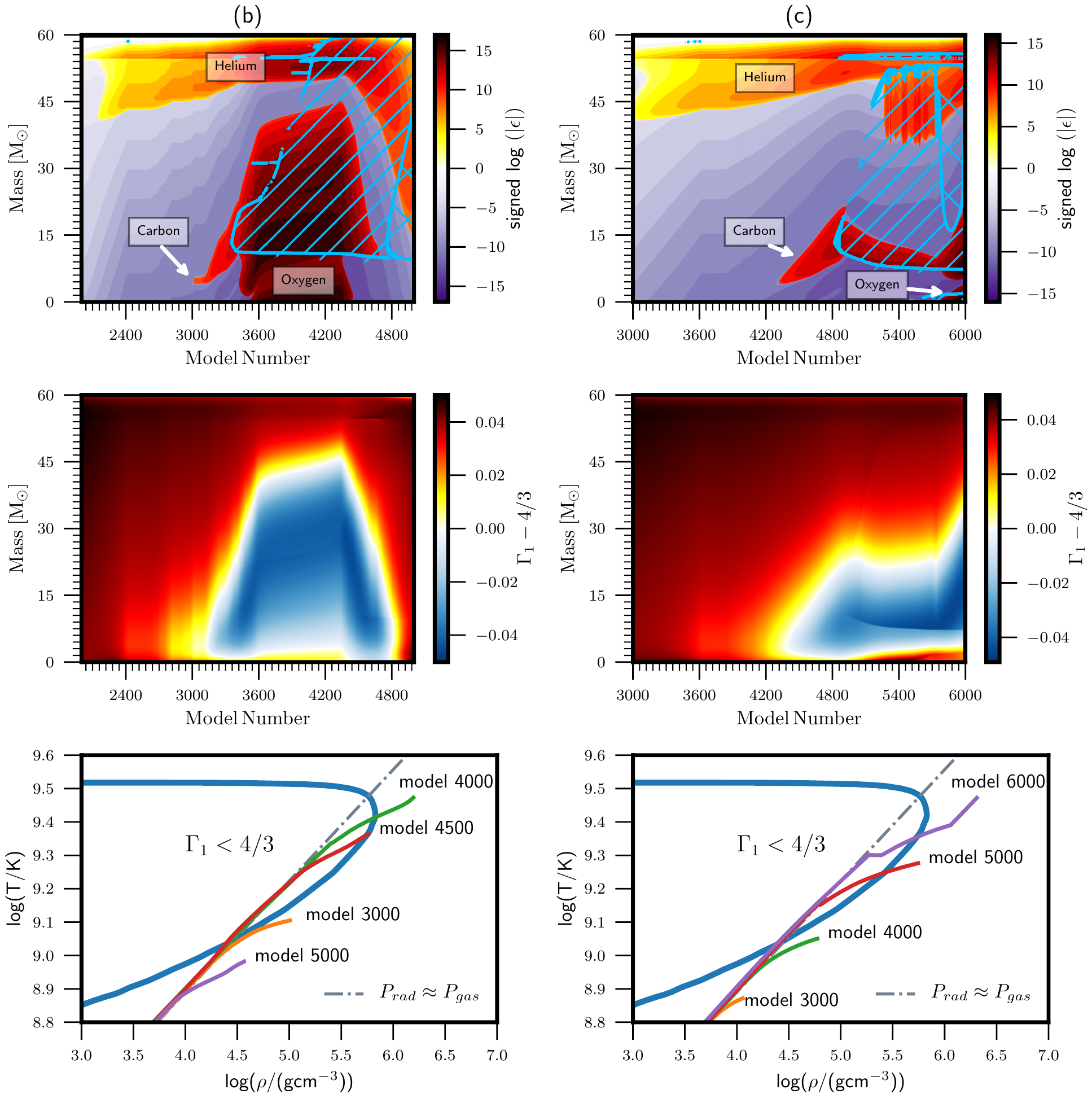}}
\caption{
Evolution of M$_{{\rm He,init}}$\,=\,60\,M$_{\odot}$ models for case (b) and case (c) time resolutions; also see Figure 4 of \citet{farmer_2020_aa}.
{The purpose of the figure is to highlight the difference in carbon ignition between two stellar models, not to show the evolution of a full pulsation cycle in each model.}
The top row shows the signed logarithm of the net specific power,
$\rm{sign}\left(\epsilon_{\rm{nuc}}-\epsilon_{\nu}\right)\log_{10}\left(\rm{max}\left(1.0,|\epsilon_{\rm{nuc}}-\epsilon_{\nu}|\right)\right)$,
where $\epsilon_{nuc}$ is the specific energy generation rate
and $\epsilon_{\nu}$ is the specific energy loss from neutrinos. Purple
regions denote strong neutrino cooling and red regions denote regions
of strong nuclear burning. {Positive sloped blue hatched regions indicate standard mixing length convection,
negatively sloped blue hatched regions indicate convection with no mixing.}
The different fuels burning are labelled.
The middle row shows the evolution of locally unstable regions with $\Gamma_{1}$ $<$ 4/3. 
The bottom row shows the density-temperature structure for different model numbers.
The dashed line shows where the gas
pressure is equal to the radiation pressure. The solid
blue curve encloses the $\Gamma_{1} < 4/3$ region.
{When pressure-weighted volume average $\langle\Gamma_{1}\rangle$ drops below 4/3 the model becomes dynamically unstable.}
}
\label{fig:60msol_panel}
\end{figure*}

\subsection{Measurement of mass gap with upcoming LIGO-Virgo observations}\label{sec:mass_gap_measurement}

Figure~\ref{fig:bh_mass_gap} shows that at the median ($\sigma=0$) of the
\textsuperscript{12}C($\alpha$,$\gamma$)\textsuperscript{16}O reaction rate,
the mass gap would typically fall in the range $60\mbox{--}130M_{\odot}$ considering the overlapping parts of the range from the two
\textsuperscript{12}C($\alpha$,$\gamma$)\textsuperscript{16}O rates. Including such a range in 
Figures~\ref{fig:q4_width_2} and \ref{fig:q1_25_width_2} (see shaded gray region), 
we find that the uncertainties in the primary-mass measurements for
asymmetric IMBHB systems are $\sim 17-25\%$, while for nearly symmetric
systems they are $\sim 30-50\%$. For the latter systems, the secondary
mass also quite often falls in the mass gap, and can be constrained
with an uncertainty of $\sim 40-70\%$.

\begin{figure*}[htb] 
	\centering
		\includegraphics[width=2.\apjcolwidth]{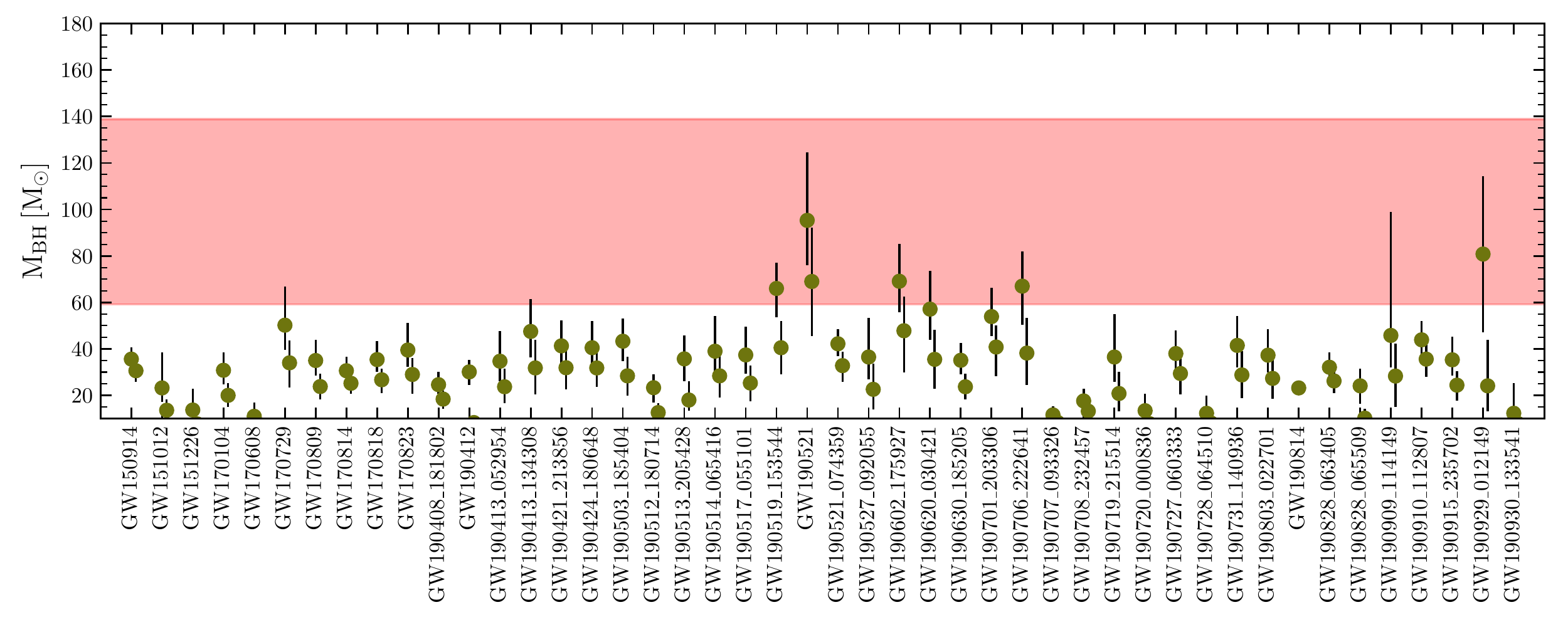}
		\caption{The source-frame component masses and their associated $90\%$ credible intervals of all events from GWTC-1 and GWTC-2 with the restriction that the median estimated mass of the primary is $\geq 10 M_{\odot}$. The red region shows the mass gap at the $\sigma=0$ for the updated \textsuperscript{12}C($\alpha$,$\gamma$)\textsuperscript{16}O rate (see Figure \ref{fig:c12ag}). The primary mass of the GW190521 event along with its associated $90\%$ credible interval lies well inside the red region indicating that this could be a BH in the mass gap.  Our re-analysis of GW190521 in Section \ref{GW90521_analysis}, confirms this result with the waveform models used in this work. We note that {there are also 5 more component masses, including  the secondary mass of GW190521, whose posterior-distribution medians lie in the mass gap.}}
	\label{fig:GW_events_in_mass_gap}
\end{figure*}

How confidently the upcoming O4 run will be able to identify 
that the component masses of IMBHB systems lie in the mass gap? To address this question we need to account
for the uncertainties in the boundaries of the mass gap shown in
Figure~\ref{fig:bh_mass_gap}, which are caused by uncertainties in the
\textsuperscript{12}C($\alpha$,$\gamma$)\textsuperscript{16}O
rate. The probability that a component mass of the IMBHB system lies
in the mass gap can be computed as:
\begin{align}
    P(\rm{MG}) &=P(m_{\rm{MG}}^{\rm{L}} < m_{i}^{s}<m_{\rm{MG}}^{\rm{U}})\,, \nonumber \\
    &= \int P(x) dx \int_{m_{\rm{MG}}^{\rm{L}}(x)}^{m_{\rm{\rm{MG}}}^{\rm{U}}(x)} P(m_{i}^{s}|d) dm_{i}^{s}\,,
\label{eq:mg_prob}
\end{align}
where $i=1,2$, $x$ denotes the \textsuperscript{12}C($\alpha$,$\gamma$)\textsuperscript{16}O rate, $m_{\rm{MG}}^{\rm{L}}$ and $m_{\rm{MG}}^{\rm{U}}$ denote the lower and upper edges of the mass gap and $d$ represents the data (i.e., the simulated GW signal). 
We can approximate the above equation as a discrete sum in $x$:
\begin{align}
    P(\rm{MG})&=P(m_{\rm{MG}}^{\rm{L}} < m_{i}^{s}<m_{\rm{MG}}^{\rm{U}})\,, \nonumber \\ 
    & = \sum_{j=1}^{N=13} P(x_{j}) \int_{m_{\rm{MG}}^{\rm{L}}(x_{j})}^{m_{\rm{\rm{MG}}}^{\rm{U}}(x_{j})} \,P(m_{i}^{s}|d) dm_{i}^{s}\,,
\label{eq:mg_prob_discrete}
\end{align}
where $j$ runs from 1 to the 13 grid points of the \textsuperscript{12}C($\alpha$,$\gamma$)\textsuperscript{16}O reaction rate shown in Figure~\ref{fig:bh_mass_gap} and $P(x_{j})$ is the corresponding probability~\footnote{ We compute these probabilities from the standard normal distribution, evaluated at the given $\sigma$ uncertainty and then re-normalize them so that they add up to 1. Alternatively, one could interpolate between these grid points and make the approximation in Equation~\ref{eq:mg_prob_discrete} more and more precise by increasing N, but we do not expect that the results will change very significantly.}. 

If $P(\rm{MG})$ is large, then the source could be 
a mass-gap event. However, events that are not in the mass gap will
occasionally have large $P(\rm{MG})$ simply due to noise fluctuations. 
Thus, the probability necessary to claim a confident detection of a
mass-gap event depends on the relative rate of mass-gap and
no-mass-gap events. To test the hypothesis that there are events with
components in the mass gap, we should compute the Bayesian evidence for
the no-mass-gap versus mass-gap hypotheses, marginalized over
the uncertain proportion of events that are in the mass gap. This
approach leverages information from all events, not just those
with high $P(\rm{MG})$, and therefore has greater statistical
power. However, such an analysis is beyond the scope of the current
paper. Here we will only compute $P(\rm{MG})$ for noise-free data with
a variety of injected IMBHB systems. Signals with high $P(\rm{MG})$ are more
likely to be robustly identified as mass-gap events, but we warn the reader 
against over-interpreting our numbers.

\begin{figure*}[htb] 
	\centering
		\includegraphics[width=2.15\apjcolwidth]{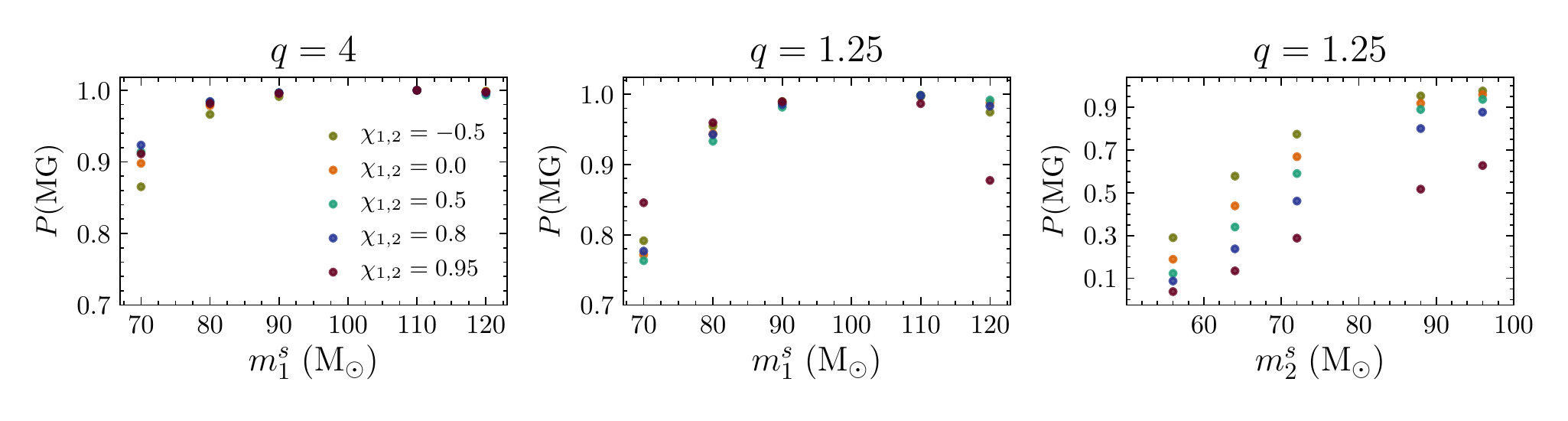}
		
	\caption{Probability of being in the BH's mass gap (see  Equation~(\ref{eq:mg_prob_discrete})) as fucntion of source-frame component masses, for  
		several IMBHB systems with primary mass $m_{1}^{s}=[70, 80, 90, 110, 120] M_{\odot}$, different spin values (as illustrated in the legend in the left panel), at inclination angle of $\pi/3$ and SNR of 20. Results are obtained using the spinning, nonprecessing ${\tt SEOBNRHM}$ model. Independently of the mass ratio, when IMBHBs have primary mass in the range $[80, 120]M_{\odot}$, the signal can be identified as a BH in the mass gap with probability $>90\%$.}
	\label{fig:mass_gap_prob}
\end{figure*}

We show in Figure~\ref{fig:mass_gap_prob} $P(\rm{MG})$ for IMBHB systems
with primary mass $m_{1}^{s}=[70, 80, 90, 110, 120] M_{\odot}$ and several spin values, 
observed at inclination angle of $\pi/3$, computed using the updated
${}^{12}\text{C}(\alpha, \gamma) {}^{16} \text{O}$ rate. As we can
see, for asymmetric IMBHB systems (i.e., $q> 1.25$), if the primary mass is
well within the median mass gap (e.g., $m_{1}^{s} \in
[80,120]M_{\odot}$), then the primary mass has probability of being in 
the mass gap $>95 \%$ (i.e., a single observation could be sufficient to robustly 
identify the existence of sources in the mass gap). For systems with primary mass close to the
lower edge of the median mass gap (e.g., $m_{1}^{s}\sim 70 M_{\odot}$),
$P(\rm{MG})$ reduces to $\sim 0.85$ for anti-aligned systems. For
nearly symmetric IMBHB systems (i.e., $q = 1.25$), $P(\rm{MG}) > 95\%$ for systems with
somewhat higher primary mass, namely, $m_{1}^{s}\gtrsim 85 M_{\odot}$.
Symmetric systems can also have high probability that the
secondary mass lies in the mass gap. The right panel in 
Figure~\ref{fig:mass_gap_prob} indicates that for IMBHB systems with
$m_{2}^{s}\in [90, 120] M_{\odot}$, the posterior probability that the
secondary mass lies within the mass gap exceeds $90\%$ for systems
with $\chi_{1,2}<0.80$. By contrast, high spin systems with $\chi_{1,2}>
  0.80$ exhibit bi-modality which worsens the precision of the mass
  measurement, as discussed in Section~\ref{sec:bimodality}.

For zero inclination-angle (face-on) IMBHBs, our study shows that $P(\rm{MG})$ is generally
lower for measurements of the primary mass, reaching values as low as
$\sim 0.6$ in some cases (e.g., at the upper edge of the mass gap
$\sim 120 M_{\odot}$, where bi-modality occurs). Nonetheless, systems
with $m_{1}^{s} \in [80,110]M_{\odot}$, would still yield $P(\rm{MG})$
exceeding $\sim 0.8$. Measurements of the secondary mass only
provide $P(\rm{MG})> 0.8$ when the binary system has $m_{2}^{s}> 90
M_{\odot}$.

\begin{figure*}[!htb] 
	\centering
	\includegraphics[width=2.15\apjcolwidth]{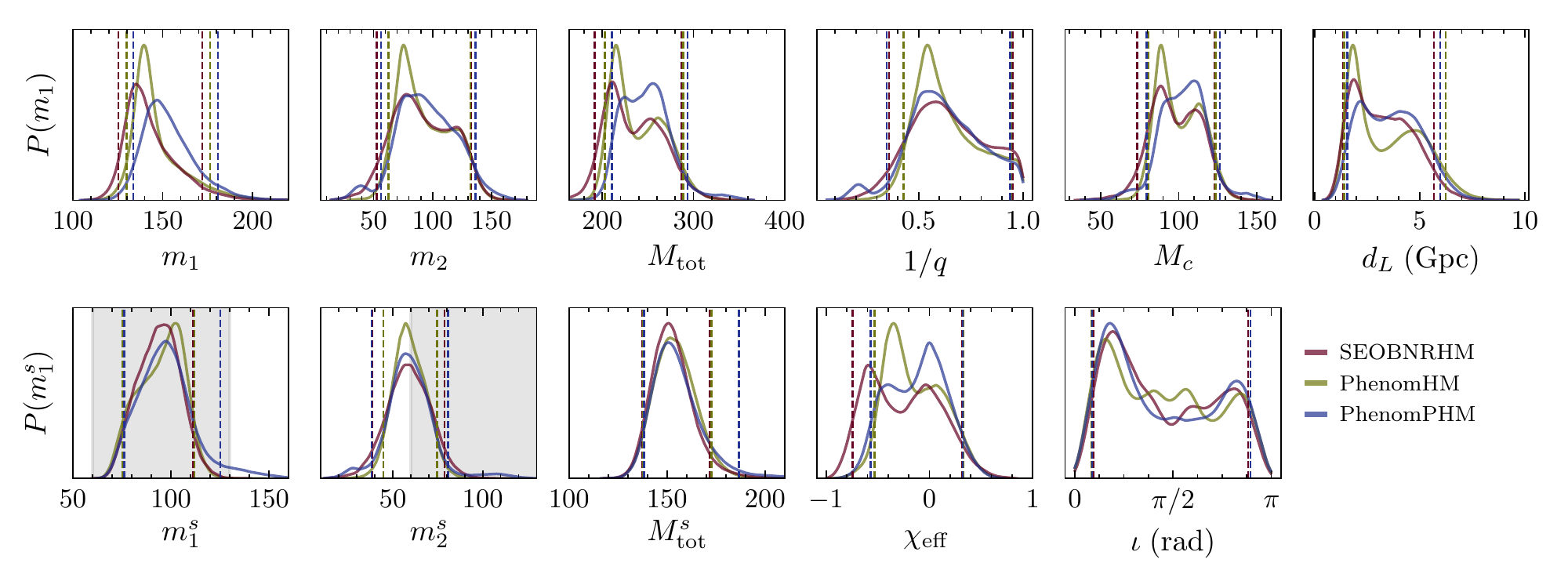}
	\vspace*{-3mm}
	\caption{Posterior distributions of the  parameters for the GW190521 event observed by LIGO and Virgo detectors~\citep{Abbott:2020tfl,Abbott:2020mjq}. The parameter $M_{f}^{s}$ and $a_{f}$ denote the mass and spin of the remnant BH, while the other parameters have been introduced in Section~\ref{sec:waveforms}. The vertical dashed lines in each plot indicate the $90\%$ credible interval for each posterior shown by the same color. The shaded region represents the BH's mass gap $[60, 130]M_{\odot}$ derived in Section~\ref{sec:c12ag} and computed at the median of the \textsuperscript{12}C($\alpha$,$\gamma$)\textsuperscript{16}O reaction rate (see Figure~\ref{fig:bh_mass_gap}). The non-precessing ${\tt SEOBNRHM}$ and ${\tt PhenomHM}$ models do not show bi-modality in the source-frame masses $m_{1,2}^{s}$, while the precessing ${\tt PhenomPHM}$ model shows additional small bumps in the secondary-mass posterior at $m_{2}^{s}\sim 30 M_{\odot}$ and $m_{2}^{s}\sim 110 M_{\odot}$. Except for the latter, there is good agreement between the results of non-precessing and precessing waveforms, and, in all cases, the 90$\%$ credible interval of the primary-mass posterior lies inside the mass gap.}
	\label{fig:Gw190521_plot}
\end{figure*}

\subsection{Re-analysis of GW190521}
\label{GW90521_analysis}

As discussed in Section~\ref{sec:intro}, \cite{Abbott:2020tfl}
reported the detection of the first IMBHB system, namely,
GW190521. Using spinning, precessing waveform models, the primary mass
of GW190521 was estimated to lie in the mass gap $[65, 120] M_{\odot}$
with probability above $99\%$, i.e., $P(\rm{MG})> 0.99$ (see
  this event in the Figure~\ref{fig:GW_events_in_mass_gap}), when
using the ${\tt NRSurPHM}$ model. When taking into account the
uncertainty in the mass gap boundaries itself (see
Equation~\ref{eq:mg_prob_discrete}), we still find that the
probability is above $98\%$. Recently, \cite{Nitz_2021}, using the
spinning, precessing waveform model {\tt
  PhenomPHM}~\citep{Pratten2020}~\footnote{In LAL this waveform
  model is denoted ${\tt IMRPhenomXPHM}$~\citep{Pratten2020}.}, which was not employed in
\cite{Abbott:2020tfl}, re-analysed GW190521. They extended the priors on
the mass ratio, observed multi-modality in the component
masses, and also found that the maximum likelihood parameter lies at high 
mass ratio, $q\sim10$ indicating that the event might be an intermediate 
mass-ratio inspiral. Using their public data, namely, the results with the prior
flat in the source-frame component masses, we find that their results
would lead to a probability that the primary mass of GW190521 is in the mass
gap of $55\%$. We shall comment again on the results of \cite{Nitz_2021} at the end of this section.

Here, we re-analyze GW190521, but mainly with the spinning,
  non-precessing ${\tt SEOBNRHM}$ and ${\tt PhenomHM}$ waveform models
  used in this work. We employ the same settings as used for the results 
  publicly released by LIGO and
  Virgo~\citep{GWOSCGW190521, 2021SoftX..1300658A,2015JPhCS.610a2021V}, except for two
  modifications: i) we extend the prior ranges in the component masses
  and the mass ratio; and ii) we use a luminosity-distance prior
  uniform in comoving volume (see Equation~\ref{eq:comov_vol}). We
  show the results in Figure~\ref{fig:Gw190521_plot}. With 
  non-precessing waveforms, we find bi-modality in the posterior
  distributions of some parameters --- for example, in the
  secondary-component mass ($m_{2}$), the luminosity distance
  ($d_{L}$), and the total mass ($M_{\rm{tot}}$). However, the
  posteriors of the source-frame component masses ($m_{1,2}^{s}$) do
  not show bi-modality. Our non-precessing analysis also shows
  that irrespective of the waveform used, the posterior probability
  for the primary mass to lie in the mass gap is $\sim 99\%$, while
  the posterior probability that the secondary mass lies in the mass
  gap is $\sim 52\%$. 

Furthermore, to contrast the results from the non-precessing waveforms 
to the precessing case, we analyze GW190521 with the ${\tt PhenomPHM}$ model. 
We note that after the paper \cite{Nitz_2021} came out, it was realized 
that the ${\tt PhenomPHM}$ model had an issue in modeling properly the merger-ringdown 
waveforms of spinning BBHs when the primary-spin is (close to) anti-aligned 
with the orbital angular momentum at merger. Here, we first perform our analysis with 
the same version of the ${\tt PhenomPHM}$ model used by \cite{Nitz_2021}, and find 
full agremeent with their results. Then, we re-analyze GW190521 with the new 
publicly released version of ${\tt PhenomPHM}$. We display the results 
in Figure~\ref{fig:Gw190521_plot}. As it can be seen, the posteriors with the 
precessing ${\tt PhenomPHM}$ waveforms show an additional peak, though small, at, e.g.,
  $q\sim 5$ in the mass ratio. The 
  secondary source-frame mass posterior, however, has two additional small bumps at $m_{2}^{s}\sim 30 M_{\odot}$ and $m_{2}^{s}\sim 110 M_{\odot}$. Besides those differences, the agreement between the posteriors
  from the precessing and non-precessing waveforms is quite good. 
  With the precessing ${\tt PhenomPHM}$ model in Figure~\ref{fig:Gw190521_plot}, we find that $P(\rm{MG})$ for the
  primary mass is $\sim 98\%$, while for the secondary mass it is $\sim 54\%$. To further understand the robustness of these findings, we are 
  currently finalizing a comprehensive analysis with other waveform models with spin-precession, namely,
  the time-domain spinning, precessing model from the EOB family~\citep{Ossokine:2020kjp}. Moreover, such an analysis has also been carried out with a new time-domain phenomenological IMR model in \cite{IMRPhenomTPHMGW190521}. With their default version of the new time domain IMR model, \cite{IMRPhenomTPHMGW190521} reports 91.6\% probability for the primary mass to be in the mass gap by simply integrating the primary mass posterior in the range $[70, 161]$. Their analysis employs a similar prior setting like ours. Our posteriors yield $\sim 99\%$ probabilities when we also integrate them in the mass gap range $[70, 161]$. Thus, the frequency domain waveforms used in this section also do a good job as far as the interpretation of the primary mass is concerned.

\section{Conclusions}\label{sec:conclusion}

In this paper, we used the spinning, non-precessing ${\tt SEOBNRHM}$ waveform model 
to estimate the precision with which the parameters of
non-precessing IMBHBs could be estimated in the upcoming LIGO-Virgo
O4 and O5 runs. We simulated IMBHBs for total detector-frame masses in the range
$M_{\rm{tot}}=[50, 500] M_{\odot}$, mass ratios $q=\{1.25, 4, 10\}$ and
component spins $\chi_{1}=\chi_{2}= [-0.8, 0.95]$ at a SNR of 20. 
We showed that for high mass-ratio binaries with relatively high
inclination ($\iota = \pi/3$), the mass of the heavier component (i.e., the primary BH) 
can be constrained with an uncertainty $\sim 11-25\%$. These precisions are much 
better than what is expected from electromagnetic observations. We also showed that 
the total source mass of IMBHBs can be constrained with uncertainties $\sim 10-30\%$,
independently of the parameters of the IMBHB systems. These results
suggest that future LIGO and Virgo observations hold the potential to
measure the mass function of IMBHs and IMBHBs, which remains an important open question 
in astrophysics.

We also focused on IMBHB systems whose component BHs
fall in the upper--stellar-mass gap, which is predicted from 
stellar-evolution theory for massive stars. We first studied 
the sensitivity of the mass-gap edges to the uncertainties in 
the most relevant nuclear rates, such as
${}^{12}\text{C}(\alpha, \gamma) {}^{16} \text{O}$, ${}^{12}\text{C} +
{}^{12}\text{C}$, and ${}^{16}\text{O} + {}^{16}\text{O}$, using  \MESA. 
We confirmed the results of \citet{farmer_2020_aa} that the boundaries of
the mass gap are dependent on the ${}^{12}\text{C}(\alpha,
\gamma) {}^{16} \text{O}$ rate, e.g., the lower edge of the mass gap
can vary between $\sim 40-90 M_{\odot}$ while the upper edge can lie
anywhere between $\sim 125-170 M_{\odot}$.% depending on the uncertainty in the ${}^{12}\text{C}(\alpha, \gamma) {}^{16} \text{O}$ rate.  
The other nuclear rates, $^{12}$C\,+$^{12}$C, $^{12}$C\,+$^{16}$O, and
$^{16}$O\,+$^{16}$O  move the BH mass gap boundary only by
$\lesssim$\,1\,\Msun.  
The main difference between our results (Figure~\ref{fig:bh_mass_gap}) and \citet{farmer_2020_aa} is at the
$\simeq$\,20\% level for the lower edge of the BH mass gap and at the $\simeq$\,5\% level for the upper edge of the BH mass gap.
This is primarily due to the increased resolution of the tabulated ${}^{12}\text{C}(\alpha, \gamma) {}^{16} \text{O}$ reaction rate 
and the increased temporal resolution of our calculations.

Having updated the boundaries of the mass gap, 
we analyzed a few IMBHB systems with component masses in the most 
probable range of the mass gap, and computed the posterior probability of them being in the mass gap based on their
single-event observations. 
We found that for asymmetric inclined IMBHBs
whose primary component mass $m_{1}^{s}$ lies in the range 
$[80,120]M_{\odot}$, the posterior probability, $P(\rm{MG})$, that the
primary mass lies in the mass gap is $\gtrsim 95\%$. For
symmetric inclined IMBHBs, the same holds when $m_{1}^{s}\in [85, 120]
M_{\odot}$. Lowering the inclination reduces the precision of the mass
measurement and hence the posterior probability of being in the mass
gap reduces. However, face-on IMBHBs with $m_{1}^{s} \in
[80,110]M_{\odot}$, would still yield $P(\rm{MG})$ exceeding $\sim
0.8$; systems with $m_{1}^{s}\sim 120 M_{\odot}$ and high spins fall
within the parameter space where bi-modality occurs (see section ~\ref{sec:bimodality}) and hence the
precision (width) of the mass posteriors decreases (increases).  The
secondary mass can also fall in the mass gap especially when the mass
ratio of the IMBHBs is near one ($q\sim 1$). However, since the
measurement of the secondary mass is relatively poor, $P(\rm{MG})>
0.9$ can be achieved only when the secondary mass lies far above the lower edge of the mass gap, $m_{2}^{s}\in [90, 120] M_{\odot}$, given
also that the inclination is higher. 

Within this context, we re-analysed the GW190521 event with spinning, non-precessing waveforms ${\tt SEOBNRHM}$ and ${\tt PhenomHM}$ and also with the latest version of (frequency-domain) precessing ${\tt PhenomPHM}$ waveform. We found that there is a very good agreement between the results of non-precessing and precessing waveforms, at least for the source frame component masses, which we are interested in. However, the precessing ${\tt PhenomPHM}$ waveform shows two additional, though really small, bumps in the secondary source-frame component mass ($m_{2}^{s}$). We showed that the $P(\rm{MG})$ for the primary mass ($m_{1}^{s}$) with the non-precessing waveforms is $\sim 99\%$ while with the precessing waveform is $\sim98\%$. The $P(\rm{MG})$ for the secondary mass is $\sim 52\%$ with the non-precessing waveforms while with the precessing waveform it increases a bit, $\sim 54\%$ because one of the additional bumps occurs in the mass gap. To complete the re-analysis of GW190521 with state-of-the-art precessing waveforms, we are further investigating this event with more accurate (when compared to numerical-relativity simulations) waveforms, notably the time-domain spinning, precessing model from the EOB family~\citep{Ossokine:2020kjp} (see also \cite{IMRPhenomTPHMGW190521}, which uses a new time-domain phenomenological model). We further note that the posterior probabilities were obtained using our default priors (see  Section~\ref{sec:priors}), notably priors flat in detector-frame component masses and uniform in comoving volume for the luminosity distance.  Employing different priors may change the results.

Furthermore, the probability $P(\rm{MG})$ is a guide to how confidently an event can be identified as being a mass-gap event, but the number should not be over-interpreted, as noise fluctuations can lead to large $P(\rm{MG})$'s for events that are not in the mass gap. The value required for an event to be confidently identified as being a mass gap event depends on the relative, and unknown, rates of events inside and outside the mass gap. To properly address the question of how many events would be required for a robust identification of a mass-gap population, would require a Bayesian model comparison between a ``mass-gap'' model and a ``no mass-gap model'', using all observed events. We leave this work for future studies. However, systems with high $P(\rm{MG})$ are more likely to be robustly identified as mass-gap events.

We note that in this analysis we have not included the KAGRA detector, as it is currently uncertain at what sensitivity KAGRA will be contributing to the network during the O4 and O5 runs~\citep{Abbott:2020qfu}. Adding an additional detector of comparable sensitivity will improve parameter estimation by increasing the observed SNR and breaking parameter degeneracies. In the context of IMBH observations, we expect that KAGRA will bring a modest improvement in the precision of the luminosity distance, allowing a slightly better determination of the intrinsic mass, which could facilitate the identification of sources in the mass gap. However, such improvements will be much smaller than the uncertainties in the location of the mass gap described in this manuscript.

The inference studies in this work were limited to multipolar spinning, non-precessing quasi-circular 
waveform models, although we performed an analysis where we injected mildly precessing quasi-circular 
signals and found that the measurement uncertainties on the component masses only changed by $\sim 5\%$ 
when recovering those signals with non-precessing waveforms. 
In view also of possible multi-modal posterior distributions for IMBHB systems, 
we plan in the future to carry out a comprehensive investigation using a larger set 
of injections of spinning, precessing signals and recover them with precessing (instead of 
non-precessing) waveforms. Furthermore, it will be crucial to extend the parameter-estimation 
study to IMBHBs moving on eccentric orbits. Indeed, given the shortness of IMBHB signals, all physical effects 
need to be included to avoid misinterpreting the properties of the source \citep{CalderonBustillo:2020odh,Romero-Shaw:2020thy}. 
Recently, a few examples of multipolar spinning, non-precessing waveform models with mild 
eccentricity have been developed~\citep{Nagar:2021gss,Liu:2021pkr,Yun:2021jnh, Khalil:2021txt} that could be used for such studies.

\acknowledgements

We are grateful to Andrew Matas for providing us with comments on this manuscript. 
We thank Rob Farmer for sharing his \MESA\ version r11701 materials and his insights on our mass gap models.
We also thank Pablo Marchant for useful discussions of PISN. 
The \MESA\ project is supported by the National Science Foundation (NSF) under the Software Infrastructure for Sustained Innovation
program grants (ACI-1663684, ACI-1663688, ACI-1663696).
This research was also supported by the NSF under grant PHY-1430152 for the Physics Frontier Center 
``Joint Institute for Nuclear Astrophysics - Center for the Evolution of the Elements'' (JINA-CEE).
A.T. is a Research Associate at the Belgian Scientific Research Fund (F.R.S-FNRS).

This research has made use of data, software and/or web
tools obtained from the Gravitational Wave Open Science
Center (\href{https://www.gw-openscience.org/about/}{https://www.gw-openscience.org}), a service of LIGO
Laboratory, the LIGO Scientific Collaboration and the Virgo
Collaboration. LIGO is funded by the U.S. National Science
Foundation. Virgo is funded by the French Centre National de
Recherche Scientifique (CNRS), the Italian Istituto Nazionale
della Fisica Nucleare (INFN) and the Dutch Nikhef, with contributions by Polish and Hungarian institutes. The authors are grateful for computational resources at the AEI, specifically the Hypatia cluster where all the the computations were carried out.}
The research of R.J.D utilized resources from the Notre Dame Center for Research Computing. 
% This acknowledgment is said above, do we need it twice?
%RJD and MW were supported by the National Science Foundation through Grant No. Phys-0758100, and the Joint Institute for Nuclear Astrophysics through Grant %No. Phys-0822648 and PHY-1430152 (JINA Center for the Evolution of the Elements).
This research made extensive use of the SAO/NASA Astrophysics Data System (ADS).This material is based upon work supported by NSF’s LIGO Laboratory which is a major facility fully funded by the National Science Foundation.

\software{
\texttt{AZURE2} \citep[][\url{http://azure.nd.edu}]{azure, azure2},
\MESA\ \citep[][\url{http://mesa.sourceforge.net}]{paxton_2011_aa,paxton_2013_aa,paxton_2015_aa,paxton_2018_aa,paxton_2019_aa},
\texttt{MESASDK} 20190830 \citep{mesasdk_linux,mesasdk_macos},
\texttt{matplotlib} \citep{hunter_2007_aa}, and
\texttt{NumPy} \citep{der_walt_2011_aa}.
         }

%\appendix

\begin{table*}[tbh]
	\caption{The $95 \%$ width (in percentage) of one-dimensional marginalized posteriors scaled by the true value for injections and recoveries with ${\tt SEOBNRHM}$ waveforms. For all runs, $q=1.25$, $\iota=\pi/3$ and $\text{SNR}=20$. The spin column represents the equal component spins (i.e, $\chi_{1}=\chi_{2}$) along the orbital angular momentum $\mathbf{L}$.  The results in the parenthesis denote the corresponding face-on case (i.e., $\iota=0$).} 
	\resizebox{7.1in}{!}  {%
		\hspace*{-2cm}\begin{tabular}{ ccccccccc} 
			\toprule[1pt]
			\addlinespace[2pt]
			$\text{Spin}$& $M_{\rm{tot}}=50M_{\odot}$ & $100M_{\odot}$ & $150M_{\odot}$ & $200M_{\odot}$ & $250M_{\odot}$ & $300M_{\odot}$  & $400M_{\odot}$ & $500M_{\odot}$ \\
			\addlinespace[2pt]
			\midrule[1pt]
			\addlinespace[2pt]
			\multicolumn{9}{c}{$\Delta M_{c}/M_{c}$} \\ 
			\addlinespace[2pt]

 -0.80&	6.53	 (6.69)&	 13.76	 (14.85)&	 23.65	 (25.91)&	 38.73	 (38.79)&	 44.03	 (40.90)&	 47.66	 (44.36)&	 43.84	 (39.59)&	 35.97	 (36.33)\\	 
 -0.50&	6.08	 (6.11)&	 21.45	 (20.69)&	 22.99	 (23.35)&	 24.03	 (26.60)&	 27.97	 (30.15)&	 28.52	 (32.96)&	 22.75	 (34.83)&	 18.11	 (31.63)\\	 
 0.00&	4.20	 (4.18)&	 16.02	 (16.33)&	 20.32	 (20.85)&	 20.22	 (22.43)&	 21.07	 (22.91)&	 23.56	 (25.01)&	 23.64	 (25.95)&	 21.00	 (26.02)\\	 
 0.50&	3.33	 (3.30)&	 13.43	 (13.44)&	 20.55	 (20.62)&	 23.49	 (23.58)&	 23.93	 (24.49)&	 22.59	 (23.68)&	 22.52	 (23.53)&	 23.66	 (24.29)\\	 
 0.80&	3.08	 (3.02)&	 11.25	 (10.95)&	 19.18	 (18.55)&	 22.26	 (22.20)&	 23.64	 (24.55)&	 24.26	 (25.92)&	 21.96	 (25.76)&	 19.83	 (25.99)\\	 
 0.95&	2.34	 (2.33)&	 8.49	 (8.56)&	 16.37	 (16.82)&	 20.90	 (22.05)&	 24.29	 (25.20)&	 26.79	 (26.26)&	 30.81	 (30.68)&	 34.79	 (40.10)\\	 
			 
			\addlinespace[2pt]
			
			\multicolumn{9}{c}{$\Delta \nu/\nu$} \\ 	
			\addlinespace[2pt]

 -0.80&	13.99	 (14.95)&	 8.43	 (7.59)&	 17.36	 (17.43)&	 30.87	 (26.84)&	 34.16	 (29.24)&	 39.47	 (32.44)&	 36.78	 (28.54)&	 27.04	 (25.92)\\	 
 -0.50&	12.33	 (12.60)&	 8.02	 (7.42)&	 9.50	 (10.35)&	 11.41	 (15.06)&	 15.56	 (18.93)&	 17.12	 (22.32)&	 9.05	 (24.82)&	 4.48	 (24.23)\\	 
 0.00&	8.87	 (8.50)&	 9.14	 (8.04)&	 11.67	 (11.21)&	 12.77	 (13.82)&	 13.59	 (14.68)&	 14.86	 (17.10)&	 12.12	 (17.61)&	 7.49	 (17.78)\\	 
 0.50&	10.59	 (10.10)&	 10.28	 (9.98)&	 14.09	 (12.99)&	 16.75	 (16.13)&	 18.04	 (17.57)&	 17.50	 (17.58)&	 15.28	 (17.45)&	 13.99	 (17.17)\\	 
 0.80&	12.37	 (11.36)&	 12.23	 (11.90)&	 15.76	 (15.13)&	 18.34	 (17.92)&	 20.63	 (21.55)&	 22.24	 (23.61)&	 22.50	 (26.68)&	 19.50	 (27.49)\\	 
 0.95&	13.13	 (12.36)&	 12.22	 (11.84)&	 17.53	 (18.10)&	 21.45	 (21.85)&	 24.48	 (25.48)&	 28.69	 (28.25)&	 35.40	 (35.79)&	 41.52	 (49.53)\\	 
			
			\addlinespace[2pt]
			
			\multicolumn{9}{c}{$\Delta d_{L}/d_{L}$} \\ 
			\addlinespace[2pt]

 -0.80&	109.24	 (48.12)&	 124.30	 (51.67)&	 114.53	 (51.11)&	 114.46	 (63.14)&	 125.25	 (66.61)&	 127.59	 (71.34)&	 121.86	 (66.87)&	 117.86	 (68.45)\\	 
 -0.50&	103.65	 (48.38)&	 129.83	 (54.84)&	 131.48	 (51.71)&	 130.28	 (52.81)&	 135.20	 (54.66)&	 136.64	 (55.23)&	 124.22	 (56.76)&	 109.45	 (57.42)\\	 
 0.00&	99.20	 (49.37)&	 113.16	 (50.23)&	 114.13	 (48.00)&	 112.25	 (47.21)&	 108.44	 (45.96)&	 110.75	 (47.10)&	 107.76	 (46.74)&	 97.85	 (47.04)\\	 
 0.50&	103.31	 (48.77)&	 110.10	 (49.15)&	 110.33	 (47.05)&	 106.26	 (45.79)&	 104.54	 (46.31)&	 104.99	 (44.27)&	 102.17	 (45.73)&	 97.40	 (44.49)\\
 0.80&	103.08	 (49.66)&	 105.15	 (47.99)&	 103.31	 (44.34)&	 100.17	 (42.86)&	 104.40	 (42.41)&	 101.70	 (42.34)&	 103.96	 (42.39)&	 100.91	 (42.72)\\
 0.95&	102.05	 (47.09)&	 99.08	 (44.33)&	 88.38	 (36.40)&	 85.40	 (36.85)&	 88.10	 (39.02)&	 88.58	 (39.61)&	 88.39	 (42.79)&	 91.10	 (51.23)\\	 
			
			\addlinespace[2pt]
			
			\multicolumn{9}{c}{$\Delta m_{1}^{s}/m_{1}^{s}$} \\	
			\addlinespace[2pt]

 -0.80&	40.98	 (45.62)&	 31.75	 (30.72)&	 35.04	 (39.82)&	 39.54	 (50.16)&	 42.04	 (54.91)&	 49.19	 (60.53)&	 48.16	 (59.19)&	 38.07	 (58.29)\\	 
 -0.50&	37.89	 (41.78)&	 29.41	 (30.81)&	 30.78	 (33.11)&	 31.96	 (38.60)&	 35.49	 (45.52)&	 36.42	 (49.73)&	 33.02	 (53.81)&	 28.95	 (55.42)\\	 
 0.00&	30.70	 (32.47)&	 29.80	 (30.15)&	 30.29	 (32.31)&	 31.57	 (35.59)&	 31.46	 (37.38)&	 31.79	 (40.52)&	 28.29	 (43.06)&	 23.98	 (44.56)\\	 
 0.50&	34.68	 (36.27)&	 32.27	 (34.38)&	 33.18	 (36.65)&	 34.18	 (38.88)&	 34.82	 (41.31)&	 35.44	 (42.00)&	 33.88	 (43.33)&	 28.58	 (43.98)\\	 
 0.80&	37.22	 (38.46)&	 34.15	 (37.63)&	 34.39	 (39.69)&	 34.34	 (42.72)&	 36.02	 (47.98)&	 37.41	 (51.04)&	 38.86	 (57.85)&	 37.79	 (61.08)\\	 
 0.95&	37.64	 (39.72)&	 33.05	 (36.85)&	 32.69	 (43.14)&	 35.02	 (46.72)&	 37.59	 (52.18)&	 40.99	 (58.19)&	 48.21	 (71.12)&	 54.54	 (92.24)\\	  
			
			\addlinespace[2pt]
			
			\multicolumn{9}{c}{$\Delta m_{2}^{s}/m_{2}^{s}$} \\	
			\addlinespace[2pt]

 -0.80&	37.94	 (41.76)&	 37.66	 (39.26)&	 52.09	 (56.24)&	 63.16	 (66.60)&	 66.53	 (68.92)&	 72.07	 (71.17)&	 74.50	 (66.15)&	 69.50	 (62.66)\\	 
 -0.50&	33.88	 (37.77)&	 40.83	 (42.93)&	 45.74	 (48.60)&	 49.70	 (55.92)&	 56.81	 (60.14)&	 59.23	 (62.33)&	 49.08	 (62.41)&	 36.17	 (59.93)\\	 
 0.00&	29.69	 (34.13)&	 38.23	 (40.26)&	 45.75	 (48.84)&	 47.74	 (51.72)&	 50.02	 (52.34)&	 53.09	 (54.47)&	 50.45	 (53.06)&	 43.21	 (51.57)\\	 
 0.50&	32.82	 (36.51)&	 37.25	 (41.91)&	 45.75	 (49.90)&	 50.89	 (55.81)&	 52.98	 (57.25)&	 53.51	 (55.19)&	 51.94	 (53.01)&	 50.32	 (51.12)\\	 
 0.80&	35.37	 (38.93)&	 38.37	 (44.42)&	 45.55	 (51.92)&	 50.55	 (56.62)&	 55.09	 (60.72)&	 57.34	 (61.86)&	 59.69	 (63.35)&	 58.78	 (62.21)\\	 
 0.95&	34.17	 (37.98)&	 35.93	 (42.42)&	 41.92	 (48.82)&	 48.28	 (54.91)&	 54.21	 (61.09)&	 58.86	 (64.34)&	 65.04	 (69.39)&	 70.40	 (80.83)\\	 	 	 
			 
			\addlinespace[2pt]
			
			\multicolumn{9}{c}{$\Delta M_{tot}^{s}/M_{tot}^{s}$} \\ 
			\addlinespace[2pt]

 -0.80&	12.96	 (12.72)&	 19.25	 (16.84)&	 21.39	 (17.91)&	 22.59	 (19.26)&	 24.02	 (19.86)&	 24.56	 (19.20)&	 26.05	 (18.56)&	 25.47	 (18.92)\\	 
 -0.50&	12.53	 (12.35)&	 20.71	 (19.89)&	 23.45	 (20.43)&	 24.91	 (20.24)&	 26.94	 (20.46)&	 27.89	 (19.28)&	 28.43	 (18.05)&	 25.42	 (18.20)\\
 0.00&	11.54	 (11.66)&	 19.06	 (17.57)&	 22.15	 (19.70)&	 22.99	 (19.14)&	 24.28	 (18.83)&	 24.71	 (17.86)&	 24.89	 (17.10)&	 23.40	 (16.57)\\	 
 0.50&	13.03	 (12.68)&	 18.43	 (17.24)&	 21.77	 (19.41)&	 23.44	 (19.16)&	 24.66	 (19.09)&	 26.28	 (17.62)&	 26.56	 (17.07)&	 25.75	 (16.66)\\	 
 0.80&	13.57	 (13.25)&	 18.27	 (17.51)&	 21.49	 (18.76)&	 23.30	 (19.23)&	 25.58	 (19.08)&	 26.69	 (17.83)&	 29.45	 (17.68)&	 30.97	 (17.81)\\
 0.95&	13.72	 (13.18)&	 17.70	 (16.37)&	 18.37	 (15.27)&	 19.96	 (15.18)&	 22.76	 (16.32)&	 23.48	 (16.41)&	 25.38	 (18.74)&	 27.09	 (24.16)\\	 
			
			\addlinespace[2pt]
			
			\multicolumn{9}{c}{$\Delta \chi_{1}/\chi_{1}$} \\ 
			\addlinespace[2pt]
			
 -0.80&	57.79	 (59.56)&	 62.54	 (68.11)&	 69.60	 (84.77)&	 113.42	 (119.57)&	 125.66	 (125.30)&	 134.07	 (124.58)&	 142.31	 (131.19)&	 145.65	 (134.80)\\	 
 -0.50&	167.26	 (167.97)&	 190.27	 (193.72)&	 191.59	 (192.84)&	 200.30	 (202.10)&	 203.99	 (203.28)&	 212.54	 (207.80)&	 206.81	 (211.35)&	 199.01	 (222.16)\\	 
 0.50&	142.76	 (146.91)&	 161.14	 (165.53)&	 155.77	 (170.61)&	 164.73	 (171.53)&	 171.11	 (176.87)&	 176.26	 (180.27)&	 188.26	 (198.70)&	 187.56	 (201.88)\\	 
 0.80&	38.25	 (38.19)&	 48.82	 (50.35)&	 53.32	 (54.15)&	 54.50	 (54.87)&	 57.82	 (55.19)&	 59.03	 (57.63)&	 65.26	 (65.47)&	 71.83	 (70.79)\\	 
 0.95&	10.42	 (10.35)&	 17.02	 (17.86)&	 20.70	 (21.17)&	 21.32	 (22.72)&	 21.71	 (23.13)&	 21.20	 (22.56)&	 19.83	 (23.97)&	 20.31	 (27.11)\\	  	 
			
			\addlinespace[2pt]
			
			\multicolumn{9}{c}{$\Delta \chi_{2}/\chi_{2}$} \\ 
			\addlinespace[2pt]

 -0.80&	100.77	 (114.01)&	 79.72	 (83.05)&	 117.00	 (118.74)&	 143.12	 (145.64)&	 150.22	 (156.60)&	 163.02	 (158.71)&	 167.28	 (161.52)&	 156.74	 (166.48)\\	 
 -0.50&	207.89	 (208.63)&	 203.40	 (200.99)&	 212.48	 (213.34)&	 220.37	 (227.68)&	 242.12	 (244.35)&	 248.53	 (252.90)&	 255.91	 (260.95)&	 249.87	 (258.30)\\	 
 0.50&	207.34	 (208.59)&	 222.72	 (218.71)&	 225.93	 (225.32)&	 229.50	 (225.50)&	 234.26	 (224.77)&	 232.81	 (231.76)&	 230.68	 (237.31)&	 234.28	 (237.26)\\	 
 0.80&	98.71	 (90.01)&	 118.82	 (117.89)&	 135.50	 (131.24)&	 138.07	 (136.65)&	 138.26	 (149.40)&	 140.59	 (137.77)&	 129.21	 (146.30)&	 131.95	 (143.18)\\	 
 0.95&	42.45	 (40.38)&	 68.89	 (67.14)&	 108.41	 (112.57)&	 123.05	 (124.62)&	 120.47	 (132.93)&	 126.90	 (127.62)&	 131.70	 (136.27)&	 134.21	 (136.63)\\	 	  	 
			
			\addlinespace[2pt]
			
			\bottomrule[1pt]
			
		\end{tabular}%
	}
	\label{table:q125_table}
\end{table*}

\begin{table*}[tbh]
	\centering
	\caption{The $95 \%$ width (in percentage) of one-dimensional marginalized posteriors scaled by the true value for injections and recoveries with ${\tt SEOBNRHM}$ waveforms. For all runs, $q=4$, $\iota=\pi/3$ and $\text{SNR}=20$. The spin column represents the equal component spins (i.e, $\chi_{1}=\chi_{2}$) along the orbital angular momentum $\mathbf{L}$.  The results in the parenthesis denote the corresponding face-on case (i.e., $\iota=0$).} 
	\resizebox{7.1in}{!}{%
		\hspace*{-2cm}\begin{tabular}{ ccccccccc} 
			\toprule[1pt]
			\addlinespace[2pt]
			$\text{Spin}$& $M_{\rm{tot}}=50M_{\odot}$ & $100M_{\odot}$ & $150M_{\odot}$ & $200M_{\odot}$ & $250M_{\odot}$ & $300M_{\odot}$  & $400M_{\odot}$ & $500M_{\odot}$ \\
			\addlinespace[2pt]
			\midrule[1pt]
			\addlinespace[2pt]
			
			\multicolumn{9}{c}{$\Delta M_{c}/M_{c}$} \\ 
			\addlinespace[2pt]

 -0.80&	3.06	 (5.10)&	 24.23	 (64.35)&	 38.81	 (62.92)&	 47.22	 (68.48)&	 60.64	 (84.09)&	 63.12	 (74.90)&	 57.12	 (73.01)&	 51.07	 (72.81)\\	 
 -0.50&	2.44	 (4.06)&	 35.08	 (50.82)&	 55.56	 (62.38)&	 61.89	 (58.60)&	 62.30	 (74.33)&	 59.85	 (64.66)&	 56.14	 (74.01)&	 48.41	 (60.82)\\	 
 0.00&	1.81	 (2.93)&	 11.28	 (18.71)&	 33.00	 (73.94)&	 48.90	 (59.53)&	 51.53	 (42.28)&	 55.06	 (47.46)&	 49.16	 (45.79)&	 40.34	 (45.50)\\	 
 0.50&	1.44	 (2.22)&	 7.91	 (11.24)&	 18.25	 (27.65)&	 27.16	 (46.72)&	 31.72	 (80.82)&	 33.31	 (73.86)&	 31.89	 (41.36)&	 29.96	 (36.71)\\	 
 0.80&	1.25	 (1.82)&	 5.32	 (7.62)&	 11.20	 (17.35)&	 17.42	 (27.00)&	 21.86	 (35.60)&	 25.20	 (45.22)&	 27.15	 (75.79)&	 25.94	 (54.37)\\	 
 0.95&	1.17	 (1.36)&	 4.06	 (5.82)&	 8.83	 (12.06)&	 12.99	 (16.91)&	 15.84	 (20.42)&	 17.84	 (23.91)&	 19.32	 (34.02)&	 23.16	 (42.01)\\	 
			
			\addlinespace[2pt]
			
			\multicolumn{9}{c}{$\Delta \nu/\nu$} \\ 	
			\addlinespace[2pt]

 -0.80&	25.59	 (28.17)&	 29.72	 (57.01)&	 41.66	 (51.78)&	 51.33	 (55.62)&	 57.07	 (73.90)&	 58.96	 (65.28)&	 53.84	 (65.64)&	 49.11	 (67.59)\\	 
 -0.50&	25.95	 (34.98)&	 40.70	 (56.41)&	 47.91	 (48.07)&	 52.23	 (45.70)&	 57.14	 (66.92)&	 56.23	 (55.03)&	 53.97	 (68.97)&	 47.95	 (57.68)\\	 
 0.00&	23.03	 (36.59)&	 25.14	 (33.43)&	 34.70	 (64.77)&	 45.75	 (49.93)&	 48.53	 (32.46)&	 49.03	 (39.06)&	 48.30	 (39.64)&	 42.00	 (38.73)\\	 
 0.50&	25.20	 (35.75)&	 21.09	 (24.83)&	 21.08	 (27.57)&	 24.84	 (41.79)&	 29.09	 (80.86)&	 33.28	 (72.85)&	 35.44	 (35.61)&	 35.33	 (31.75)\\	 
 0.80&	18.84	 (24.80)&	 16.15	 (18.05)&	 14.86	 (18.50)&	 16.11	 (22.68)&	 18.42	 (31.15)&	 21.69	 (43.34)&	 25.10	 (90.22)&	 25.53	 (67.91)\\	 
 0.95&	15.18	 (16.45)&	 11.74	 (13.57)&	 13.71	 (15.90)&	 16.10	 (18.48)&	 18.62	 (21.99)&	 19.80	 (25.81)&	 20.08	 (40.85)&	 30.16	 (54.17)\\	  
			
			\addlinespace[2pt]
			
			\multicolumn{9}{c}{$\Delta d_{L}/d_{L}$} \\ 
			\addlinespace[2pt]

 -0.80&	80.23	 (26.19)&	 79.94	 (113.39)&	 85.19	 (118.92)&	 88.45	 (132.54)&	 151.00	 (163.15)&	 144.86	 (151.74)&	 118.05	 (158.19)&	 103.42	 (180.17)\\	 
 -0.50&	87.40	 (25.58)&	 123.58	 (90.76)&	 150.47	 (119.08)&	 152.48	 (114.32)&	 144.31	 (142.43)&	 132.20	 (126.29)&	 117.51	 (146.26)&	 102.48	 (143.31)\\	 
 0.00&	85.53	 (25.60)&	 81.41	 (42.10)&	 96.18	 (128.58)&	 107.68	 (107.21)&	 103.42	 (91.97)&	 106.08	 (92.65)&	 93.52	 (92.39)&	 80.20	 (98.80)\\	 
 0.50&	88.50	 (25.37)&	 80.90	 (31.63)&	 77.94	 (50.67)&	 79.40	 (74.99)&	 78.03	 (121.11)&	 75.25	 (115.71)&	 71.23	 (81.83)&	 64.99	 (76.73)\\	 
 0.80&	86.90	 (24.40)&	 80.97	 (26.64)&	 75.31	 (35.61)&	 73.42	 (46.28)&	 71.99	 (56.41)&	 70.28	 (67.01)&	 68.52	 (108.58)&	 66.31	 (86.28)\\	 
 0.95&	91.11	 (24.53)&	 81.52	 (24.88)&	 73.04	 (28.32)&	 66.87	 (32.59)&	 62.08	 (35.68)&	 59.68	 (38.67)&	 55.85	 (49.72)&	 53.80	 (58.03)\\

			\addlinespace[2pt]
			
			\multicolumn{9}{c}{$\Delta m_{1}^{s}/m_{1}^{s}$} \\	
			\addlinespace[2pt]

 -0.80&	28.61	 (30.60)&	 18.60	 (35.26)&	 17.46	 (37.51)&	 18.79	 (39.56)&	 23.28	 (48.99)&	 27.40	 (47.92)&	 26.40	 (48.43)&	 25.64	 (47.64)\\	 
 -0.50&	24.55	 (34.49)&	 24.39	 (32.05)&	 25.06	 (35.54)&	 24.97	 (36.08)&	 25.56	 (45.59)&	 25.68	 (41.53)&	 25.83	 (48.42)&	 25.85	 (46.06)\\
 0.00&	20.64	 (33.46)&	 19.94	 (25.12)&	 19.14	 (41.07)&	 20.05	 (37.02)&	 19.59	 (30.61)&	 18.69	 (34.05)&	 18.43	 (35.20)&	 16.94	 (34.39)\\
 0.50&	23.68	 (33.91)&	 21.06	 (23.13)&	 19.97	 (20.39)&	 20.07	 (21.57)&	 20.00	 (48.40)&	 20.48	 (48.48)&	 20.80	 (32.73)&	 18.87	 (30.40)\\	 
 0.80&	18.45	 (24.90)&	 18.34	 (18.24)&	 18.01	 (17.19)&	 17.80	 (16.71)&	 17.97	 (17.91)&	 18.26	 (22.75)&	 18.79	 (60.33)&	 18.89	 (50.79)\\	 
 0.95&	16.46	 (17.03)&	 15.82	 (14.01)&	 16.84	 (14.48)&	 16.90	 (15.31)&	 17.05	 (16.40)&	 17.10	 (18.18)&	 16.96	 (27.45)&	 17.88	 (37.37)\\	 
			
			\addlinespace[2pt]
			
			\multicolumn{9}{c}{$\Delta m_{2}^{s}/m_{2}^{s}$} \\	
			\addlinespace[2pt]

 -0.80&	18.69	 (20.44)&	 45.03	 (146.80)&	 69.64	 (153.41)&	 81.87	 (150.82)&	 83.61	 (169.60)&	 79.82	 (154.00)&	 70.84	 (155.98)&	 65.17	 (158.58)\\ 
 -0.50&	20.53	 (25.88)&	 59.73	 (99.74)&	 91.40	 (153.57)&	 96.89	 (143.49)&	 97.26	 (165.00)&	 91.11	 (145.36)&	 84.40	 (159.96)&	 72.80	 (145.10)\\	 
 0.00&	19.50	 (28.76)&	 28.04	 (35.48)&	 52.90	 (169.22)&	 76.55	 (155.11)&	 80.97	 (124.01)&	 81.61	 (129.71)&	 76.62	 (124.83)&	 64.40	 (122.44)\\
 0.50&	20.67	 (27.67)&	 22.16	 (21.72)&	 28.85	 (33.26)&	 38.10	 (55.15)&	 45.59	 (165.68)&	 51.95	 (168.77)&	 55.17	 (116.65)&	 55.18	 (109.47)\\	 
 0.80&	16.46	 (18.60)&	 19.14	 (15.52)&	 21.64	 (20.14)&	 25.62	 (28.14)&	 29.83	 (38.69)&	 34.78	 (53.36)&	 40.35	 (167.81)&	 39.92	 (146.18)\\	 
 0.95&	16.01	 (12.87)&	 18.15	 (12.95)&	 23.36	 (17.06)&	 27.19	 (20.32)&	 30.24	 (24.29)&	 31.58	 (27.97)&	 32.24	 (42.50)&	 39.73	 (55.75)\\	 
			
			\addlinespace[2pt]
			
			\multicolumn{9}{c}{$\Delta M_{tot}^{s}/M_{tot}^{s}$} \\ 
			\addlinespace[2pt]

-0.80&	19.86	 (20.83)&	 15.32	 (18.74)&	 17.34	 (18.87)&	 18.85	 (19.11)&	 19.63	 (19.91)&	 19.72	 (19.39)&	 18.53	 (19.57)&	 18.23	 (19.92)\\	 
 -0.50&	16.36	 (22.69)&	 16.73	 (16.98)&	 21.36	 (20.11)&	 21.45	 (19.54)&	 21.09	 (19.25)&	 20.73	 (18.17)&	 20.92	 (18.55)&	 21.39	 (18.63)\\	 
 0.00&	13.90	 (21.34)&	 14.07	 (15.35)&	 15.61	 (17.69)&	 17.98	 (18.57)&	 18.08	 (18.44)&	 17.90	 (16.97)&	 16.82	 (16.23)&	 14.80	 (16.03)\\	 
 0.50&	16.09	 (21.84)&	 16.26	 (15.65)&	 17.38	 (14.79)&	 19.10	 (14.29)&	 19.81	 (15.09)&	 20.16	 (16.45)&	 19.97	 (15.94)&	 18.11	 (15.28)\\	 
 0.80&	13.25	 (16.61)&	 15.32	 (12.96)&	 16.36	 (12.98)&	 17.08	 (13.02)&	 17.66	 (12.65)&	 18.30	 (12.67)&	 18.84	 (21.24)&	 18.74	 (18.72)\\	 
 0.95&	13.21	 (11.76)&	 14.63	 (10.25)&	 16.24	 (10.73)&	 16.89	 (11.19)&	 17.27	 (11.41)&	 17.40	 (11.86)&	 16.95	 (15.60)&	 16.54	 (20.55)\\	
			
			\addlinespace[2pt]
			
			\multicolumn{9}{c}{$\Delta \chi_{1}/\chi_{1}$} \\ 
			\addlinespace[2pt]

 -0.80&	68.73	 (57.12)&	 50.48	 (89.33)&	 49.23	 (101.19)&	 54.29	 (114.19)&	 115.62	 (119.20)&	 119.17	 (120.38)&	 112.42	 (125.58)&	 109.84	 (127.39)\\	 
 -0.50&	100.66	 (115.24)&	 122.17	 (140.52)&	 151.05	 (184.85)&	 164.38	 (190.73)&	 158.42	 (196.81)&	 150.45	 (196.82)&	 148.16	 (206.80)&	 151.68	 (208.22)\\	 
 0.50&	46.73	 (48.52)&	 46.42	 (51.89)&	 58.35	 (68.90)&	 67.49	 (81.77)&	 69.50	 (102.69)&	 69.81	 (168.79)&	 71.77	 (197.82)&	 69.04	 (211.44)\\	 
 0.80&	14.41	 (15.49)&	 14.98	 (17.20)&	 17.47	 (23.66)&	 22.27	 (29.43)&	 25.49	 (31.29)&	 26.77	 (30.52)&	 26.04	 (40.26)&	 23.95	 (62.71)\\	 
 0.95&	4.36	 (4.54)&	 4.40	 (5.49)&	 4.95	 (7.04)&	 5.51	 (8.29)&	 6.14	 (8.50)&	 6.54	 (8.38)&	 6.91	 (8.43)&	 7.15	 (8.11)\\	 	 	 
			
			\addlinespace[2pt]
			
			\multicolumn{9}{c}{$\Delta \chi_{2}/\chi_{2}$} \\ 
			\addlinespace[2pt]

 -0.80&	159.54	 (162.59)&	 152.68	 (146.97)&	 148.79	 (149.74)&	 156.49	 (161.73)&	 165.62	 (168.52)&	 171.08	 (169.93)&	 171.03	 (169.06)&	 170.25	 (174.42)\\	 
 -0.50&	262.59	 (270.20)&	 256.04	 (258.78)&	 245.29	 (243.53)&	 257.00	 (252.92)&	 264.46	 (260.38)&	 273.12	 (271.42)&	 271.56	 (266.43)&	 271.45	 (268.27)\\	 
 0.50&	266.38	 (257.87)&	 259.36	 (265.83)&	 256.61	 (268.90)&	 255.57	 (273.41)&	 245.65	 (277.63)&	 250.18	 (239.98)&	 244.83	 (248.33)&	 238.23	 (252.89)\\	 
 0.80&	143.92	 (149.29)&	 148.14	 (157.34)&	 152.42	 (165.12)&	 153.04	 (170.84)&	 151.81	 (170.30)&	 149.25	 (172.16)&	 145.39	 (173.50)&	 142.27	 (152.74)\\	 
 0.95&	67.77	 (73.63)&	 83.15	 (97.37)&	 103.08	 (124.20)&	 114.54	 (139.45)&	 121.39	 (144.29)&	 117.70	 (143.50)&	 109.63	 (147.88)&	 115.18	 (145.90)\\	  	 	 
			
			\addlinespace[2pt]
			
			\bottomrule[1pt]
			
		\end{tabular}%
	}
	\label{table:q4_table}
\end{table*}

\begin{table*}[tbh]
	\centering
	\caption{The $95 \%$ width (in percentage) of one-dimensional marginalized posteriors scaled by the true value for injections and recoveries with ${\tt SEOBNRHM}$ waveforms. For all runs, $q=10$, $\iota=\pi/3$ and $\text{SNR}=20$. The spin column represents the equal component spins (i.e, $\chi_{1}=\chi_{2}$) along the orbital angular momentum $\mathbf{L}$.  The results in the parenthesis denote the corresponding face-on case (i.e., $\iota=0$).} 
	\resizebox{6.8in}{!}{%
		\hspace*{-2cm}\begin{tabular}{ccccccccc} 
			\toprule[1pt]
			\addlinespace[2pt]
			$\text{Spin}$& $M_{\rm{tot}}=50M_{\odot}$ & $100M_{\odot}$ & $150M_{\odot}$ & $200M_{\odot}$ & $250M_{\odot}$ & $300M_{\odot}$  & $400M_{\odot}$ & $500M_{\odot}$ \\
			\addlinespace[2pt]
			\midrule[1pt]
			\addlinespace[2pt]
			
			\multicolumn{9}{c}{$\Delta M_{c}/M_{c}$} \\ 
			\addlinespace[2pt]

 -0.80&	1.83	 (3.14)&	 12.19	 (72.67)&	 83.57	 (130.25)&	 219.60	 (147.28)&	 123.69	 (136.57)&	 50.09	 (137.15)&	 30.90	 (152.80)&	 26.94	 (117.98)\\	 
 -0.50&	2.48	 (2.79)&	 10.31	 (30.93)&	 49.03	 (163.31)&	 65.12	 (122.41)&	 68.00	 (127.08)&	 66.58	 (126.37)&	 55.53	 (134.03)&	 48.18	 (105.43)\\	 
 0.00&	1.49	 (1.95)&	 4.72	 (12.00)&	 13.82	 (54.59)&	 25.68	 (129.93)&	 36.25	 (92.34)&	 38.47	 (87.81)&	 35.76	 (92.42)&	 31.84	 (72.31)\\	 
 0.50&	1.22	 (1.64)&	 3.08	 (5.80)&	 7.63	 (18.03)&	 14.83	 (42.78)&	 21.30	 (82.63)&	 26.75	 (148.46)&	 25.19	 (64.02)&	 22.69	 (62.52)\\	 
 0.80&	0.99	 (1.36)&	 2.36	 (4.26)&	 4.92	 (10.41)&	 8.16	 (17.92)&	 11.66	 (26.27)&	 16.05	 (40.96)&	 22.34	 (116.26)&	 27.49	 (136.58)\\	 
 0.95&	0.64	 (0.81)&	 1.68	 (3.20)&	 3.05	 (6.38)&	 4.31	 (9.82)&	 5.28	 (12.81)&	 6.26	 (17.34)&	 7.83	 (36.98)&	 9.82	 (62.95)\\	  
			
			\addlinespace[2pt]
			
			\multicolumn{9}{c}{$\Delta \nu/\nu$} \\ 	
			\addlinespace[2pt]

 -0.80&	28.31	 (36.62)&	 25.15	 (103.99)&	 95.19	 (136.93)&	 198.58	 (160.93)&	 120.07	 (148.38)&	 61.28	 (147.45)&	 37.00	 (174.71)&	 32.79	 (143.79)\\	 
 -0.50&	53.84	 (50.44)&	 36.63	 (55.97)&	 59.75	 (177.13)&	 73.93	 (131.25)&	 78.65	 (136.07)&	 77.11	 (135.84)&	 67.40	 (148.57)&	 60.67	 (126.30)\\	 
 0.00&	41.79	 (49.66)&	 24.27	 (36.95)&	 26.34	 (72.55)&	 35.73	 (136.06)&	 47.88	 (94.84)&	 51.30	 (91.34)&	 50.82	 (100.71)&	 51.37	 (73.80)\\	 
 0.50&	41.66	 (72.24)&	 24.14	 (35.21)&	 18.86	 (31.66)&	 20.92	 (49.91)&	 24.17	 (94.04)&	 28.02	 (188.81)&	 26.42	 (59.79)&	 25.21	 (58.16)\\	 
 0.80&	32.96	 (47.31)&	 18.67	 (26.10)&	 14.35	 (20.61)&	 13.06	 (22.18)&	 13.93	 (27.57)&	 16.08	 (43.81)&	 20.59	 (160.46)&	 25.10	 (205.87)\\	 
 0.95&	14.88	 (16.97)&	 10.56	 (13.90)&	 9.65	 (12.63)&	 9.70	 (14.14)&	 9.57	 (16.04)&	 9.91	 (20.75)&	 10.24	 (47.39)&	 11.32	 (87.60)\\	 
			
			\addlinespace[2pt]
			
			\multicolumn{9}{c}{$\Delta d_{L}/d_{L}$} \\ 
			\addlinespace[2pt]

 -0.80&	73.11	 (25.06)&	 65.60	 (154.89)&	 231.74	 (311.80)&	 779.20	 (361.17)&	 396.09	 (359.12)&	 107.18	 (374.55)&	 54.23	 (448.51)&	 36.62	 (443.06)\\	 
 -0.50&	72.74	 (25.24)&	 66.74	 (65.21)&	 104.59	 (373.35)&	 123.77	 (306.22)&	 125.95	 (326.62)&	 124.00	 (329.27)&	 103.44	 (383.32)&	 93.69	 (363.36)\\	 
 0.00&	74.63	 (24.48)&	 63.99	 (34.19)&	 59.96	 (103.53)&	 66.07	 (286.97)&	 75.47	 (224.17)&	 74.46	 (215.62)&	 68.73	 (240.12)&	 60.79	 (224.08)\\	 
 0.50&	78.26	 (24.18)&	 68.11	 (26.58)&	 62.94	 (40.44)&	 62.63	 (76.50)&	 62.80	 (139.50)&	 64.71	 (270.90)&	 57.53	 (163.59)&	 54.76	 (160.13)\\	 
 0.80&	76.81	 (23.87)&	 69.44	 (24.91)&	 64.64	 (29.68)&	 62.21	 (38.27)&	 58.95	 (47.82)&	 59.18	 (66.28)&	 58.95	 (188.20)&	 63.14	 (243.47)\\	 
 0.95&	76.34	 (21.90)&	 69.09	 (24.14)&	 64.40	 (24.37)&	 59.30	 (27.87)&	 56.62	 (30.44)&	 52.64	 (33.92)&	 48.95	 (58.46)&	 49.68	 (96.24)\\	 
			
			\addlinespace[2pt]
			
			\multicolumn{9}{c}{$\Delta m_{1}^{s}/m_{1}^{s}$} \\	
			\addlinespace[2pt]

 -0.80&	24.98	 (34.85)&	 13.70	 (26.92)&	 13.84	 (34.98)&	 34.42	 (39.55)&	 14.71	 (40.06)&	 9.44	 (40.44)&	 5.93	 (47.07)&	 4.87	 (43.39)\\	 
 -0.50&	47.98	 (41.47)&	 25.17	 (24.95)&	 20.87	 (43.39)&	 19.63	 (37.73)&	 18.56	 (38.88)&	 17.75	 (39.69)&	 17.34	 (42.20)&	 17.63	 (39.50)\\	 
 0.00&	29.90	 (35.83)&	 17.34	 (22.64)&	 13.98	 (21.61)&	 12.42	 (38.02)&	 12.10	 (32.02)&	 11.74	 (31.30)&	 11.83	 (33.34)&	 12.78	 (28.82)\\	 
 0.50&	28.84	 (44.84)&	 18.27	 (26.33)&	 14.86	 (21.08)&	 13.98	 (18.85)&	 13.02	 (21.13)&	 12.80	 (49.78)&	 11.73	 (23.69)&	 11.33	 (23.00)\\	 
 0.80&	24.22	 (31.94)&	 15.52	 (21.40)&	 13.52	 (17.59)&	 12.97	 (16.11)&	 12.57	 (14.71)&	 12.73	 (14.82)&	 11.96	 (39.63)&	 11.48	 (58.17)\\	 
 0.95&	12.57	 (13.74)&	 11.36	 (12.32)&	 11.96	 (11.08)&	 12.39	 (11.18)&	 12.77	 (10.91)&	 12.55	 (11.05)&	 12.80	 (18.07)&	 13.71	 (31.41)\\	 
			
			\addlinespace[2pt]
			
			\multicolumn{9}{c}{$\Delta m_{2}^{s}/m_{2}^{s}$} \\	
			\addlinespace[2pt]

 -0.80&	14.96	 (18.74)&	 22.50	 (123.95)&	 130.83	 (346.02)&	 403.11	 (353.57)&	 166.47	 (323.09)&	 75.83	 (311.73)&	 45.95	 (359.18)&	 41.34	 (330.96)\\	 
 -0.50&	27.13	 (24.63)&	 23.81	 (51.06)&	 73.94	 (421.33)&	 98.56	 (336.48)&	 101.92	 (328.70)&	 98.20	 (318.44)&	 80.91	 (338.26)&	 69.20	 (322.25)\\	 
 0.00&	20.90	 (23.89)&	 15.61	 (24.89)&	 24.92	 (78.17)&	 40.38	 (367.98)&	 55.33	 (293.82)&	 58.81	 (277.45)&	 56.95	 (283.17)&	 54.70	 (249.85)\\	 
 0.50&	21.44	 (34.68)&	 14.79	 (17.94)&	 15.59	 (24.15)&	 22.33	 (51.83)&	 29.00	 (105.19)&	 35.61	 (363.76)&	 33.97	 (217.18)&	 31.52	 (210.56)\\	 
 0.80&	18.00	 (23.10)&	 13.30	 (12.75)&	 13.55	 (13.28)&	 15.24	 (19.26)&	 17.35	 (27.98)&	 21.03	 (46.00)&	 26.23	 (173.56)&	 30.84	 (347.95)\\	 
 0.95&	10.50	 (9.01)&	 11.94	 (8.56)&	 13.74	 (9.74)&	 15.11	 (12.40)&	 16.27	 (15.10)&	 16.88	 (19.80)&	 17.79	 (43.84)&	 19.40	 (77.47)\\	 
			
			\addlinespace[2pt]
			
			\multicolumn{9}{c}{$\Delta M_{tot}^{s}/M_{tot}^{s}$} \\ 
			\addlinespace[2pt]

 -0.80&	21.50	 (30.09)&	 11.63	 (19.21)&	 12.20	 (19.37)&	 15.49	 (18.28)&	 11.32	 (18.91)&	 10.10	 (18.95)&	 7.65	 (20.27)&	 7.20	 (20.36)\\	 
 -0.50&	41.23	 (35.53)&	 21.40	 (20.46)&	 19.88	 (19.78)&	 20.04	 (19.07)&	 18.74	 (18.07)&	 17.64	 (17.82)&	 15.60	 (18.67)&	 15.06	 (18.98)\\	 
 0.00&	25.48	 (30.52)&	 14.95	 (18.87)&	 12.22	 (16.48)&	 11.54	 (18.14)&	 11.09	 (17.21)&	 10.44	 (16.04)&	 9.66	 (16.23)&	 9.58	 (15.87)\\	 
 0.50&	24.46	 (37.56)&	 15.99	 (22.54)&	 13.41	 (18.22)&	 13.03	 (16.10)&	 12.67	 (14.84)&	 12.97	 (18.99)&	 11.90	 (14.01)&	 11.40	 (14.05)\\	 
 0.80&	20.81	 (27.06)&	 13.89	 (18.54)&	 12.56	 (15.57)&	 12.42	 (14.39)&	 12.23	 (13.27)&	 12.53	 (12.34)&	 11.98	 (21.73)&	 11.60	 (27.24)\\	 
 0.95&	11.20	 (11.86)&	 10.79	 (10.84)&	 11.68	 (9.89)&	 12.21	 (10.13)&	 12.73	 (9.80)&	 12.58	 (9.62)&	 12.91	 (13.58)&	 13.74	 (22.32)\\

			\addlinespace[2pt]

			\multicolumn{9}{c}{$\Delta \chi_{1}/\chi_{1}$} \\ 
			\addlinespace[2pt]

 -0.80&	53.72	 (68.55)&	 33.18	 (64.17)&	 51.25	 (97.30)&	 125.65	 (110.64)&	 96.31	 (114.81)&	 54.26	 (116.72)&	 33.30	 (118.64)&	 28.77	 (116.68)\\ 
 -0.50&	137.67	 (126.95)&	 81.82	 (87.06)&	 105.28	 (167.70)&	 112.60	 (186.32)&	 117.31	 (187.19)&	 123.53	 (192.92)&	 137.55	 (194.92)&	 140.38	 (197.06)\\	 
 0.50&	36.01	 (54.61)&	 25.42	 (33.53)&	 25.75	 (39.63)&	 32.19	 (54.47)&	 37.47	 (64.16)&	 41.37	 (124.42)&	 36.68	 (206.31)&	 38.38	 (216.32)\\	 
 0.80&	7.57	 (7.38)&	 7.90	 (11.10)&	 8.45	 (14.28)&	 9.75	 (17.74)&	 11.05	 (19.76)&	 13.72	 (20.67)&	 16.18	 (22.60)&	 18.33	 (49.87)\\	 
 0.95&	2.39	 (3.19)&	 2.09	 (3.95)&	 1.95	 (4.50)&	 1.82	 (5.46)&	 1.74	 (5.25)&	 1.72	 (5.00)&	 1.71	 (5.34)&	 1.88	 (5.57)\\	 
			
			\addlinespace[2pt]
			
			\multicolumn{9}{c}{$\Delta \chi_{2}/\chi_{2}$} \\ 
			\addlinespace[2pt]

 -0.80&	170.25	 (172.21)&	 170.05	 (170.02)&	 171.44	 (161.66)&	 170.63	 (167.33)&	 170.89	 (173.95)&	 174.03	 (173.59)&	 169.61	 (173.30)&	 169.13	 (173.63)\\	 
 -0.50&	271.18	 (276.01)&	 272.48	 (277.39)&	 275.31	 (260.76)&	 270.29	 (269.17)&	 273.27	 (271.54)&	 276.18	 (272.73)&	 277.80	 (275.65)&	 278.82	 (271.75)\\	 
 0.50&	281.33	 (264.94)&	 268.18	 (274.95)&	 261.38	 (270.31)&	 252.48	 (273.60)&	 245.65	 (274.66)&	 246.45	 (266.32)&	 250.92	 (253.02)&	 269.66	 (256.35)\\	 
 0.80&	151.86	 (124.07)&	 143.18	 (164.41)&	 140.32	 (163.40)&	 142.57	 (166.37)&	 141.21	 (168.46)&	 145.89	 (170.90)&	 144.07	 (177.06)&	 148.05	 (161.97)\\	 
 0.95&	86.72	 (104.97)&	 82.09	 (103.09)&	 96.21	 (117.98)&	 100.94	 (136.49)&	 104.87	 (144.37)&	 106.65	 (143.13)&	 105.35	 (146.09)&	 108.04	 (146.61)\\	 	 
			
			\addlinespace[2pt]
			
			\bottomrule[1pt]
			
		\end{tabular}%
	}
	\label{table:q10_table}
\end{table*}

\clearpage

\bibliographystyle{aasjournal}
\bibliography{references,references_fxt,references_rjd}

\end{document}